\documentclass[12pt]{article}
\usepackage{amsfonts, amsmath, amssymb, amsthm, bbm, bm, color, enumitem, forloop, framed, graphicx, lscape, mathrsfs, pdfpages, rotating, sectsty, setspace, threeparttable, tocloft,multirow,verbatim,rotating,caption,subcaption}
\usepackage{algorithm,algpseudocode}
\usepackage{hyperref}
\usepackage[T1]{fontenc}
\usepackage{microtype}
\usepackage{tikz}
\usetikzlibrary{matrix,chains,positioning,decorations.pathreplacing,arrows}

\usepackage[round]{natbib}  
\usepackage[title,titletoc]{appendix}

\hypersetup{pdfborder = {0 0 0},colorlinks=true,linkcolor=blue,citecolor=blue}

\newtheorem{theorem}{Theorem}
\newtheorem{corollary}{Corollary}

\newtheorem{lemma}{Lemma}
\newtheorem{assumption}{Assumption}

\numberwithin{equation}{section}

\theoremstyle{definition}
\newtheorem{remarkTmp}{Remark}
\newenvironment{remark}
	{\medskip  \begin{remarkTmp} 	}
	{ 
		\renewcommand{\qedsymbol}{$\clubsuit$}
		\qed 
		\end{remarkTmp} 
	}

\newtheorem{egTmp}{Example}

\allowdisplaybreaks[1]

\setenumerate[1]{label=\bf(\arabic*)}
\setenumerate[2]{label=\bf(\roman*)}

	\newcommand{\sumi}{\sum_{i=1}^n}

	\DeclareMathOperator*{\argmin}{arg\,min}

	\DeclareMathOperator*{\supess}{ess\,sup}

	\newcommand{\E}{\mathbb{E}}
	
	\newcommand{\V}{\mathbb{V}}
	\renewcommand{\P}{\mathbb{P}}

	\newcommand{\R}{\mathbb{R}}

	\newcommand{\One}{\mathbbm{1}}

    \def\ddefloop#1{\ifx\ddefloop#1\else\ddef{#1}\expandafter\ddefloop\fi}
    \def\ddef#1{\expandafter\def\csname c#1\endcsname{\ensuremath{\mathcal{#1}}}}
    \ddefloop ABCDEFGHIJKLMNOPQRSTUVWXYZ\ddefloop
    \def\ddef#1{\expandafter\def\csname s#1\endcsname{\ensuremath{\mathsf{#1}}}}
    \ddefloop ABCDEFGHIJKLMNOPQRSTUVWXYZ\ddefloop
	\def\ddef#1{\expandafter\def\csname b#1\endcsname{\ensuremath{\bm{#1}}}}
	\ddefloop aAbBcCgGhHjJMpPqQrRsStTwWxXzZyY\ddefloop
	
	\newcommand{\bmu}{\boldsymbol{\mu}}
	\newcommand{\btheta}{\boldsymbol{\theta}}	
		
	\newcommand{\bbeta}{\boldsymbol{\beta}}

	\newcommand{\bpsi}{\boldsymbol{\psi}}	
	\newcommand{\bPsi}{\boldsymbol{\Psi}}	
	\newcommand{\bL}{\boldsymbol{\Lambda}}

    \newcommand{\mtrue}{\mu^\star}
	\newcommand{\bmtrue}{\bmu^\star}
	\newcommand{\ttrue}{\theta^\star}
	\newcommand{\bttrue}{\btheta^\star}
	\newcommand{\atrue}{\alpha^\star}
	\newcommand{\btrue}{\beta^\star}
	\newcommand{\bbtrue}{\bbeta^\star}
	\newcommand{\dtrue}{\delta^\star}
	
	\newcommand{\ztrue}{\zeta^\star}
	\newcommand{\bLtrue}{\bL^\star}
	\newcommand{\ltrue}{\lambda^\star}

    \newcommand{\btstar}{\tilde{\bt}}
    \newcommand{\tstar}{\tilde{t}}

    \newcommand{\ropt}{r_{\rm{opt}}}

	\newcommand{\mhat}{\widehat{\mu}}
	\newcommand{\bmhat}{\bm{\widehat{\mu}}}
	\newcommand{\that}{\widehat{\theta}}
	\newcommand{\bthat}{\bm{\widehat{\theta}}}
	\newcommand{\Psihat}{\widehat{\Psi}}
	\newcommand{\bPsihat}{\bm{\widehat{\Psi}}}
	
	\newcommand{\ahat}{\widehat{\alpha}}
	\newcommand{\bhat}{\widehat{\beta}}
	\newcommand{\bbhat}{\widehat{\bbeta}}
	\newcommand{\lhat}{\widehat{\lambda}}
	\newcommand{\bLhat}{\widehat{\bL}}

	\newcommand{\dy}{d_{\bY}}  
	\newcommand{\dx}{d_{\bX}}  
	\newcommand{\dc}{d_c}  
	\newcommand{\dt}{d_{\bT}}  
	\newcommand{\dmu}{d_{\bmu}}  
	\newcommand{\dtheta}{d_{\btheta}}  

\begin{document}

\title{
	Deep Learning for Individual Heterogeneity 
	\thanks{
         We thank Whitney K. Newey for helpful comments and suggestions, as well as participants at several seminars. Janani Sekar and Kirill Skobelev provided outstanding research assistance.
    }
}

\author{
	Max H. Farrell\thanks{Department of Economics, UC Santa Barbara} \and Tengyuan Liang\thanks{Booth School of Business, University of Chicago} \and
	Sanjog Misra\thanks{Booth School of Business, University of Chicago} \and
}


\date{\today}

\maketitle

\begin{abstract} 
 This paper integrates deep neural networks (DNNs) into structural models to increase flexibility and capture rich heterogeneity while preserving interpretability. Economic (or scientific or domain-restricted) structure and machine learning are complements in empirical modeling, not substitutes: DNNs provide the capacity to learn complex, nonlinear heterogeneity, while the structure ensures the estimates remain interpretable and suitable for decision-making and policy analysis. We start with a standard parametric structural model and then enrich its parameters into fully flexible functions, which are estimated using a DNN with the model structure built in. We illustrate our framework with an application to demand estimation in consumer choice. We show that by enriching a demand model we can capture rich heterogeneity exploit it to create personalized pricing. Optimization is not possible without structure, but cannot be heterogeneous without machine learning. The same lessons apply to precision dosing, adaptive treatment, educational testing, and other targeting settings. We provide theoretical justification for our proposed methodology: nonasymptotic bounds and a novel and general influence function for feasible inference via double machine learning, so that the latter can be easily applied in numerous new contexts. These results may be of interest in other contexts as they generalize prior work.
\end{abstract}

\bigskip


\setcounter{page}{0}
\thispagestyle{empty}

\newpage
\doublespacing
\addtocontents{toc}{\protect\setcounter{tocdepth}{-1}}

\section{Introduction}
	\label{sec:intro}

Structural models are designed to obey theoretical and domain-specific restrictions emanating from the discipline. As a consequence, when taken to data, estimates and inferences from structural models are scientifically interpretable and directly useful for decisions, counterfactuals, and policy making. Crucially, in many decision problems, the structure of the model and the discipline it imposes afford meaningful interpolation and extrapolation which, in turn, enables the construction of counterfactuals and ultimately optimization. 

Structural models are, however, potentially incomplete. Theory often does not specify every aspect of the framework we need to conduct empirical analysis. One critical element, and our focus in this paper, is the specification of heterogeneity. Even in cases where the presence of heterogeneity is implied by scientific reasoning, the form and type is often unknown. As such, researchers must choose, or search for, a specification for heterogeneity that respects the structural assumptions of the model, as well as the practical considerations imposed by the data. 

Often, the choice of the manner in which heterogeneity is modeled reflects the underlying objectives of the researcher. There are scenarios where one wishes to merely ``control'' for heterogeneity, since the key constructs of interest are not directly tied to it. In these cases, heterogeneity is treated as a nuisance parameter (or function). More recently, however, there has been a push to ``exploit'' heterogeneity by constructing individualized or personalized policies. Here, heterogeneity itself becomes a key quantity of interest. In these cases, we also will need to predict heterogeneity (for a new observation), and, as such, usual control-type methods (e.g. fixed effects) become less relevant. Targeting and personalization must be done using observables.

We argue that both the discipline of structure and the flexibility of machine learning (ML) are essential for heterogeneity-dependent constructs such as personalization, targeting, and counterfactuals. Both have their strengths and weaknesses, but, by combining them appropriately, one can use the strengths of each to remedy the shortcomings of the other. On one hand, theory and structure make optimization feasible and reliable, but do not dictate heterogeneity patterns. On the other hand, ML cannot learn structure with finite data (as demonstrated below). Therefore, ML cannot take the place of domain knowledge, theory, and constraints, but flexibly learning patterns and heterogeneity is precisely the strength of modern ML. Thus, while ML cannot replace structural modeling, it can be useful to augment and enrich structure. We argue that structure and machine learning are \emph{complements} in empirical modeling, not \emph{substitutes}. 

We show how to embed machine learning, in the form of modern deep neural networks (DNNs), into structural, scientific, domain-restricted statistical models. Our approach begins with a structural model imposed by the researcher. The model relates the outcomes $\by_i$ to the covariates of interest $\bt_i$ (e.g., treatments or policy-relevant variables) and depends on parameters of interest $\btheta$, which are to be estimated from the data. We suppose this structural model is captured by a loss function $\ell(\by_i, \bt_i, \btheta)$. The structural model encodes restrictions and constraints on the data generating process and the parameters $\btheta$ have direct economic interpretations. Our main focus is economic modeling, but structural models are common in many contexts, such as structural causal models where $\bt_i$ is a dose or time-varying exposure.

We then enrich the model by changing the parameters $\btheta$ into \emph{parameter functions} $\btheta(\bx)$, which are fully flexible functions of a vector of observed characteristics $\bx_i$. The new model is then $\ell(\by_i, \bt_i, \btheta(\bx_i))$. See Figure \ref{fig:arch-general}.   This adds flexibility while maintaining all the structure and meaning of the original model. The incompleteness of theory is reflected in treating $\btheta(\bx_i)$ as an unknown function. This is exactly where the strength of ML is exploited, and how it complements economic structure. Either one alone is insufficient: structure without ML would lose the richness and fail to capture important patterns in the data, while naively applying ML without structure would lose the interpretability and the utility in policy/decision making. As an aside, our framework nests some prior approaches to controlling for heterogeneity, but that is not our focus.

The structural model allows for optimization, but the ML-enriched structure allows for individual-level optimization (i.e. for each unique $\bx$). This is useful for decision making and policy analysis at the individual level, answering ``who gets what'' questions regarding targeting and ``what who gets'' questions of personalization. The former is useful, for example, in deciding to which type of individuals we allocate a scarce resource or treatment, while the latter focuses on the assignment of different treatments to each individual.

\begin{figure}
	\centering
    \scalebox{0.9}{			\def\layersep{2cm}
			\begin{tikzpicture}[shorten >=1pt,->,draw=black!50, node distance=\layersep]
			    \tikzstyle{every pin edge}=[<-,shorten <=1pt]
			    \tikzstyle{neuron}=[circle, fill=black!25, minimum size=20pt, inner sep=0pt]
			    \tikzstyle{input neuron}=[neuron,  minimum size=25pt, fill=blue!50];
			    \tikzstyle{hidden neuron}=[neuron, fill=gray!50];
			    \tikzstyle{parameter neuron}=[neuron, minimum size=40pt, fill=green!50];
			    \tikzstyle{output neuron}=[neuron, minimum size=25pt, fill=red!50];
			
			    \tikzstyle{annot} = [text width=4em, text centered]
			
				\node[input neuron] (I-1) at (0,-1.8) {$x_1$};
				\node[input neuron] (I-2) at (0,-3) {};
				\node[input neuron] (I-3) at (0,-4.2) {$x_{\dx}$};
				
			    \foreach \name / \y in {1,...,6}
			        \path[yshift=0.5cm]
			            node[hidden neuron] (Ha-\name) at (\layersep,-\y cm) {};
			
			    \foreach \name / \y in {1,...,6}
			        \path[yshift=0.5cm]
			            node[hidden neuron,right of=Ha-\name] (Hb-\name) at (\layersep,-\y cm) {};
			
				\node[yshift=15pt, parameter neuron, right of=Hb-2] (P-1) { \ $\that_1(\bx)$ \ };
				\node[parameter neuron, right of=I-2, xshift=4cm] (P-2) { \  \ };
				\node[yshift=-15pt, parameter neuron, right of=Hb-5] (P-3) { \ $\that_{\dtheta}(\bx)$ \ };

			    \node[output neuron, right of=Hb-2, xshift=\layersep] (Y) { \ $\by$ \ };
   			    \node[output neuron, right of=Hb-5, xshift=\layersep] (T) { \ $\bt$ \ };
			    \node[output neuron, pin={[pin edge={->}]right:$\ell\big(\by,\bt, \bthat(\bx)\big) $}, right of=I-2, xshift=6cm] (O) {$\ell$};
			
			    \foreach \source in {1,...,3}
			        \foreach \dest in {1,...,6}
			            \path (I-\source) edge (Ha-\dest);
			            
			    \foreach \source in {1,...,6}
			        \foreach \dest in {1,...,6}
			            \path (Ha-\source) edge (Hb-\dest);

			    \foreach \source in {1,...,6}
			        \foreach \dest in {1,2,3}
			            \path (Hb-\source) edge (P-\dest);
			
			        
				\path (P-1) edge (O);
				\path (P-2) edge (O);
				\path (P-3) edge (O);
				\path (Y) edge (O);
				\path (T) edge (O);
				
			    \node[annot,above of=Ha-1, xshift=\layersep/2, node distance=1cm] {Hidden layers};
			    \node[annot,above of=Ha-1, node distance=1cm] (hal) {};
			    \node[annot,above of=Hb-1, node distance=1cm] (hbl) {};
			    \node[annot,left of=hal] {Inputs};
			    \node[annot,right of=hbl] (pl) {Parameter layer};
			    \node[annot,right of=pl] {Model layer};
			
			\end{tikzpicture}}
	\caption{{\bf Structural deep learning.} A schematic depiction of the deep neural network architecture for estimating the parameter functions $\btheta(\bx)$ in the structural economic model $\ell(\bY, \bT, \btheta(\bX))$ defined in \eqref{eqn:first stage}.}
	\label{fig:arch-general}
\end{figure}

We introduce a novel yet simple structural deep learning approach to estimate the parameter functions. The key to this approach is the architecture depicted in Figure \ref{fig:arch-general}. Neural networks allow the structure of the model to be directly encoded in the estimation through the network architecture. The expressive power of DNNs comes from the hidden layers. In standard approaches this flexibility is used to learn regression functions or form predictions, which determine both the target and the loss functions. In our approach, the hidden layers are directed through a \emph{parameter layer} so that the power of ML is used entirely for learning the parameter functions. These then enter the \emph{model layer} according to the structural model in order to optimize the loss. The required change to the architecture is intuitive and simple to implement. The neural network optimizes the parameter functions to minimize the same structural loss, not prediction loss. (See Figure B.1 in the online appendix for a direct comparison to prediction.) This formalizes exactly how ML and structure are complementary. The ML enriches the economic model, filling in gaps left by theory. Simultaneously the economics aids the implementation of ML to estimate structural objects with meaning and interpretation. Our method enables learning any parameters, such as coefficients, variances, elasticities, et cetera, as rich, heterogeneous functions.

For more intuition, consider optimization of the network in Figure \ref{fig:arch-general}. Neural network estimation relies on using gradients to find optima, proceeding by back propagation through the network. Our idea is simply to structure the final layers of this network, as shown in Figure \ref{fig:arch-general}, optimizing the loss through the parameter functions, instead of optimizing prediction directly, but the computation is the same. This intuition also goes the other direction: parametric structural estimation would be optimized in exactly the same way, but the gradients would stop at the parameter layer. Our idea simply extends this through additional layers to add heterogeneity. \looseness=-1

We theoretically justify this part of the methodology by proving nonasymptotic bounds, and implied convergence rates, for the structured deep neural networks (Theorem \ref{thm:dnn rates}). This result generalizes \cite{Farrell-Liang-Misra2021_Ecma}. These rates depend only on the dimension of the heterogeneity $\bx_i$, because the relationship of $\bt_i$ to $\by_i$ is not learned from the data. Our bounds apply to many settings of interest, requiring only mild conditions on the loss function $\ell(\cdot)$ and standard smoothness requirements, and thus may be of interest outside structural modeling.

We obtain valid inference by applying the double machine learning (DML) methodology of \cite{Chernozhukov-etal2018_EJ}, where the requisite orthogonal score is obtained from a novel and general influence function calculation (Theorem \ref{thm:normality}). Our insight here exploits the fact that we explicitly enrich a parametric model to obtain a two-step semiparametric setting, and in this case we show that \emph{ordinary} derivatives characterize the influence function, just as in the parametric case. The influence function is thus known for a wide class of models and parameters, and need not be newly derived in each case. The only thing required is the value of these derivatives at the data points, not as functions per se. These may be known in advance or, importantly for practice, obtained using automatic differentiation built into neural network estimation without any human derivations. Collectively, this means we can deliver the influence function, and thus obtain valid inference via DML, in more contexts than previously possible. For example, in our empirical application the influence function is not available in closed form, but inference is feasible. This hopefully lessens the barriers to obtaining post-ML inference in practice.

We illustrate our methods and results by revisiting and extending the analysis of \cite{Bertrand-etal2010_QJE}. The data is from a large scale field experiment run on behalf of a financial institution in South Africa. Consumers were sent marketing material for short-term loans where a number of features of the advertising content and the interest rate offered were randomized. As in the original paper, we employ a binary choice model, one of the workhorse models in applied economics, but we enrich the model to capture heterogeneity. We then model demand as a function of prices and marketing efforts, and once the demand function is estimated, we compute optimal prices as a function of characteristics (third degree price discrimination) and conduct inference on quantities such as counterfactual profits. This involves applying DML to objects not available in closed form (the solution to a fixed point problem), which illustrates the generality of our inference method. Though pricing is an economics and marketing illustration, the same ideas apply to precision medicine, educational interventions, and other targeting contexts.

Our work touches on several areas, to which we cannot hope to do justice in a few paragraphs. We will give an overview here, with some further discussion is given throughout the paper. First, at a broad methodological level, our work is related to the recent interest in integrating ML into economic research. A large part, if not all, of this work has used ML to learn regression functions or make predictions. ML is often a substitute for nonparametric regression, and succeeds where classical methods fail due to complexities or limitations of the data. Outside of pure prediction, causal inference is often the end goal, but our work, being motivated by structure, extends to a breadth of contexts in economics and beyond.

On the theoretical front, the structured deep learning approach we propose, and the accompanying nonasymptotic bounds, connect our work to the recent literature studying the statistical properties of ML methods. Our theoretical results build directly upon and generalize \cite{Farrell-Liang-Misra2021_Ecma}. Conceptually, \cite{Athey-Tibshirani-Wager2019_AoS} and \cite{Foster-Syrgkanis2023_AoS} may be the most closely related, as both focus on quantities other than prediction functions, use models beyond regression, and utilize orthogonal scores as key ingredients. The contexts and results are different, as we focus on deep learning and semiparametric inference. \cite{Athey-Tibshirani-Wager2019_AoS} study random forests and nonparametric inference, while \cite{Foster-Syrgkanis2023_AoS} is concerned with risk bounds under weak conditions and greater generality.

The use of influence functions explicitly with the goal of obtaining valid inference under weaker conditions is not new, whether for classical nonparametrics \cite{Cattaneo2010_JoE} or machine learning \cite{Farrell2015_arXiv}. Our work contributes most directly to the recent literature applying DML \cite{Chernozhukov-etal2018_EJ}. DML combines sample splitting with an orthogonal score, which an influence function provides. Our contribution here is the derivation of a generic influence function that covers a broad class of models and can be computed automatically. We share this goal with the ``auto-DML'' method (see Remark \ref{rem:auto DML}); our approach is different and has relative strengths and weaknesses.

The rest of the paper is organized as follows. Section \ref{sec:structure} shows a simple, motivating example demonstrating why neither structural models nor machine learning individually suffice. Our framework for structured deep learning and subsequent inference is described in Section \ref{sec:framework}. We then apply the method to data in Section \ref{sec:application}. Section \ref{sec:theory} gives the theoretical results and Section \ref{sec:conclusion} concludes. The online appendix has proofs, a simulation study, discussion of related contexts and examples, and treats in detail enriched generalized linear models.

\section{The Importance of Structure}
	\label{sec:structure}

Most modern applications of personalized policy require the ability to recover heterogeneous responses to treatment(s) from data and the facility to use those in an optimization framework to obtain counterfactuals. While the former can be done via ML, the latter requires that the estimated response functions obey certain domain-specific restrictions. As such, our choices are to either learn these constraints from data or impose them via structure. We demonstrate, with a simple example, that learning structure from finite data can be impossible but that structure is invaluable for decision making problems. To make the point clear we ignore heterogeneity and rely on the reader to extrapolate to the fact that the lessons here become even more important when recovering heterogeneity or personalization, because this creates greater demands on data. Our key point is that structure imposes appropriate (e.g., economic) restrictions and constraints, and consequently data is used more efficiently to learn the heterogeneity. 

The value of structure in economics is well understood. Indeed, the half-century old Lucas Critique is a (major) example of the value of structural thinking. With the power of ML, there is a renewed temptation to learn everything from data alone. This view has its origins in the ML literature, which is focused on prediction, but has seeped into other contexts. \cite{Zadrozny2004_ICML} writes, in the context of classification, that the ML community is ``interested in the predictive performance of the model and not in making conclusions about the underlying mechanisms that generate the data.'' So successful is ML, and deep learning in particular, at prediction tasks, that it can be tempting to assume that any problem can be tackled with accurate enough predictions. But in contrast with those contexts, where prediction is sufficient, decision making cannot proceed without the ``underlying mechanisms'', that is, without structure and science. The danger in naively applying ML in decision making is that \emph{economic} assumptions and structure are replaced with the \emph{statistical and computational} assumptions and structure of estimators. 

To illustrate, we will use the data of \cite{Bertrand-etal2010_QJE}, which is analyzed in Section \ref{sec:application}. This context is ideal because, by focusing on a subset of the data, we find that this is simultaneously a straightforward demand estimation problem and straightforward prediction task, making both structural modeling and ML well-motivated. The context is demand for a short-term loan given its interest rate, and our goal is to estimate demand as a function of price and then derive the optimal interest rate offer. For 40,507 individuals we observe the binary outcome $y_i \in \{0,1\}$ indicating a loan application decision and the policy relevant variable $t_i = r_i$ giving the interest rate offered. The good being ``purchased'' is the loan and its price is captured by the interest rate. The interest rate $t_i = r_i$ is randomized, so there are no endogeneity concerns, and is furthermore continuous, taking on 46 unique values. With 40,507 observations on two variables, this is a tailor-made classification task, and the naive ML view would hold that if conversion for any interest rate offered can be predicted accurately, then demand function can be recovered, and finally the interest rate can be optimized. However, as we will see, neglecting the ``underlying mechanisms that generate the data'' will lead to failure. If the unstructured approach fails in this context, adding heterogeneity is only more difficult.

\begin{figure}[h]
    \begin{subfigure}[t]{0.32\columnwidth}
        \centering	
        \caption{Random Forests}
        \includegraphics[trim={0cm 0cm 0cm 1.5cm}, clip, scale=.28]{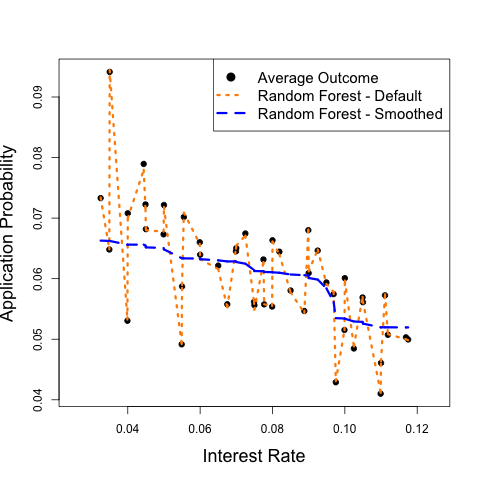}
    \end{subfigure}
    \begin{subfigure}[t]{0.32\columnwidth}
        \centering	
        \caption{Neural Networks}
        \includegraphics[trim={0cm 0cm 0cm 1.5cm}, clip, scale=.28]{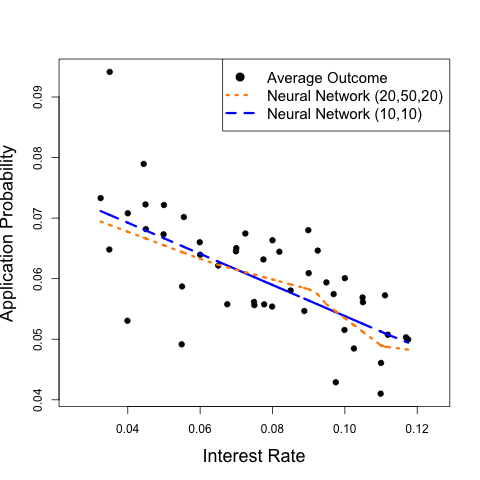}
    \end{subfigure}
    \begin{subfigure}[t]{0.32\columnwidth}
        \centering	
        \caption{Structural Logit}
        \includegraphics[trim={0cm 0cm 0cm 1.5cm}, clip, scale=.28]{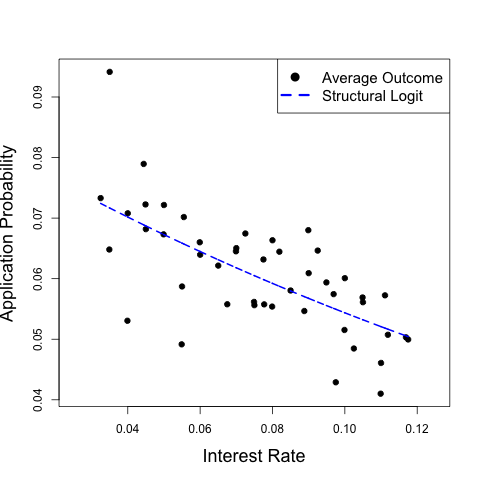}
    \end{subfigure}\\[.5 ex]
    \begin{subfigure}[t]{0.32\columnwidth}
        \centering	
        \caption{Extrapolating to [0,20]\%}
        \includegraphics[trim={0cm 0cm 0cm 1.5cm}, clip, scale=.28]{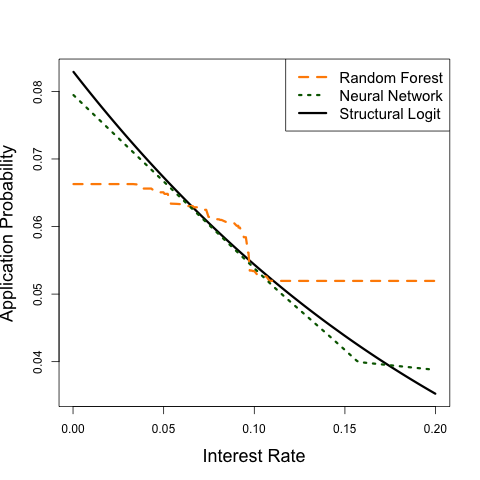}
    \end{subfigure}
    \begin{subfigure}[t]{0.32\columnwidth}
        \centering	
        \caption{Extrapolating to [0,200]\%}
        \includegraphics[trim={0cm 0cm 0cm 1.5cm}, clip, scale=.28]{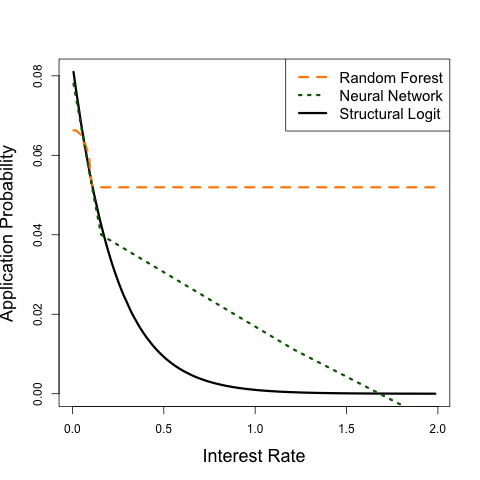}
    \end{subfigure}
    \begin{subfigure}[t]{0.32\columnwidth}
        \caption{Implied Revenue Functions}
        \centering	
        \includegraphics[trim={0cm 0cm 0cm 1.5cm}, clip, scale=.28]{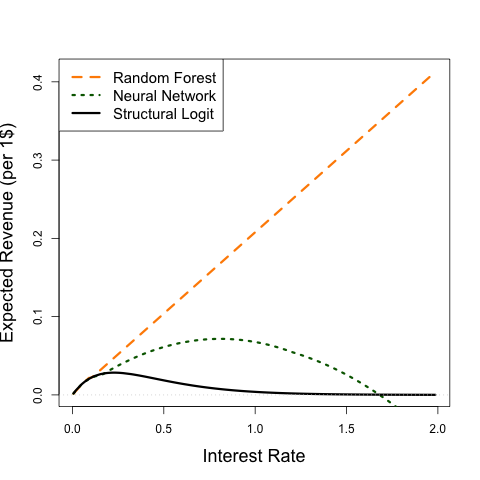}
    \end{subfigure}
    \caption{{\bf Structural modeling and machine learning for demand functions.} Panels (a), (b), and (c) show estimated demand functions using random forests, neural networks, and a structural binary choice model, respectively, using the data of \cite{Bertrand-etal2010_QJE}. Panels (d) and (e) show the extrapolated demand functions and (f) the implied revenue functions.}
    \label{fig:simple eg}
\end{figure}

Figure \ref{fig:simple eg} shows three different approaches to estimating the demand function in this data. In each case, the dots show the empirical application rate (average purchasing decision) at each offered interest rate. First we consider two ML approaches which treat this as purely a predication/classification task, and are thus based on the model $\P[ Y \!=\! 1 \mid R \!=\! r] = \eta(r)$, 
for an unknown function $\eta(r)$ to be nonparametrically estimated. Panel (a) shows two random forest estimates using the {\tt ranger} package in {\sf R}. The dotted orange line uses the default settings. This appears to be undersmoothed and thus the blue dashed line forces a smoother fit using an ad-hoc restriction on the tree depth. It is easy (by eye) to reject the default random forest as an unreasonable estimate of the demand function, although it should be noted that even this judgment requires economic theory, not statistical/ML criteria which do not encode that curve should be smooth or downward-sloping. Panel (b) shows two neural network fits. The dotted orange line uses three hidden layers with 20, 50, and 20 nodes, respectively, for a flexible fit. Note that despite the richness, the fit does not appear to be overly complex. The blue dashed line uses two layers with 10 nodes each for a smoother fit. Finally, panel (c) applies the workhorse structural model in industrial organization, namely a linear randomly utility model with a logistic errors, so that $\P[ Y \!=\! 1 \mid R \!=\! r] = G\left(\ttrue_1 + \ttrue_2 r \right) := [1+\exp\left(-\left[\ttrue_1 + \ttrue_2 r\right]\right)]^{-1}$, where $G(u)$ is the logit function and $\ttrue_1$ and $\ttrue_2$ are the intercept and slope, respectively.

The smoothed-out forest and both neural networks produce curves that could reasonably be demand functions. However, as we continue with the pricing problem, it becomes clear how different are these fits. Panels (d) and (e) show the extrapolated estimated demand functions of the smoothed forest, the (10,10) neural network, and the binary choice model to interest rates of zero to 20\% (panel (d)) and zero to 200\% (panel (e)). Panel (f) then shows the implied revenue for each demand function estimate. While this does not include costs, it is clear what each estimate implies for an optimal price (see Section \ref{sec:application} for more realistic profit optimization).

The forest fit is perhaps the most striking. Demand is completely flat for any higher interest rate, and therefore revenue grows without bound, and so the optimal price is infinite. This is both mechanical and a general phenomenon: a forest is an average of trees, which are piecewise constant, and therefore all extrapolation is based on a flat line, no matter the context or data. The neural network closely resembles the binary choice model on the support of the data, but when we extrapolate we see that the invisible complexities of the network yield very different demand and revenue curves. In contrast to both of these, the structural model gives well-behaved demand and revenue curves and yields a reasonable optimal price. 

To conclude this example, several remarks are in order. (i) Students of all fields are taught never to extrapolate a statistical fit outside the support of the data, and here we see why in dramatic fashion. Note that the same logic applies to interpolation. However, our goal of price optimization \emph{requires} inter- and extra-polation, which by definition is based on assumptions and not the data. (ii) One can dismiss this example with the argument that no reasonable decision maker sets an infinite price or that the demand/revenue functions look ``obviously wrong'', but again, these conclusions come from using economic structure to inform the ML, not the other way around. (iii) The lessons in this example are unrelated to statistical uncertainty. Different data would yield different fits, but the issues would persist, or perhaps manifest differently (except the forest, which always yields infinite prices, and thus has zero statistical uncertainty). (iv) Finally, it is worth noting that structural reasoning is commonly used in ML to improve prediction tasks. The structure of image recognition tasks is directly encoded into convolutional neural networks. The transformer architecture is behind the recent success of large language models. In sum, structure at its core means constraints and restrictions, and this example shows that statistical structure may not be appropriate or sufficient for decision making.

\section{Embedding Deep Learning in Structural Models}
	\label{sec:framework}

\subsection{Enriched Structural Model}
	\label{sec:model}

We now turn to our approach to enrich structural models. We describe our structured deep neural network architecture, which bakes the model into the ML, and how second-step inference can be done using DML. The prior section shows that ML is not suitable for learning structure, which is by definition the opposite of flexibility. Our goal here is to use ML for what it is good at, recovering complex heterogeneity, while retaining all the advantages of the structural model. 

The starting point is a standard parametric structural model, described by
\begin{equation}
    \label{eqn:theta parametric}
    \bttrue = \argmin_{\btheta \in \bm{\Theta}} \E \left[ \ell(\bY,\bT,\btheta) \right],
\end{equation}
where the loss function $\ell(\by, \bt, \btheta)$ encodes the researcher's economic restrictions on how the outcomes $\bY \in \R^{\dy}$ relate to the variables of interest $\bT \in \R^{\dt}$, depending on parameters $\btheta \in \bm{\Theta} \subset \R^{\dtheta}$.\footnote{
{\it Notation.} Vectors and matrices will be written in boldface. Capital letters are used for population random variables; lower case for realizations. The expectation operator with respect to the true data generating process is denoted $\E[\cdot]$. True values use a superscript $\star$. The $L_2$ norm for a function $g(\bx)$ is $\| g \|_2 = \E[g(\bX)^2]^{1/2}$.
}
The variables $\bT$ can be randomized or not and can be continuous, discrete, or mixed; the model need not be causal. In the first step the structural parameters $\btheta$ are estimated from the data by solving the empirical analogue of \eqref{eqn:theta parametric} over the appropriate parameter space $\bm{\Theta}$.

In a second step, inference is conducted on an object of the form
\begin{equation}
    \label{eqn:mu parametric}
    \E[\bH(\bX, \btheta, \btstar)],
\end{equation}
for a known function $\bH$, that depends on the parameters, covariates, and possibly some value of interest $\btstar$ for the policy relevant variables. The function $\bH$ can be the parameters themselves, quantities such as marginal effects, elasticities, measures of surplus, and can encompass optimization and other such operations. For example, $\btstar$ can be an optimal price, as in Section \ref{sec:application}. The definition of $\bH$ can subsume $\btstar$, but it is expositionally useful to make explicit.

Other than restricting to per-observation losses and smooth functions (so that derivatives exist), the combination of Equations \eqref{eqn:theta parametric} and \eqref{eqn:mu parametric} encompass a wide variety of two-step models, including many M- and Z- estimation problems, such regression models, quasi/pseudo-likelihoods, or generalized estimating equations and accompanying second step objects like marginal effects, average elasticities, and other economic quantities.

The positives of structural modeling are widely appreciated, and demonstrated above. The model incorporates theory-based restrictions and constraints. This means that the estimated parameters, or transformations thereof, are interpretable and directly useful in policy analysis, decision making, and the formation of counterfactuals. The two-step nature of the problem reflects exactly these types of analyses, where second step counterfactuals are obtained after estimating primitives of the model in the first stage. It also means that the data is used more efficiently, since structure implies restrictions, which will be crucial when embedding ML.

The major drawback of \eqref{eqn:theta parametric} is rigidity: it does not allow for any heterogeneity in the relationship between $\bY$ and $\bT$. We want to allow for heterogeneity in a way that is flexible but maintains the economic structure. This could be for robustness, as results will be biased and subsequent decisions or analyses will be erroneous if heterogeneity exists but is neglected. This type of concern has been the focus of much recent research, particularly in program evaluation. Further, as mentioned earlier, capturing and exploiting heterogeneity is key in modern targeting and personalization contexts, which is also an active area of research.

Our approach recasts the parameters $\btheta$ as \emph{parameter functions} $\btheta(\bX)$ that are fully flexible in observed covariates $\bX \in \R^{\dx}$. That is, we assume the true first stage structural model is 
\begin{equation}
    \label{eqn:theta flm2}
    \bttrue(\cdot) = \argmin_{\btheta \in \cF} \E \big[ \ell\big(\bY,\bT,\btheta(\bX)\big) \big],
\end{equation}
for a function class $\cF$, which obeys standard restrictions (Section \ref{sec:dnn theory}). Equation \eqref{eqn:theta flm2} is a specific, though quite general, formulation of nonparametric M-estimation \cite{Gallant-Nychka1987_Ecma}. Crucially all the economic structure is maintained: whatever the interpretation of the parameter $\btheta$, the same holds for $\bttrue(\bx)$ for individuals of ``type'' $\bX = \bx$. These are not ``nuisance'' functions. The view, particularly common in the realm of inference after ML, is that first step functions are literally a nuisance, i.e. something annoying that must be dealt with, but not of interest. We object to this view: in many applications the learned heterogeneity is actually the most interesting part, and because the $\bttrue(\bx)$ are interpretable functions, using them is straightforward.

The structure of the model is crucial for interpretation and decision-making, as illustrated above. This is common in structural modeling. Robustness to misspecification varies from case to case, but is a potential concern in applications that must be balanced against the need for structure to make policies or decisions. Although it is beyond the scope of this work, our theory could be used as a building block for a specification test. For example, one could consider testing the enriched linear (i.e., varying-coefficient) model $\E[Y \mid \bt, \bx] = \bttrue(\bx)'(1,\bt')'$ against the fully nonparametric alternative $\E[Y \mid \bt, \bx] = g(\bx,\bt)$. Developing such theory would be an interesting extension that we leave to future research. However, it is important to remember that such flexibility can be detrimental, as in Sections \ref{sec:structure} and \ref{sec:application}, even when supported by the data. Structure is not typically diagnosed with the data: structure comes from economic theory.

To complete the framework, the second step parameter of interest is correspondingly enriched:
\begin{equation}
    \label{eqn:mu flm2}
    \bmtrue = \E\big[\bH\big(\bX, \bttrue(\bX), \btstar\big)\big].
\end{equation}
To keep exposition simple we focus on the case where the parameter of interest of this form. Replacing this step with generalized method of moments is straightforward but notationally cumbersome (Remark \ref{rem:gmm}). Equations \eqref{eqn:theta flm2} and \eqref{eqn:mu flm2} together define a broad class of two-step semiparametric settings, matching the generality of \eqref{eqn:theta parametric} and \eqref{eqn:mu parametric}. Appendix \ref{appx:glm} and \ref{appx:examples} show some examples, both familiar and new. Finally, note that $\bmtrue$ is defined using the true $\bttrue(\bX)$, but for some decision problems this may not be appropriate (Remark \ref{rem:implementation uncertainty}).

\subsection{Structural Deep Learning}
	\label{sec:dnn}

To estimate the parameter functions $\bttrue(\bx)$ we solve the empirical analogue of \eqref{eqn:theta flm2} where minimization is over a class of \emph{structural} deep neural networks, i.e. those with architecture shown in Figure \ref{fig:arch-general} as discussed above. We define 
\begin{equation}
    \label{eqn:first stage}
    \bthat(\cdot) = \argmin_{\btheta \in \cF_{\textsc{dnn}}} \frac{1}{n} \sumi \ell(\by_i,\bt_i,\btheta(\bx_i)),
\end{equation}
where $\cF_{\textsc{dnn}}$ is the class of deep neural networks that encodes not only the overall architecture, but also tuning parameters such as the width and depth, and other features of the network. We focus on the standard fully connected feedforward neural network with ReLU activation for the hidden layers, but our ideas apply to other cases. To save space, we will not review deep learning basics here. Recent textbook treatments include \cite{Roberts-Yaida-Hanin2022_book} for theory and \cite{Keydana2023_book} for implementation. Theorem \ref{thm:dnn rates} in Section \ref{sec:theory} gives the convergence of these $\bthat(\cdot)$.

Our structural deep learning approach can be applied to any model/loss $\ell(\by, \bt, \btheta(\bx))$ and is easy to implement. Only a few lines of code need be changed relative to standard implementations to enforce the structural model and use the power of the network to learn the parameter functions rather than optimize the loss directly. This ``structural compatibility'' is one argument in favor of DNNs. There are several further reasons why neural networks are well suited to this task. (1) Perhaps most obviously, deep learning is a state of the art method exhibiting great expressive power, the capacity for more inputs, and the ability to use novel data such as images or text. (2) Practically, neural networks handle discrete covariates seamlessly without affecting the convergence rate or the implementation. While in theory including discrete covariates does not impact the rate for many nonparametric estimators, obtaining these estimates requires special care and custom methods. (3) Deep learning is built on automatic differentiation, which allows us to obtain the necessary influence function computationally for free for any $\bmtrue$ from \eqref{eqn:mu flm2} based on any first step \eqref{eqn:theta flm2}, making DML more widely applicable.

DNNs are not the only method that could be used to recover the parameter functions, nor do we claim any formal optimality property. The economic model holds globally, and we wish to match this in estimation, as it is important to learn counterfactual quantities in the second step, but other methods have the same property, including series methods such as splines or polynomials and basis-function penalization methods like ridge regression and lasso. All these estimators impose the structural model globally and in a computationally simple way. This may be one reason why series methods were often used in classic nonparametric M estimation (see \cite{Gallant-Nychka1987_Ecma} for pioneering early work and \cite{Chen2007_handbook} for further theory). But these methods lack advantages (1), (2), and (3) from the prior paragraph. Classical methods often do not work for modern applications with more than a couple covariates (we do not treat the diverging dimension case). Selection methods require pre-specification of bases and interaction effects.

Although it is not always as straightforward, it is possible to use ``local'' methods to learn the value of the parameter functions at a point, i.e. $\bttrue(\bx)$ for some $\bx$, and much recent work has been done in this area. \cite{Fan-Zhang2008_SII} discusses kernels while \cite{Zeileis-Hothorn-Hornik2008_JCGS,Athey-Tibshirani-Wager2019_AoS,Nekipelov-Novosad-Ryan2019_WP,Chatla-Shmueli2020_JCGS} use trees and random forests; all share our goal of learning non-prediction functions. \cite{Athey-Tibshirani-Wager2019_AoS} and \cite{Foster-Syrgkanis2023_AoS} are the most recent closest antecedents to our work, as both move ML away from prediction and both use orthogonal scores as a key ingredient. \cite{Athey-Tibshirani-Wager2019_AoS} studies random forests in a similar class of problems to \eqref{eqn:theta flm2}. Random forests also handle higher dimensional inputs, discreteness, and flexible interactions, but are less common for novel data types. The forests of \cite{Athey-Tibshirani-Wager2019_AoS} are probably the closest substitute for neural networks in our setting. The goal in \cite{Athey-Tibshirani-Wager2019_AoS} is inference on the nonparametric object $\bttrue(\bx)$ at a point, which is different than our two-step problem. Influence functions are used, though in the estimation step to aid with tree splitting, rather than for inference per se. \cite{Foster-Syrgkanis2023_AoS} is focused on risk minimization, as opposed to inference, but consider a similar class of models and use orthogonal scores to obtain improved properties. Their estimation target more closely resembles our two step problem, though a key innovation in their case is performing this in one step, rather than estimating the primitives and then studying different counterfactual questions. Their first and second step parameters can also be more general objects than ours.

\subsection{Inference}
	\label{sec:inference}

For inference on $\bmtrue = \E[\bH(\bX, \bttrue(\bX), \btstar)]$, we apply and contribute to the double machine learning (DML) method \cite{Chernozhukov-etal2018_EJ}. DML is a now-common, generic method for obtaining valid inference that combines two ingredients: sample splitting and a Neyman orthogonal score. Asymptotic normality then follows under weak conditions on first step estimators; typically only $L_2$ rates are need. This is important for ML first steps, as less is known of their properties compared to classical nonparametrics. 

Our contribution is to show that an orthogonal score for $\bmtrue$ is automatically available for any combination of $\ell(\cdot)$ and $\bH(\cdot)$ in Equations \eqref{eqn:theta flm2} and \eqref{eqn:mu flm2} provided their ordinary derivatives exist. We give the generic form of this score and discuss how it can be automatically computed, even when it cannot be derived in closed form. A key insight is that the score need not be known as a function per se, we only require its values at the data points. This makes it easy to deploy, because $\ell(\cdot)$ and $\bH(\cdot)$ are defined by the researcher, and the rest can proceed automatically. Our goal and results here are closest to the series of work on ``auto-DML'' (Remark \ref{rem:auto DML}). In applications restricted to familiar settings, known scores can be used (or are easy to calculate). For example, our simulation study (Appendix \ref{appx:simuls}) considers the case of the average treatment effect to verify the automatic differentiation approach. However, this is not always the case. In our empirical application (Section \ref{sec:application}) the required derivatives are not known a priori and, moreover, are not available in closed form. This highlights the utility of our ideas in real world data.

Though more general, orthogonal scores often stem from influence functions, and this is the case for our work. To state the result, we first define relevant derivatives. Let $\bH_{\btheta}(\bx, \cdot; \btstar)$ be the $\dmu \times \dtheta$ Jacobian of $\bH$ with respect to $\btheta$. Let $\bm{\ell}_{\btheta}(\by, \bt, \cdot)$ be the gradient of $\ell$ with respect to $\btheta$ and $\bm{\ell}_{\btheta\theta}(\by, \bt, \cdot)$ be the matrix of second derivatives. These are \emph{ordinary} derivatives, not functional derivatives, and are thus computable automatically, even if they are evaluated at the value of the function $\bttrue(\bx)$.\footnote{To be precise, $\bm{\ell}_{\btheta}(\by, \bt, \btheta(\bx))$ is the $\dtheta$-vector of first derivatives with respect to the parameter, evaluated at the number $\btheta(\bx)$, as in $\bm{\ell}_{\btheta}(\by,\bt, \btheta(\bx)) = \left. \partial \ell\big(\by,\bt, \bb \big) / \partial \bb \right|_{\bb = \btheta(\bx)}$.
The Jacobian $\bH_{\btheta}(\bx, \cdot; \btstar)$ is similar. Then, $\bm{\ell}_{\btheta\btheta}(\by,\bt, \btheta(\bx))$ is the $\dtheta \times \dtheta$ matrix of second derivatives, with $\{j, m\}$ element given by $[\bm{\ell}_{\btheta\btheta}(\by,\bt, \btheta(\bx))]_{j,m} = \left.  \partial^2 \ell\left(\by,\bt, \bb \right) / \partial b_{j}\partial b_{m} \right \vert_{\bb = \btheta(\bx)}$, where $b_{j}$ and $b_{m}$ are the respective elements of the place-holder $\bb$. The use of standard differentiation in these contexts has been used in some prior work, though to our knowledge not paired with automatic differentiation to obtain feasible inference in such a broad set of models.
} 
For a generic $\btheta(\bx)$, define $\bL(\bx; \btheta) = \E [  \bm{\ell}_{\btheta\btheta}(\by,\bt, \btheta(\bx)) \mid \bX = \bx ]$ and $\bLtrue(\bx) = \bL(\bx; \bttrue)$. Then an influence function that applies to any combination of enriched model \eqref{eqn:theta flm2} and parameter of interest \eqref{eqn:mu flm2} is $\bpsi(\by,\bt,\bx, \bttrue, \bLtrue) - \bmtrue$, with
\begin{equation}
    \label{eqn:if flm2}
    \bpsi(\by, \bt, \btheta, \bL) =  \bH\left(\bx, \btheta(\bx); \btstar\right)   - \bH_{\btheta}(\bx, \btheta(\bx ); \btstar) \bL(\bx; \btheta)^{-1} \bm{\ell}_{\btheta}(\by,\bt, \btheta(\bx)).
\end{equation}
Here $\bL(\bx; \btheta)$ is the ``other'' nonparametric object that generally arises in the correction term. The propensity score is perhaps the most familiar example, and our requirements on $\bLtrue(\bx)$ mirror that case exactly. Beyond two-step inference, influence functions are used in other contexts where our result may be useful (see Remark \ref{rem:other uses}).

To build intuition, note that \eqref{eqn:if flm2} has the same form as its parametric counterpart, but appropriately generalized. For the parametric two step inference problem of \eqref{eqn:theta parametric} and \eqref{eqn:mu parametric}, the influence function is $\bH\left(\bx, \btheta; \btstar\right)  - \bH_{\btheta}(\bx, \btheta; \btstar) \E [  \bm{\ell}_{\btheta\btheta}(\by, \bt, \btheta)]^{-1} \bm{\ell}_{\btheta}(\by, \bt, \btheta)$ \cite[\S 6]{Newey-McFadden1994_handbook}. Equation \eqref{eqn:if flm2} is the same, and equally general, but enriched and hence conditional on $\bX = \bx$. This tight connection helps with implementation, because if the original structural model is understood by the researcher, so is the enriched version. If the derivatives are known, they can be used in the enriched model. Identification in the parametric case requires nonsingularity of $\E [  \bm{\ell}_{\btheta\btheta}(\by, \bt, \btheta)]$, and in the enriched version this must hold for all ``types'' $\bx$. This is often a matter of assuming a positive conditional variance, rather than marginal variance. Appendix \ref{appx:glm} and \ref{appx:examples} give examples.

With the structural deep learning and influence function in hand, obtaining second-step point estimates and standard errors is done using DML \cite{Chernozhukov-etal2018_EJ}. DML is now standard, so we keep the description brief (see \cite{Ahrens-Chernozhukov-Hansen-Kozbur-Schaffer-Wiemann2026_JEL} for an introduction). See Algorithm \ref{alg:crossfitting} for an overview. First, the data is divided into $K$ disjoint subsets denoted by $\cI_k$, each with $|\cI_k|$ observations (assumed proportional to $n$) and complement $\cI_k^c$. For each fold $k$, we learn $\bthat_{k}(\cdot)$ and $\bLhat_{k}(\cdot)$ using $i \in \cI_k^c$. In general $\bL(\bx; \btheta)$ depends on $\btheta(\bx)$, so we split $\cI_k^c$ and obtain $\bthat_{k}(\cdot)$ and $\bLhat_{k}(\cdot)$ on the separate subsamples to help establish validity. Then for $i \in \cI_k$, compute $\widehat{\bpsi}_i := \bpsi(\by_i,\bt_i, \bthat_{k}(\bx_i), \bLhat_{k}(\bx_i))$. Once done for all $k \leq K$,  estimates of the parameter $\bmtrue$ and asymptotic variance $\bPsi = \V[\bpsi(\bY,\bT,\bX, \bttrue, \bLtrue)]$ are obtained as 
\begin{equation}
    \label{eqn:flm2 estimates}
    \bmhat = \frac{1}{n}\sumi \widehat{\bpsi}_i  		\qquad \text{ and } \qquad  		  \bPsihat = \frac{1}{n}\sumi   \left( \widehat{\bpsi}_i  - \bmhat \right)^2.
\end{equation}
Theorem \ref{thm:normality} validates \eqref{eqn:if flm2} and asymptotic inference. As is standard, we state inference results for generic first-step estimators. Our inference contribution is the novel generic influence function, and not related to deep learning per se. However, we verify the requirements using neural networks and note that automatic differentiation may be convenient for computing the terms when not known from prior work.

Sample splitting, including the potential three-way splitting, is used in general in all DML inference as a key ingredient in the proving of validity, as it allows for simple and tractable sufficient conditions. Obtaining asymptotic normality without sample splitting is possible but requires stronger conditions on the first-stage and case-specific theory (see, e.g., \cite{Newey1994_Ecma,Cattaneo2010_JoE,Cattaneo-Farrell2011_chapter,Farrell2015_arXiv,Chen-Syrgkanis-Austern2022_NIPS}). The splitting of $\cI_k^c$ can be skipped if it is known that $\bL(\bx; \btheta)$ does not depend on $\btheta$. Remark \ref{rem:lambda}, Section \ref{sec:if theory}, and the Appendix give further discussion, including proving valid the three-way splitting approach to estimating $\bLtrue(\bx)$. 

Note that throughout, particularly when paired with automatic differentiation, we do not need to know the \emph{functions} $\bH_{\btheta}(\cdot, \btheta(\cdot); \btstar)$, $\bm{\ell}_{\btheta}(\by, \bt, \btheta(\cdot))$, and $\bm{\ell}_{\btheta\btheta}(\by, \bt, \btheta(\cdot))$; it suffices to know the \emph{values} $\bH_{\btheta}(\bx_i, \bthat(\bx_i); \btstar)$, $\bm{\ell}_{\btheta}(\by_i, \bt_i, \bthat(\bx_i))$, and $\bm{\ell}_{\btheta\btheta}(\by_i, \bt_i, \bthat(\bx_i))$, which are readily available. For estimation of $\bLtrue(\bx)$, we obtain $\bm{\ell}_{\btheta\btheta}(\by_i, \bt_i, \bthat(\bx_i))$ automatically from the optimization of \eqref{eqn:first stage}, and we then nonparametrically regress these columns of data onto $\bx_i$ to obtain $\bLhat(\bx_i; \bthat)$ (elementwise). This is identical to the more standard practice of obtaining the functional derivatives, characterizing the nonparametric function and then estimating it: we are not doing numerical approximation. The same ideas apply to all other unknown functions.

\begin{algorithm}[ht]
	\caption{Structural Deep Learning -- Cross-Fitting for Estimation and Inference}
	\label{alg:crossfitting}
	\footnotesize
	\setlength{\abovedisplayskip}{3pt}
	\setlength{\belowdisplayskip}{3pt}
	\begin{algorithmic}[1]
		\Require Data $\{(\by_i, \bt_i, \bx_i)\}_{i=1}^n$; loss $\ell(\by, \bt,\btheta)$; target $H(\bx,\btheta; \btstar)$; class $\cF_{\textsc{dnn}}$; folds $K$; split rule $\mathsf{Spl}\in\{\mathrm{two},\mathrm{three}\}$.
		\Ensure Estimate $\bmhat$, estimate $\bPsihat$.
		
		\State Partition $\{1,\ldots,n\}$ into folds $\cI_1,\ldots,\cI_K$ of equal size.
		
		\For{$k=1,\ldots, K$}
			\State Let $\cI_k$ be the evaluation fold and its complement $\cI_k^c$ the auxiliary sample.
			
			\State If $\mathsf{Spl}=\mathrm{three}$, split $\cI_k^c=\cA_k\cup\cB_k$ disjointly; otherwise set $\cA_k=\cB_k=\cI_k^c$.
				
			\State Using $i\in\cA_k$, estimate $\bthat_k(\cdot) = \argmin_{\btheta \in \cF_{\textsc{dnn}}} \sum_{i\in\cA_k} \ell(\by_i,\bt_i,\btheta(\bx_i)) / |\cA_k|$.
				
			\State Using $i\in\cB_k$, estimate $\bLhat_k(\bx_i)$ by regressing $\bm{\ell}_{\btheta\btheta}(\by,\bt, \bthat_k(\bx))$ on $\bx_i$.
				
			\State For $i\in\cI_k$, compute the orthogonal score
				    	\[
				        	\widehat{\bpsi}_i := \bpsi(\by_i, \bt_i, \bthat_k(\bx_i), \bLhat_k(\bx_i)) =  \bH(\bx_i, \bthat_k(\bx_i); \btstar)   - \bH_{\btheta}(\bx, \bthat_k(\bx_i); \btstar) \bLhat_k(\bx_i)^{-1} \bm{\ell}_{\btheta}(\by_i, \bt_i, \bthat_k(\bx_i)).
				    	\]
		    	
		\EndFor
		
		\State Aggregate the fold estimates, report estimate of parameter and asymptotic variance:
			\[
				\bmhat = \frac{1}{n}\sumi \widehat{\bpsi}_i, \qquad \bPsihat = \frac{1}{n} \sumi (\widehat{\bpsi}_i - \bmhat)^2.
			\]
		
	\end{algorithmic}
\end{algorithm}

\begin{remark}[Notes on $\bLtrue(\bx)$]
    \label{rem:lambda}
    The function $\bLtrue(\bx)$ is a nuisance in the truest sense: it is required only because we use influence functions as a tool to obtain valid inference. It is always low-dimensional and consists only of regressions, not conditional densities. Inverse functions are standard in semiparametric inference, and in practice this can be difficult. The challenge is generally model- and data-specific, not specific to the choice of nonparametric/ML method. Often regularization is used, either explicitly (e.g. trimming the propensity score \cite{Ma-Wang2020_JASA}) or implicit (Remark \ref{rem:auto DML}). In some cases $\bLtrue(\bx)$ is simplified. If $\bT$ is independent of $\bX$, then $\bLtrue(\bx)$ can often be computed or estimated more simply, though it may remain a function of $\bx$ and $\bttrue(\bx)$. If $\bT$ is known to depend on a subvector of $\bX$, such as in targeting problems, this can be imposed. See Appendix \ref{appx:proofs} for estimation and Appendix \ref{appx:glm} and \ref{appx:examples} for examples.
\end{remark}

\begin{remark}[Other Second Steps]
    \label{rem:gmm}
    Our first-step correction can also be used in generalized method of moments (GMM) estimation where \eqref{eqn:mu flm2} is replaced by a set of moment conditions $\E[\bm{\widetilde{H}}(\bX, \bttrue(\bX), \bmtrue, \btstar)] = 0$ for an $\bm{\widetilde{H}}$. The correction factor then takes the form $\phi(\by,\bt,\bx, \bLtrue, \bttrue) = \bm{\widetilde{H}}_{\btheta}(\bx, \bttrue(\bx ),\bmtrue, \btstar) \bLtrue(\bx)^{-1} \bm{\ell}_{\btheta}(\by,\bt, \bttrue(\bx))$. Following \cite{Chernozhukov-etal2022_Ecma}, adding this correction to the original moments yields the orthogonal moment conditions $\E[\bm{\widetilde{H}}(\bX, \bttrue(\bX), \bmtrue, \btstar) - \phi(\by,\bt,\bx, \bLtrue, \bttrue)] = 0$. \cite{Chernozhukov-etal2022_Ecma}, extending \cite{Chernozhukov-etal2018_EJ}, show that the advantages of DML carry over to GMM based on these moments. Our methodology applies here, including the use of automatic differentiation if needed. Asymptotic normality will follow by applying \cite{Chernozhukov-etal2022_Ecma} instead of \cite{Chernozhukov-etal2018_EJ}.
\end{remark}

\begin{remark}[Auto-DML]
    \label{rem:auto DML}
    The ``auto-DML'' method shares our motivation of enabling DML-based inference without having to derive a new orthogonal score each time. This work \cite[and references therein]{Chernozhukov-etal2024_WP} exploits the fact that in many settings the correction term of the influence function is of the form $R(\bx,\bt) (\by - \E[\by \mid \bt, \bx])$, and the insight used is that the function $R(\cdot)$, the Riesz representer, obeys a certain moment condition that can be fit on the data, thus allowing the entire correction term to be estimated. In the notation of \eqref{eqn:if flm2}, the Riesz representer includes $\bH_{\btheta}(\bx, \bttrue(\bx ); \btstar) \bLtrue(\bx)^{-1}$ and generally a factor from $\bm{\ell}_{\btheta}(\by,\bt, \btheta(\bx))$. The originating motivation of our work is enriching structural models with ML to capture heterogeneity, and the influence function \eqref{eqn:if flm2} is a by-product, whereas the auto-DML method began explicitly with the goal of semiparametric inference (with regression as the first stage). The end results of both, looking only at the inference piece of our work, are methods that are roughly speaking equally widely-applicable are share the goal of avoid hand-derivation of influence functions. Each has pros and cons. An advantage of our approach is the close connection of the first and second stage. Both methods require a second nonparametric estimation of a nuisance function. Our correction term need only be estimated once for different inference targets, whereas the function $R(\cdot)$ is specific to each estimand. However, this comes at the cost of estimating the matrix $\bLtrue$, whereas auto-DML estimates vector. We estimate $\bLtrue$ then invert, whereas the inverse itself is (a piece of) the auto-DML target. That can mean more stable performance since it implicitly regularizes away small denominators. On the other hand, examining $\bLhat(\bx)$ is often a key step in the analysis and helps diagnose identification, such as examining propensity scores to evaluate overlap, and forces the researcher to be transparent about trimming or regularization \cite{Ma-Wang2020_JASA,Hetzenecker-Osterhaus2024_WP}.  The simulations (Appendix \ref{appx:simuls}) contain details in the concrete example of average treatment effects and we find similar performance. 
\end{remark}

\section{Application: Advertising and Personalized Interest Rates}
	\label{sec:application}

\subsection{Empirical Context}

In this section we use our framework to extend \cite{Bertrand-etal2010_QJE}. The data is from a large scale field experiment run on behalf of a financial institution in South Africa. Consumers were sent marketing material for short-term loans where features of the advertising content and the interest rate offered were randomized (details are left to \cite{Bertrand-etal2010_QJE}). For the purposes of our analysis we focus on the interest rate offered as the treatment variable, and denote the scalar $T = R$. We treat the characteristics of the advertising assigned to customers as covariates. We denote these as $\bX_a$, with $\dim(\bX_a) = 11$, all of which are discrete. These are used along with eight customer demographics, denoted $\bX_d$, of which two are continuous. Thus we use $\bX=\{\bX_d,\bX_a\}$ to capture heterogeneity. The outcome ($Y$) is the indicator for whether or not the consumer applied for the loan. We use a binary choice model, a workhorse models in applied economics. Other relevant variables available in the data include an indicator of loan default ($D$) and the loan amount ($L$). The full data set has $53,194$ individuals, but we limit to the high-risk customers that form the bulk of the data. As such, $n=40,507$. Of these, 2,371 have $Y=1$.

First, we estimate a binary choice model of loan application allowing for heterogeneity in the parameter vector via our structured DNNs. We use these to compute the marginal effect of interest rate. We then use the results of the model (with some additional assumptions) to construct optimal personalized interest rate offers and compute the expected profits from implementing the personalization scheme.

\subsection{Model and Implementation}

Assume that consumers have utility $u = \ttrue_1(\bx_d,\bx_a) + \ttrue_2(\bx_d) r + \varepsilon$,
where $\bttrue(\bx) = (\ttrue_1(\bx_d,\bx_a),\ttrue_2(\bx_d))'$ are the parameter functions and $r$ is a realization of the interest rate, the policy relevant variable $T=R$ here. We assume that price sensitivity only varies with individual characteristics; people's sensitivity to price does not vary with randomly selected marketing material. We further assume that $\varepsilon$ is Logistic distributed, which gives the standard Logit probabilities of response (\cite{Bertrand-etal2010_QJE} use probit). Let $\br_1 = (1,r)'$. The response is assumed to be $\P[Y \!=\! 1 \mid \bX \!=\! \bx, R \!=\! r] = G\left( \bttrue(\bx)'\br_1 \right) = [1+\exp\left(-\left[\ttrue_1(\bx_d,\bx_a) + \ttrue_2(\bx_d) r\right]\right)]^{-1}$,
where $G(u)$ is the logit function. The log-likelihood is $y \log \left( \P[Y \!=\! 1 \mid \bX\!=\! \bx, R \!=\! r] \right) + (1-y) \log \left( \P[Y \!=\! 0 \mid \bX\! =\! \bx, R \!=\! r] \right)$,
which is a heterogeneity enriched version of the standard binary choice model. The negative of this is the loss \eqref{eqn:theta flm2}. One can verify the high-level assumptions in this setting, particularly given that the binary choice model is widely studied and well understood. For example, it is straightforward to show that $\bLtrue(\bx) = \E[G\left(\bttrue(\bx)'\bR_1 \right)(1-G\left(\bttrue(\bx)'\bR_1 \right)) \bR_1 \bR_1' \mid \bX \!=\! \bx]$, where $\bR_1 = (1, R)'$, and will be invertible under standard assumptions.

We implement Figure \ref{fig:arch-general} to maximize the likelihood with a simple network with two hidden layers of 10 nodes each. The simplicity of the network architecture is driven by the fact many dimensions of $\bX$ are binary, so there is less functional approximation required, and that we have a smallish dataset. We use the Torch interface in {\sf R} \cite{Keydana2023_book}. For inference we use 20-fold cross fitting.

\subsection{Parameters of Interest and Associated Results}

We study two different $\bmtrue$ of \eqref{eqn:mu flm2}. First, we examine the marginal effect of the treatment (interest rate). Second, we turn to the more ambitious goal of personalization and profit maximization, making full use of our framework. Other standard objects of interest in choice models that can be immediately used in our framework with heterogeneity enriched parameter functions $\btheta(\bx)$ include (i) the price elasticity at a set price (here, interest rate) $\tilde{r}$, which is $H = (1 - G\left( \theta_1 + \theta_2 \tilde{r} \right)) \theta_2 \tilde{r}$; (ii) a measure of willingness to pay obtained by taking $H = \theta_2 / \theta_1$; or (iii) expected consumer welfare, $H = - \log(1 + \exp(\theta_1 + \theta_2 \tilde{r})) / \theta_2 $. Importantly, without our explicit use of a structural model, characterizing these quantities and obtaining inference would be difficult.

\subsubsection{Marginal Effects}

\begin{figure}
    \centering \includegraphics[scale=0.5]{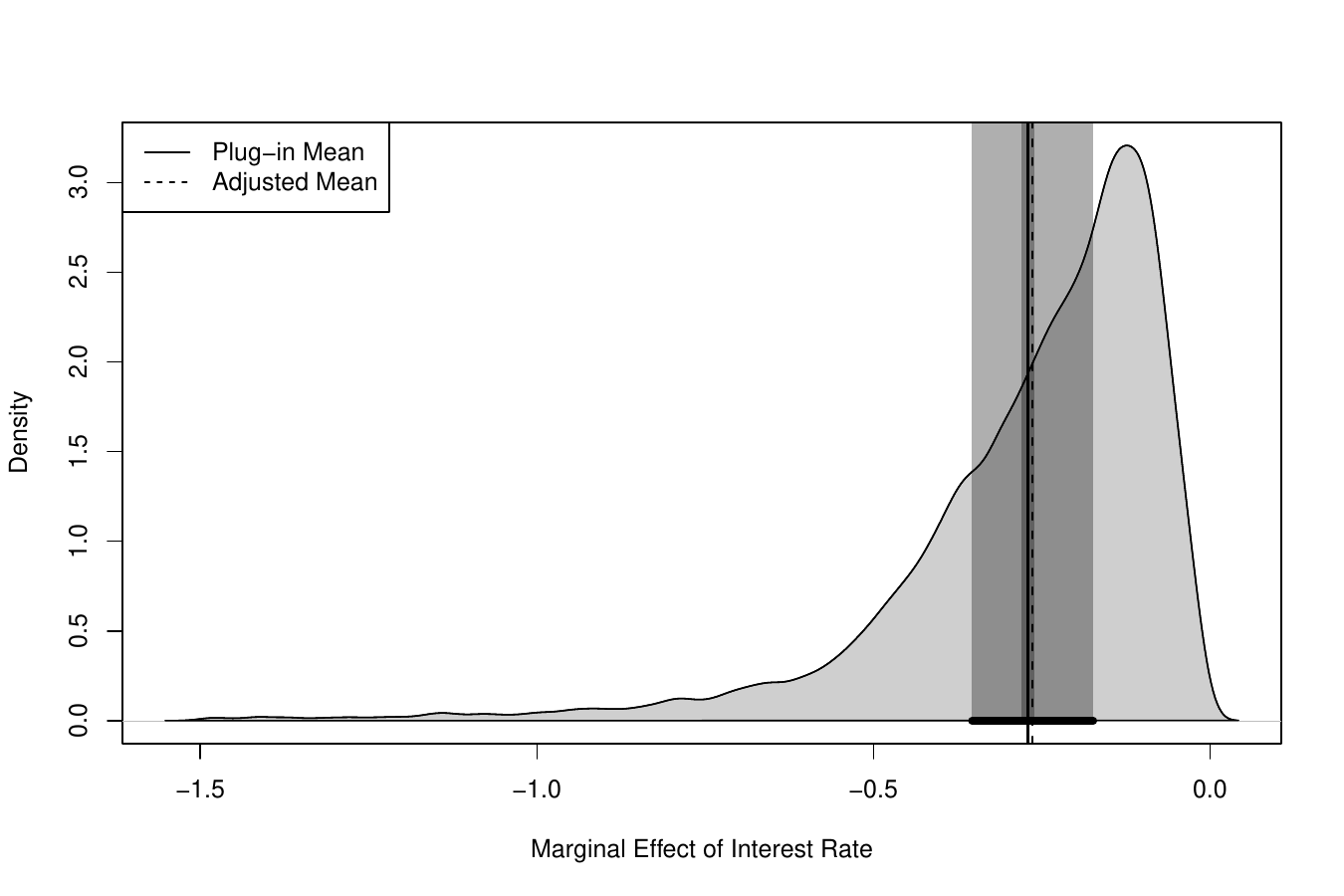}
    \caption{\label{fig:AME_r}Marginal Effect of Advertising Content}
\end{figure}

The average marginal effect (AME) can be written in closed form. Let $\tilde{r}$ be a given value of the interest rate and $\tilde{\br}_1 = (1,\tilde{r})'$. Then the parameter of interest is $\mathrm{AME}(\tilde{r}) = \E \left[ \left. \partial G(\btheta'\br_1 ) / \partial r \right|_{\bttrue(\bX)'\tilde{\br}_1} \right] = \E\left[ G(\bttrue(\bX)'\tilde{\br}_1 )\left(1-G(\bttrue(\bX)'\tilde{\br}_1)\right) \ttrue_2(\bX) \right]$. 
We set $\tilde{r}$ to the sample average ($0.084$) for simplicity. From the data we obtain the point estimate $\widehat{\mathrm{AME}}(0.084) = -0.263$ and corresponding $95\%$ confidence interval $(-0.354,-0.173)$ using \eqref{eqn:flm2 estimates}. The original analysis in \cite{Bertrand-etal2010_QJE} found a marginal effect of $-0.29$ (Table III therein). Further, since we use a subset of the data, we refit a simple model on the subset and obtain a marginal effect of $-0.2505$. Both these estimates fall within our confidence interval, which we interpret as the original findings being robust to heterogeneity.

Figure \ref{fig:AME_r} shows the distribution of the conditional average marginal effects along with the means and two confidence intervals. The lightest grey shows a kernel-smoothed density estimate. The shaded region is our 95\% confidence interval, while the darker shading shows the interval that ignores first step estimation (Remark \ref{rem:implementation uncertainty}). There is considerable heterogeneity and thus, although the average is robust to heterogeneity, there is potential for personalization.

\subsubsection{Optimal Personalized Offers}

We now demonstrate how the estimated heterogeneity can
be translated into personalized offers and the simplicity with which
one can conduct inference on quantities of interest. We examine the mean of personalized interest rate offers and the expected profits from personalization. Note that it poses no issue that the corresponding $H$ function is not available in closed form.

To construct profits we have to make some assumptions about the decision process of the firm and construct some auxiliary measures. First we will create a simple, parametric model for loan default probability. 
Given the rarity of defaults, the sample size is too small to uncover meaningful heterogeneity, and therefore we assume that the probability of default $D=1$ given an interest rate $R=r$ is $\P[D \!=\! 1 \mid R \!=\! r] = G(\dtrue_1 + \dtrue_2 r)$. 
We estimate the parameters ($\dtrue_1, \dtrue_2$) from data and, for convenience in this illustration, take these parameters as given.\footnote{With richer data, one could apply the estimation and inference framework using a bivariate outcome of application and default, $\bY = (Y,D)'$ and enrich $(\dtrue_1, \dtrue_2)$ to include heterogeneity. Accounting for the estimation of $\dtrue_1(\bx)$ and $\dtrue_2(\bx)$ would be then automatic. With only 280 defaults observed this is not possible.}

To write the firm's expected profit for a given consumer, let $L$ be the loan amount, $M$ be the loan term, and assume the outside option of the money being loaned is to obtain a rate of return $r_0$, we se set to 0.01 in the analysis. Since we focus on optimizing the interest rate for fixed values of the parameters, let $\P[Y \!=\! 1 \mid \bX \!=\! \bx, R \!=\! r] = P(r)$ and $\P[D \!=\! 1 \mid R \!=\! r] = D(r)$. Then profits are
$
\Pi(r) = 
L (
P(r) ( M (1 - D(r)) r - D(r) )
+
(1 - P(r)) M r_0
)$.
The first component is expected revenue given an initiated loan, taking into account the revenue under non-default and the loss of all funds given default. The second term reflects the opportunity cost of the loan. We do not intend for this to be a perfect representation of reality but to illustrate how our methods can be applied realistic structural settings.

To find the optimal interest rate, we obtain the first order condition as per the usual optimization machinery. This yields
\[
0=\frac{d\Pi}{dr} = 
L \left[
\dot{P}(r) \left( M(1 - D(r)) r - D(r) - M r_0 \right)
+ P(r) \left( M(1 - D(r)) - M r \dot{D}(r) - \dot{D}(r) \right)
\right],
\]
where $\dot{P}$ and $\dot{D}$ represent derivatives with respect to their scalar arguments. This profit function is smooth in $r$ but there exists the possibility that it is not uni-modal. Based on simulations we verified that, for parameters where $\dot{P}<0$ and $\dot{D}>0$ (as would be expected in this context) and for $r \in [0,r_\text{max}=0.25]$ the profit function is uni-modal and a unique $r^*$ obtains. To explore this we define a representation of the fixed point problem as $r = r + \frac{d\Pi(r)}{dr}$.

\begin{figure}
    \centering\includegraphics[scale=0.5]{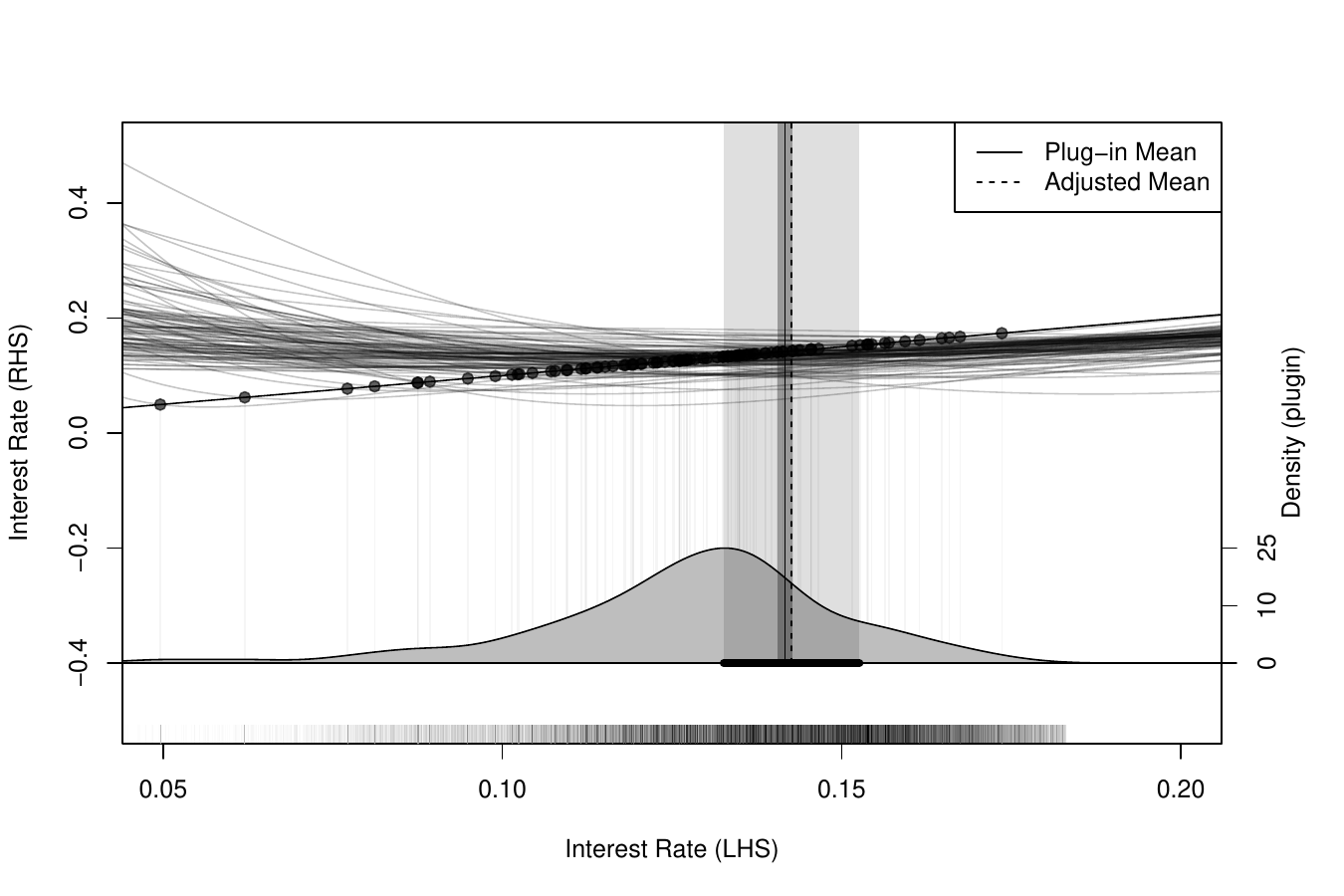}
    \caption{\label{fig:pers_r} Optimal Personalized Interest Rate Offers}
\end{figure}

This is an implicit function which will show a unique fixed point $\ropt$ if right hand side is decreasing in $r$. Figure \ref{fig:pers_r} presents a visual representation of the problem. Each light grey curve corresponds to a distinct consumer profile $\bx_i$ and its intersection with the $y=x$ line represents the fixed point
$\ropt$ (the scale of the axes is different so $y=x$ is not at $45^o$).  The density then represents the kernel density of the optimal
personalized offers $\ropt(\bx_i)$ across consumers. The fixed points are only shown for a subset of customers to avoid clutter, while the density is computed across the entire sample.

The reader should note that even though $\ropt$
is not available in closed form it remains a smooth function of the parameters $\btheta$, which is all that is required for our method to apply. We can therefore provide inference for any statistic of the form \eqref{eqn:mu flm2}. As a simple example, Figure \ref{fig:pers_r} shows estimation and inference for $\mtrue = \E[\ropt(\bttrue(\bX))]$, the average of optimal offers. We obtain a point estimate of $14.26\%$, with a 95\% confidence interval $[13.26\%, 15.26\%]$. This is shown in the figure, along with the plug-in interval as before.

Next we study expected profits from setting the optimal personalized interest rate, i.e., $\E[\Pi(\ropt(\bttrue(\bX)))] = 
\E\big[
L [
G(\ropt(\bttrue(\bX))) ( M (1 - D(\ropt(\bttrue(\bX)))) r - D(\ropt(\bttrue(\bX))) )
+
(1 - G(\ropt(\bttrue(\bX)))) M r_0 ]
\big]$. 
We can apply our framework directly to this estimand, despite the complications (again, this is a smooth function $H$ but not expressible in closed form), using automatic or numerical differentiation. We can also appeal to the envelope theorem, which ensures that $\partial\Pi / \partial r\big|_{r=\ropt}=0$. This shows the advantage of using our method over having to compute the relevant terms of the influence function by hand.

We standardize the results (for plotting) and interpret the expected profit construct as the net expected income from offering a \$1 loan at a personalized interest rate to each potential customer. We find that $\mhat=\$0.0495$ with a 95\% confidence interval of $[\$0.0461,\$0.0530]$. 

\begin{figure}
    \centering\centering\includegraphics[scale=0.5]{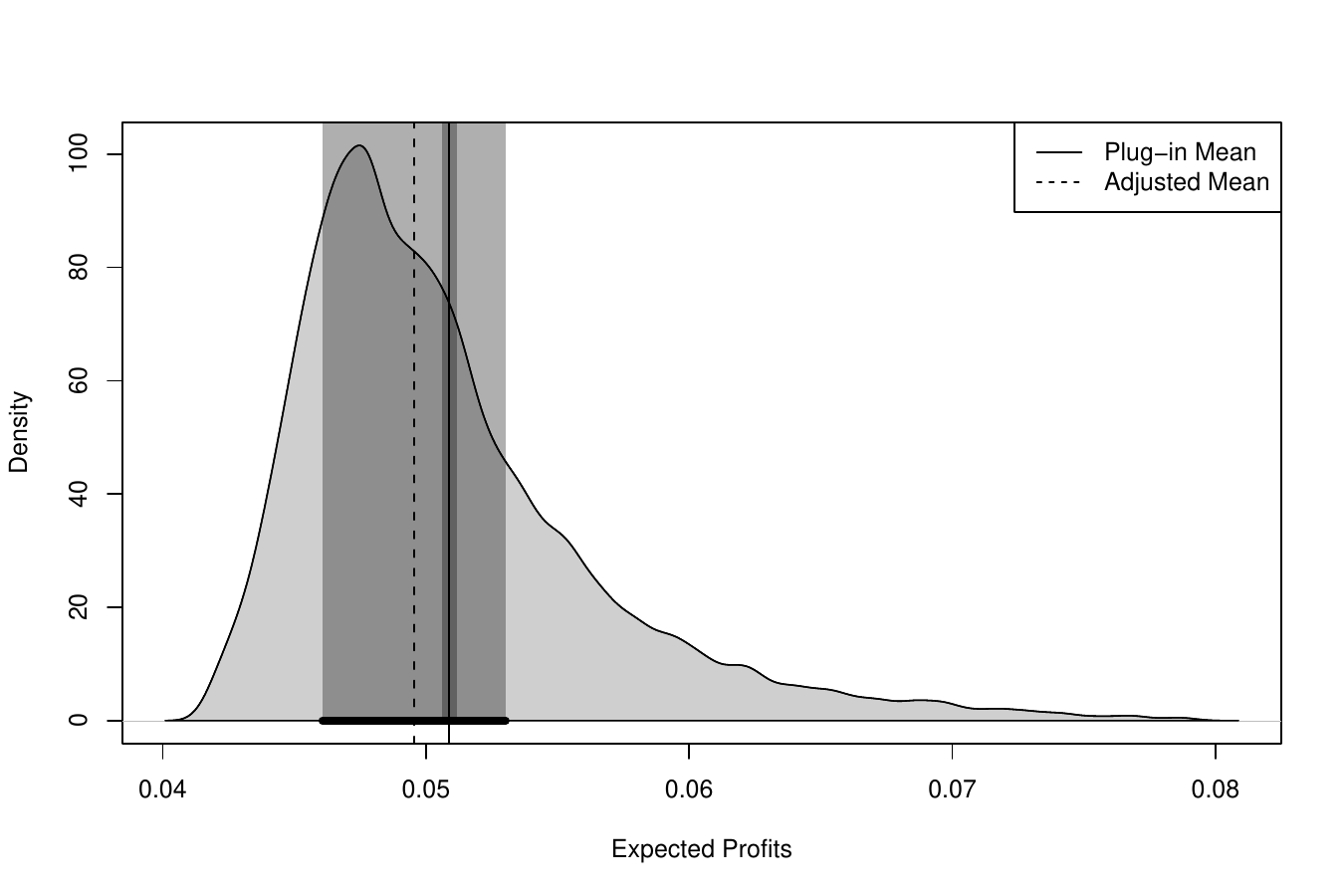}
    \caption{\label{fig:exp_profit}Expected Profits from Personalized Interest Rate Offers}
\end{figure}

Figure \ref{fig:exp_profit} depicts the density of profits across customers along with the estimate and confidence interval for the mean. Several features are notable here. The estimate $\mhat$, which includes the influence function adjustment, is outside the naive confidence interval based on the plug-in estimator. Statistically, this indicates that the bias correction from the influence function is large relative to the variance, that is, first stage noise from $\bthat(\bx)$ shifts the asymptotic distribution substantially, demonstrating the importance of DML (though see Remark \ref{rem:implementation uncertainty}).  The discrepancy between the estimates arises primarily due to the curvature of the profit function. The gradient of the profit function is steep around the estimated parameter and perturbations therein result in large changes in the influence component. The shape of the profit density is driven by a complex interaction of various components - the distribution of marginal effects of interest rate (Figure \ref{fig:AME_r}), coupled with default propensities, the optimal prices, and the formulation of profits. Even with such complexity, the profits (and prices) are well behaved and economically meaningful. We caution the reader again that our application is an illustration and ignores a number of other factors that might have bearing on the firm's and consumers' decision problems.

\begin{remark}[Implementation Uncertainty]
    \label{rem:implementation uncertainty}
    Throughout the application we use DML, building on our novel influence function, to provide valid inference for second step parameters. However, in some real-world decision making contexts, this may be inappropriate. Consider expected profits. Above, we studied $\mtrue = \E\big[\Pi(\ropt(\bttrue(\bX))) \big]$. For the firm, this corresponds to the profits they can expect from implementing \emph{true} optimal personalization, based on $\bttrue(\bx)$. We therefore require our novel influence function to account for the estimation error in the first stage, i.e., the fact that we use $\bthat(\bx)$ instead of $\bttrue(\bx)$. However, from the firm's point of view, if they choose to implement the strategy $\ropt(\bthat(\bx))$, it is more natural to consider $\bthat(\bx)$ as fixed and set the parameter of interest accordingly as $\tilde{\mu} = \E\big[\Pi(\ropt(\bthat(\bX))) \big]$, because this corresponds to the profits they can expect from what they would actually implement. In this case, DML and the influence function correction are not necessary, and the plug-in can be used directly.
\end{remark}

\section{Theoretical Results}
	\label{sec:theory}

The primary contributions of this paper are methodological, that is, showing how to enrich structural models with deep neural networks and how to conduct inference in the second step, but we accompany each step with formal theory given in this section.

\subsection{Bounds for Structural Deep Learning}
	\label{sec:dnn theory}

Our results for neural networks extend \cite{Farrell-Liang-Misra2021_Ecma} in a methodological sense, that is, to the structural and vector-valued setting. This allows for a broader range of applications. We also use recent approximation results to obtain a faster rate. Theorem \ref{thm:dnn rates} below gives the result, and to highlight the differences from \cite{Farrell-Liang-Misra2021_Ecma} we give a terse proof sketch. Full details are in Appendix \ref{appx:proofs}.

We first impose assumptions on the model/loss and the data generating process. These are standard in the nonparametric M estimation literature \cite{Chen2007_handbook} and often are implied by restrictions on the gradient and Hession of the loss known from the parametric structural model; the same conditions readily transfer to the enriched setting. Let $\bW = (\bY',\bT',\bX')'$ be the population random variables. Denote by $\bX_c$ the continuously distributed elements of $\bX$, with $\dc = \dim(\bX_c)$, and take the rest to be binary without loss of generality. 
\begin{assumption}
	\label{asmpt:dgp dnn}
    The elements of $\bW$ are bounded random variables and $\bX_c$ has compact, connected support, taken to be $[-1,1]^{\dc}$. The $\bttrue(\bx)$ are identified in \eqref{eqn:theta flm2}, uniformly bounded, and there are constants $c_1$, $c_2$, and $C_\ell$ that are bounded and bounded away from zero, such that for arbitrary $\btheta(\bx)$ and $\bm{\tilde{\theta}}(\bx)$, the loss obeys $|\ell(\by,\bt,\btheta(\bx) ) - \ell(\by,\bt,\bm{\tilde{\theta}}(\bx))| \leq C_\ell \|\btheta(\bx) - \bm{\tilde{\theta}}(\bx) \|_2$ and 
    $
        c_1 \E \left[ \|\btheta(\bX) - \bttrue (\bX) \|_2^2 \right] \leq \E [\ell(\bY,\bT,\btheta(\bX))] - \E [\ell(\bY,\bT,\bttrue(\bX))] \leq c_2 \E \left[ \|\btheta(\bX) - \bttrue (\bX) \|_2^2 \right].
    $
\end{assumption}
The assumptions we use mimic \cite{Farrell-Liang-Misra2021_Ecma} closely. Our goal is to broaden the settings to which deep learning can be applied, rather than weaken assumptions in any specific model. Differentiability is not required here and thus our results can be used in nonsmooth cases. In some application areas it might be a useful extension to weaken these assumptions in different directions. For example, in some economic or financial data it would be useful to allow for dependent data \cite{Brown2025_WP} or heavy tails \cite{Fan-Gu-Zhou2024_AoS}. We defer such technical extensions to future research.

Next we place a standard smoothness assumption on the unknown functions.
\begin{assumption}
	\label{asmpt:SMOOTH}
	As functions of $\bx_c$, $\ttrue_k(\bx) \in \cW^{p, \infty}([-1, 1]^{\dc})$, for $k=1,\ldots, \dtheta$, where for positive integers $p$ and $q$, define $\cW^{p, \infty}([-1, 1]^q)$ as the class of functions $h: \R^q \to \R$ with smoothness $p \in \mathbb{N}_+$ with
    $
    	\cW^{p, \infty}([-1, 1]^q) := \{ h:  \max_{\bm{r}, |\bm{r}| \leq p} \supess_{\bm{v} \in [-1, 1]^q}  |D^{\bm{r}} h(\bm{v})| \leq 1  \},
    $
    where $\bm{r} = (r_1, \ldots, r_q)$, $|\bm{r}| = r_1 + \ldots + r_q$ and $D^{\bm{r}} h$ is the weak derivative. (iv) $\max_{k \leq \dtheta} \sup_{\bx} |\ttrue_k(\bx)| < M$, for some positive $M$. 
\end{assumption}
Our goal is to study the standard nonparametric M estimation case for structural modeling, where the model dictates the relationship of $\bY$ and $\bT$, but $\bttrue(\bx)$ is completely nonparametric. Theoretically this serves as a benchmark and allows comparison to other estimation results. Practically, this corresponds to the researcher having no prior information on the heterogeneity, which is often the case in empirical work. At the same time, the curse of dimensionality limits the applicability of our theory in high-dimensional data and researchers may have to assume a lower-dimensional structure that is tailored to the specific context, such as sparsity, additive separability, compositions, or factor models. Other neural networks theory is available for such cases, though under different conditions than here (see \cite{Fan-Gu2024_JASA,Fan-Jana-Kulkarni-Yin2025_WP} for examples). None of these cover structural modeling, and some require additional tuning parameters such as bounds on the network weights, so a useful extension would be to combine our theory with those assumed properties. Our methodological idea of structural deep learning can be applied in conjunction with any assumptions on the nature of $\bttrue(\bx)$: if a researcher has a low-dimensional assumption in mind and a neural network tool which exploits it, this can be seamlessly built into the architecture of Figure \ref{fig:arch-general}. Also, note that our inference results in the next section are obtained under high-level conditions and can thus be applied broadly without change.

Our first result is the following high-probability, nonasymptotic bound, which yields the optimal rate up to $\log(n)$ factors. Here we use the common implementation of deep and wide multi-layer perceptrons (fully connected, feedforward neural networks) and set the width and depth to obtain the fastest rate. Equation \eqref{appxeqn:general result} in Appendix \ref{appx:proofs} shows a more general bound that is agnostic about the type of approximation, and hence the type of network, and can thus be a building block for future work based on other data structures \cite[\S2.3]{Farrell-Liang-Misra2021_Ecma}. A sketch of the proof is give next to highlight the central differences from \cite{Farrell-Liang-Misra2021_Ecma}, namely the generalization to vector-valued targets using \cite{Maurer2016VectorContraction} and using new approximation results of \cite{Tang2025_WP}.
\begin{theorem}
	\label{thm:dnn rates}
	Let $\bw_i$, $i=1,\ldots,n$, be a random sample that obeys Assumptions \ref{asmpt:dgp dnn} and \ref{asmpt:SMOOTH} and let $\cF_{\textsc{dnn}}$ of Figure \ref{fig:arch-general} have depth $L \asymp \log(n)$ and width $J \asymp n^{\dc / (4p + 2\dc)}$. Then for $n$ large enough and $k=1,\ldots,\dtheta$, with probability $1 - \exp\{-\delta\}$ it holds that
	\[
		\| \that_k - \ttrue_k \|^2_{L_2(\bX)}  + \E_n [ ( \that_k - \ttrue_k )^2 ]  \leq C  \left( n^{-2p/(2p+\dc)} \log^4 n + \frac{\delta}{n} \right),
	\]
	where $C$ may depend on the model and data-generating process but not $n$.
\end{theorem}

\subsubsection{Proof Sketch for Theorem \ref{thm:dnn rates}}
	\label{sec: dnn proof}

By Assumption \ref{asmpt:dgp dnn}, $ c_1 \E \left[ \|\bthat(\bX) - \bttrue (\bX) \|_2^2 \right]  \leq \left( \E - \E_n \right)\left[ \ell(\bY,\bT,\bthat(\bX)) - \ell(\bY,\bT,\bttrue(\bX)) \right] + \E_n \left[ \ell(\bY,\bT,\btheta_n(\bX)) - \ell(\bY,\bT,\bttrue(\bX)) \right]$. 
The second term is with probability $1-e^{-\delta}$ bounded by $c_2 \epsilon_n^2 + \epsilon_n \sqrt{ 2C_\ell^2 \delta  / n} + 7 C_\ell M \delta/n$, following \cite[Equation (A.2)]{Farrell-Liang-Misra2021_Ecma} under Assumption \ref{asmpt:dgp dnn}, where $\epsilon_n$ is the approximation error specified below. For the first term, we adapt \cite{Farrell-Liang-Misra2021_Ecma} to account for vector-valued $\bthat$. Let $\cF^0_{\textsc{dnn}} = \{ \btheta \in \cF_{\textsc{dnn}} : \E[ \|\btheta(\bX) - \bttrue (\bX) \|_2^2 ]^{1/2} \leq r_0\}$. Applying \cite{bartlett2005local},
\[
	\left( \E - \E_n \right)\left[ \ell(\bY,\bT,\bthat(\bX)) - \ell(\bY,\bT,\bttrue(\bX)) \right]
	\leq 6 \E_\eta R_n \cF^0_{\textsc{dnn}}  + \sqrt{\frac{2 C_\ell^2 r_0^2 \delta}{n}} + \frac{23 \cdot 3 M C_\ell}{3} \frac{\delta}{n},
\]
where 
$
    R_n \cF^0_{\textsc{dnn}}  = \sup_{\btheta \in  \cF^0_{\textsc{dnn}} }\sum_{i=1}^n \eta_i  \{ \ell(\by,\bt,\btheta(\bx)) - \ell(\by,\bt,\bttrue(\bx)) \} /n
$
is the empirical Rademacher complexity and $\E_\eta R_n \cF^0_{\textsc{dnn}}$ is its expectation holding fixed the data. Bounding this term is crucially different from \cite{Farrell-Liang-Misra2021_Ecma}. Here we apply \cite[Corollary 1]{Maurer2016VectorContraction}, which in our context yields (below $\eta_{ik}$'s denote i.i.d. Rademacher random variables)
\[
    \E_{\eta} \sup_{\btheta \in  \cF^0_{\textsc{dnn}} } \frac{1}{n}\sum_{i=1}^n \eta_i \left( \ell(\by,\bt,\btheta(\bx)) - \ell(\by,\bt,\bttrue(\bx)) \right) 
    \leq \sqrt{2}C_\ell \sum_{k=1}^{\dtheta} \E_{\eta} \sup_{\theta_k \in \cF^0_{\textsc{dnn},k}} \frac{1}{n} \sum_{i=1}^n \eta_{ik} \big(\theta_k(\bx_i) - \ttrue_k(\bx_i) \big).
\]
Each entry in the summation is now amenable to the techniques of Section A.2.1 and Lemmas 3 and 4 of \cite{Farrell-Liang-Misra2021_Ecma}. This yields
\[
    \E_{\eta} \sup_{\theta_k \in \cF^0_{\textsc{dnn},k}} \frac{1}{n} \sum_{i=1}^n \eta_{ik} \big(\theta_k(\bx_i) - \ttrue_k(\bx_i) \big) \leq 32 r_0 \sqrt{\frac{{\rm Pdim}(\cF_{\textsc{dnn},k})}{n} \left( \log \frac{2eM}{r_0} + \frac{3}{2} \log n \right) } ,   
\]
with probability $1 - e^{-\delta}$, where ${\rm Pdim}(\cF)$ is the pseudo-dimension of the class $\cF$. This can be plugged into the above and combined with the approximation to yield 
\[
	c_1 \E \left[ \|\bthat(\bX) - \bttrue (\bX) \|_2^2 \right]  \leq r_0  \left( K_1 \sqrt{ \frac{ Q L \log(Q) }{n}  \log n}  + \sqrt{\frac{2 C_\ell^2 \delta}{n}} \right) + c_2 \epsilon_n^2 + \epsilon_n \sqrt{\frac{2C_\ell^2 \delta }{n}} + K_2 \frac{ \delta}{n},
\]
for constants $K_1$ and $K_2$, where the final inequality applies Theorem 6 in \cite{Bartlett-etal2017_COLT} to bound the pseudo-dimension of ReLU networks in terms of their depth $L$ and total parameters $Q$. Following the critical-radius and shell-localization recursion of \cite{Farrell-Liang-Misra2021_Ecma}, adjusting the constants to account for $\dtheta \geq 1$, gives this result. Then, for $\epsilon_n$, \cite{Farrell-Liang-Misra2021_Ecma} relied on \cite{Yarotsky2017_NN} plus an embedding, yielding a suboptimal rate. Instead we use \cite{Tang2025_WP}, which shows that for approximation error $\epsilon_n$ one requires at most width $J(\epsilon_n)  \leq C \epsilon_n^{-(\dc/2p)} $ and depth $L(\epsilon_n) \leq C \cdot ( \log (1/\epsilon_n) + 1)$. To obtain $\epsilon_n = n^{-\frac{p}{2 p+\dc}}$ we set $J \asymp  n^{\frac{\dc}{4p + 2\dc}} $ and $L \asymp \log n$, yielding the final result. This network is narrower than \cite{Farrell-Liang-Misra2021_Ecma} for the same approximation, as no width is wasted in the embedding step. \hfill \qedsymbol

\subsection{Influence Function and DML-based Inference}
	\label{sec:if theory}

We now turn to our influence function and asymptotic normality. Influence functions have a long history in statistics. \cite{Newey1994_Ecma} is a seminal treatment, to which we defer for background. Our goal is not to contribute to the theory of influence functions per se, but rather to use the tools to obtain the methodological result of a broadly applicable influence function, to enable two-step inference under weak conditions. Our main result is a calculation made possible by applying \cite{Newey1994_Ecma}. We view the influence function as a tool for feasible inference, rather than an object of interest in its own right (such as for efficiency). This viewpoint is implicit in early work on inference after ML, but it is worthwhile to make it explicit to understand how this mode of thinking allows us to cover a range of applications and apply automatic differentiation in cases, such as Section \ref{sec:application}, where the requisite derivatives are not known a priori or not available in closed form.

The assumption we impose first is standard and ensures sufficient regularity for our influence function to be calculated and for asymptotic normality of the resulting estimator.
\begin{assumption}
	\label{asmpt:dgp if}
	The following hold on the distribution of $\bW$, uniformly in the conditioning elements. (i) \eqref{eqn:theta flm2} holds and identifies $\bttrue(\bx)$, where $\ell(\bw, \btheta)$ is thrice continuously differentiable in $\btheta$. (ii) $\E[ \bm{\ell}_{\btheta}(\bY, \bt, \bttrue(\bx)) \mid \bX \!=\! \bx, \bT \!=\! \bt ] = 0$. (iii) $\bLtrue(\bx)^{-1}$ exists and is bounded. (iv) $\bmtrue$ is identified and pathwise differentiable and $\bH$ is thrice continuously differentiable in $\btheta$. (v) $\bH(\bX, \btheta(\bX); \btstar)$ and $\bm{\ell}_{\btheta}(\bY,\bT, \btheta(\bX))$ have $q > 4$ finite absolute moments and positive variances. 
\end{assumption}
The most important assumptions here are that the first order condition of \eqref{eqn:theta flm2} holds, $\bttrue(\bx)$ is identified, and that $\bmtrue$ is pathwise differentiable. The latter keeps focus on regular semiparametric contexts. The former codifies our idea of taking a well-defined parametric model, for which the first-order condition and identification would hold, and enriching it with ML. This also maintains focus on structural parameters, which is crucial for interpretation and decision-making. For example, in the linear model (Appendix \ref{appx:glm} and \ref{appx:examples}), a best linear predictor can be obtained without (ii), but this parameter has a different (e.g., not ceteris paribus or noncausal) interpretation than a structural coefficient under a conditional mean restriction. Correct specification is crucial for policy- and decision-making. Condition (iii) will often be implied by assumptions on the model, such as in the case of logistic regression if $\P[Y \!=\! 1 \mid \bX \!=\! \bx, \bT \!=\! \bt]$ is bounded away from zero and one (which in turn may be implied by conditions on $\bX$, $\bT$, and the functions $\bttrue$, such as boundedness). Or, in the context of treatment effects we need the standard overlap condition. A positive variance condition, or invertibility of $\bLtrue(\bx)$, is standard in semiparametrics.

Next, following standard practice for DML, we give generic conditions on first-step estimation of $\bttrue(\bx)$ and $\bLtrue(\bx)$ that are sufficient for asymptotic normality. Let $\ltrue_{j,m}(\bx) = [\bLtrue(\bx)]_{j,m}$ denote the $(j,m)$ element and similarly for an estimator $\lhat_{j,m}(\bx)$. 
\begin{assumption}
	\label{asmpt:first step}	
	For a random sample of $n$ observations, first step estimators obey $\| \that_{j} - \ttrue_{j} \|_{L_2(\bX)} = o_P(n^{-1/4})$ and $\|\lhat_{j,m} - \ltrue_{j,m} \|_{L_2(\bX)} = o_P(n^{-1/4})$ for all $j, m \in \{1,\ldots,\dtheta\}$, and $\bLhat(\bx_i)$ is uniformly invertible. 
\end{assumption}
 The advantage of providing valid inference under general conditions is that our inference result can be used in the widest set of applications, even where neural networks are not used. For example, $\bL(\bx)$ is generic regression problem and so different tools may suit different applications. However, as above, deep neural networks are a popular and powerful method, and so we verify that they can be used to satisfy these requirements. This is a natural workflow after structural deep learning. Theorem \ref{thm:dnn rates} verifies this condition for $\bttrue$ and Lemma \ref{appxlem: lambda hat rate 2} in Appendix \ref{appx:proofs} does the same for $\bLhat$ (using further splitting in the latter case as discussed above and in Remark \ref{rem:sample splitting}).

The following  justifies our inference method. Other than the novel influence function, it is an application of DML \cite{Chernozhukov-etal2018_EJ}. Let $\bm{0}_{d}$ be the $d$-long zero vector and $\bm{I}_{d}$ be the $d$-square identity matrix.
\begin{theorem}
	\label{thm:normality}
	(i) If  Assumption \ref{asmpt:dgp if} holds, then \eqref{eqn:if flm2} gives a Neyman orthogonal score for $\bmtrue$. (ii) If $\bw_i$, $i=1,\ldots,n$, is a random sample that obeys Assumptions \ref{asmpt:dgp if} and \ref{asmpt:first step}, then $\bmhat$ and $\bPsihat$ of \eqref{eqn:flm2 estimates} obey $\sqrt{n}\bPsihat^{-1/2}(\bmhat - \bmtrue) =  \sumi \bm{\Psi}^{-1/2} \bpsi(\bw_i, \bttrue(\bx_i), \bLtrue(\bx_i))/\sqrt{n} + o_p(1) \to_d \cN(\bm{0}_{\dmu},\bm{I}_{\dmu})$.
\end{theorem}
This result gives a first-order distributional approximation allowing for feasible inference. As a word of caution, such approximations may be poor in some applications because the approximation is invariant to the first step estimator which can be unrealistic in practice. Refined results have been obtained in special cases \cite[and references therein]{Cattaneo-Farrell-Jansson-Masini2025_JoE}. Using higher order derivatives may yield further refinements or robustness. Finally, we note that in some cases our inference is efficient (Appendix \ref{appx:glm} and \ref{appx:examples}). When the original model is based on a likelihood or exponential family we conjecture that efficiency is always obtained following \cite[Remark 4.1]{Mammen-vandeGeer1997_AoS}.

\section{Conclusion}
	\label{sec:conclusion}

Structural modeling is a workhorse of empirical research. We have shown how to enrich these models with deep learning to capture rich heterogeneity, filling in the gaps left by economic theory. Our method combines the strength of structural modeling and the strength of machine learning. We established convergence rates for structured deep learning and valid inference using a novel influence function calculation. Our method represents a step toward easier and more rigorous use of machine learning in research, but is far from complete. Shape restrictions are on major form of structure arising from economic theory. Extending our methods to include impose shape constraints is an important step for future research. From an implementation point of view, there is also a lot of ground to cover for deep neural networks, including penalization and regularization, tuning parameter choices, and robust computation.

\section{References}
\singlespacing
\begingroup
\renewcommand{\section}[2]{}	
\bibliography{FLM2_2026_JASA--Bibliography}{}
\bibliographystyle{jpe}
\endgroup

\cleardoublepage

\phantomsection
\pdfbookmark[1]{Appendixes}{appendixes}

\ \bigskip

\begin{center}
\LARGE
Appendixes for ``Deep Learning for Individual Heterogeneity'' 
\end{center}
	
\ \vspace{1in}

\begin{abstract}
This supplemental appendix contains additional material for ``Deep Learning for Individual Heterogeneity''. Here we give all proofs, study the special case of generalized linear models in detail, discuss examples both old and new, and report results of a simulation study. Remarks at the end of each (sub)section give additional discussion and context. Section \ref{appx:proofs} contains proofs for structural deep neural networks for the parameter functions $\btrue(\bx)$, including a more general bound than in the main theorem, given in Equation \eqref{appxeqn:general result}. Section \ref{appx:dml} discusses second step inference, including derivation of the influence function, inference for $\bmtrue$ under general conditions, and estimation of the nuisance parameter $\bLtrue(\bx)$ using neural networks in particular. Then we discuss the special case of enriched generalized linear models, where $\E[Y \mid \bT = \bt, \bX=\bx] = G(\atrue(\bx) + \bbtrue(\bx)'\bt )$, in Section \ref{appx:glm}. Section \ref{appx:examples} then covers several examples, including the familiar (average treatment effects, partially linear models) and new (tobit, instrumental variables) to both compare our results to prior work and demonstrate the ease of application to new areas. A brief empirical application to fractional outcome models is shown. Finally, Section \ref{appx:simuls} reports details of a simulation study. 
\end{abstract}

\newpage
\tableofcontents

\newpage
\begin{appendices}
\addtocontents{toc}{\protect\setcounter{tocdepth}{3}}
\doublespacing
\numberwithin{corollary}{section}
\numberwithin{theorem}{section}
\numberwithin{lemma}{section}
\numberwithin{table}{section}
\numberwithin{figure}{section}
\numberwithin{remarkTmp}{section}
\numberwithin{assumption}{section}

\section{Proofs}
    \label{appx:proofs}

\subsection{Structural Deep Learning}
    \label{appx:dnn proofs}

Here we prove Theorem \ref{thm:dnn rates}, using the proof method of \cite{Farrell-Liang-Misra2021_Ecma}. Some details will be deferred to that paper to avoid repetition. The key changes are moving from scalar to vector-valued estimands and an improved approximation result for the fully-connected, feed-forward (multi-layer perceptron) case. 

Define $\btheta_n \in \cF_{\textsc{dnn}}$ as the best approximation to $\bttrue$ in the class of DNNs as defined in the theorem statement and let $\epsilon_n$ denote the error of the approximation:
\[
    \btheta_n = \argmin_{\btheta \in \cF_{\textsc{dnn}}} \| \btheta - \bttrue \|_{\infty}, 
    \qquad \qquad
    \epsilon_n = \| \btheta_n - \bttrue \|_\infty.
\]
For now we leave the error generic to allow for other approximation assumptions (such as other smoothness classes) and other architectures. The end of the proof treats the multi-layer perceptron case.

By Assumption \ref{asmpt:dgp dnn} and that $\bthat$ optimizes $\ell$ over $\cF_{\textsc{dnn}}$ in the data,
\begin{align*}
	c_1 \E & \left[ \|\bthat(\bX) - \bttrue (\bX) \|_2^2 \right]   			 \\ 
	& \leq \E [\ell(\bY,\bT,\bthat(\bX))] - \E [\ell(\bY,\bT,\bttrue(\bX))]   			 \\
	& \leq \E [\ell(\bY,\bT,\bthat(\bX))] - \E [\ell(\bY,\bT,\bttrue(\bX))]   -   \E_n [\ell(\bY,\bT,\bthat(\bX))] + \E_n [\ell(\bY,\bT,\btheta_n(\bX))]  			 \\
	& = \left( \E - \E_n \right)\left[ \ell(\bY,\bT,\bthat(\bX)) - \ell(\bY,\bT,\bttrue(\bX)) \right] + \E_n \left[ \ell(\bY,\bT,\btheta_n(\bX)) - \ell(\bY,\bT,\bttrue(\bX)) \right].
\end{align*}
Applying \citet[Equation (A.2)]{Farrell-Liang-Misra2021_Ecma} to the second term of the last line above, we find that with probability $1-e^{-\gamma}$
\begin{multline}
    c_1 \E \left[ \|\bthat(\bX) - \bttrue (\bX) \|_2^2 \right] 
    \\
    \leq \left( \E - \E_n \right)\left[ \ell(\bY,\bT,\bthat(\bX)) - \ell(\bY,\bT,\bttrue(\bX)) \right] + c_2 \epsilon_n^2 + \epsilon_n \sqrt{\frac{2C_\ell^2 \gamma }{n}} + \frac{7 C_\ell M \gamma}{n}.
    \label{appxeqn:decomposition}
\end{multline}

We now apply the localization-based analysis of \cite{Farrell-Liang-Misra2021_Ecma} to the first term above and then collect the results. Suppose that for some $r_0$, $\E[ \|\bthat(\bX) - \bttrue (\bX) \|_2^2 ]^{1/2} \leq r_0$, which can always be attained given the boundedness. Let $\cF^0_{\textsc{dnn}}$ be the subset of $\cF_{\textsc{dnn}}$ such that $\btheta \in \cF^0_{\textsc{dnn}}$ if $\E[ \|\btheta(\bX) - \bttrue (\bX) \|_2^2 ]^{1/2} \leq r_0$. Then by Theorem \ref{thm:normality}.1 in \cite{bartlett2005local}, for $\cG = \{g = \ell(\by,\bt,\btheta(\bx)) - \ell(\by,\bt,\bttrue(\bx)) : \btheta \in  \cF^0_{\textsc{dnn}}\}$, we find that, with probability at least $1-2e^{-\gamma}$, the empirical process term of \eqref{appxeqn:decomposition} is bounded as
\begin{equation}
	\label{appxeqn:apply-sym-one-step}
	\left( \E - \E_n \right)\left[ \ell(\bY,\bT,\bthat(\bX)) - \ell(\bY,\bT,\bttrue(\bX)) \right]
	\leq 6 \E_\eta R_n \cG  + \sqrt{\frac{2 C_\ell^2 r_0^2 \gamma}{n}} + \frac{23 \cdot 3 M C_\ell}{3} \frac{\gamma}{n},
\end{equation}
where 
\[
    R_n \cG = \sup_{g \in \cG} \frac{1}{n}\sum_{i=1}^n \eta_i g(\bw_i) = \sup_{\btheta \in  \cF^0_{\textsc{dnn}} } \frac{1}{n}\sum_{i=1}^n \eta_i \left( \ell(\by,\bt,\btheta(\bx)) - \ell(\by,\bt,\bttrue(\bx)) \right) 
\]
is the empirical Rademacher complexity and $\E_\eta R_n \cG$ is its expectation holding fixed the data, i.e. over the i.i.d. Rademacher variables $\eta_i$. The argument given in Section A.2.2 of \cite{Farrell-Liang-Misra2021_Ecma} does not apply directly to $\E_\eta R_n \cG$ because $\btheta$ is vector valued. Instead, we replace Lemma 2 therein with \citet[Corollary 1]{Maurer2016VectorContraction}, which in our context yields (below $\eta_{ik}$'s denote i.i.d. Rademacher random variables)
\begin{align*}
    \E_{\eta} \sup_{\btheta \in  \cF^0_{\textsc{dnn}} } \frac{1}{n}\sum_{i=1}^n \eta_i \left( \ell(\by,\bt,\btheta(\bx)) - \ell(\by,\bt,\bttrue(\bx)) \right) 
    &\leq \sqrt{2} C_\ell \E_{\eta} \sup_{\btheta \in  \cF^0_{\textsc{dnn}}} \sum_{k=1}^{\dtheta} \frac{1}{n} \sum_{i=1}^n \eta_{ik} \big(\theta_k(\bx_i) - \ttrue_k(\bx_i) \big)       \\
    &\leq \sqrt{2}C_\ell \sum_{k=1}^{\dtheta} \E_{\eta} \sup_{\theta_k \in \cF^0_{\textsc{dnn},k}} \frac{1}{n} \sum_{i=1}^n \eta_{ik} \big(\theta_k(\bx_i) - \ttrue_k(\bx_i) \big),
\end{align*}
with the second inequality following because the class of DNNs $\cF_{\textsc{dnn}}$ we use is decomposable with respect to each coordinate, and therefore we can bound one coordinate at a time. 

We then apply Section A.2.1 and Lemmas 3 and 4 of \cite{Farrell-Liang-Misra2021_Ecma} to the term for each component function $\theta_k$, $k=1,\ldots,\dtheta$, yielding
\[
    \E_{\eta} \sup_{\theta_k \in \cF^0_{\textsc{dnn},k}} \frac{1}{n} \sum_{i=1}^n \eta_{ik} \big(\theta_k(\bx_i) - \ttrue_k(\bx_i) \big) \leq 32 r_0 \sqrt{\frac{{\rm Pdim}(\cF_{\textsc{dnn},k})}{n} \left( \log \frac{2eM}{r_0} + \frac{3}{2} \log n \right) } ,   
\]
with probability $1 - \exp^{-\gamma}$, where ${\rm Pdim}(\cF)$ is the pseudo-dimension of the class $\cF$. Therefore, whenever $r_0 \geq 1/n$ and $n \geq (2eM)^2$, 
\begin{equation*}
    \E_{\eta} \sup_{\btheta \in  \cF^0_{\textsc{dnn}} } \frac{1}{n}\sum_{i=1}^n \eta_i \left( \ell(\by,\bt,\btheta(\bx)) - \ell(\by,\bt,\bttrue(\bx)) \right) 
    \leq
    K r_0 \sqrt{ \frac{{\rm Pdim}(\cF_{\textsc{dnn}}) }{n}  \log n},
\end{equation*}
for a constant $K$ that depends on $C_\ell$ and $\dtheta$. 

This last bound is then combined with \eqref{appxeqn:apply-sym-one-step} and put into \eqref{appxeqn:decomposition} and we find that
\begin{align}
    c_1 & \E \left[ \|\bthat(\bX) - \bttrue (\bX) \|_2^2 \right]        \nonumber \\
    & \leq 6 K r_0 \sqrt{ \frac{{\rm Pdim}(\cF_{\textsc{dnn}}) }{n}  \log n}  + \sqrt{\frac{2 C_\ell^2 r_0^2 \gamma}{n}} + \frac{23 \cdot 3 M C_\ell}{3} \frac{\gamma}{n} + c_2 \epsilon_n^2 + \epsilon_n \sqrt{\frac{2C_\ell^2 \gamma }{n}} + \frac{7 C_\ell M \gamma}{n}         \nonumber \\
    & \leq r_0  \left( 6 K \sqrt{ \frac{{\rm Pdim}(\cF_{\textsc{dnn}}) }{n}  \log n}  + \sqrt{\frac{2 C_\ell^2 \gamma}{n}} \right) + c_2 \epsilon_n^2 + \epsilon_n \sqrt{\frac{2C_\ell^2 \gamma }{n}} + K_2 \frac{ \gamma}{n}         \nonumber \\
    & \leq r_0  \left( K_1 \sqrt{ \frac{ Q L \log(Q) }{n}  \log n}  + \sqrt{\frac{2 C_\ell^2 \gamma}{n}} \right) + c_2 \epsilon_n^2 + \epsilon_n \sqrt{\frac{2C_\ell^2 \gamma }{n}} + K_2 \frac{ \gamma}{n},         \label{appxeqn:something} 
\end{align}
for constants $K_1$ and $K_2$, where the final inequality applies Theorem 6 in \cite{Bartlett-etal2017_COLT} to bound the pseudo-dimension of ReLU networks in terms of their depth $L$ and total parameters $Q$.

The bound of Equation \eqref{appxeqn:something}, reached under the assumption that $\E[ \|\bthat(\bX) - \bttrue (\bX) \|_2^2 ]^{1/2} \leq r_0$, provides the key input into Sections A.2.3 and A.2.4 of \cite{Farrell-Liang-Misra2021_Ecma}, which now go through with only change to the constants to capture the dependence on $\dtheta$. Following those steps exactly we find that with probability $1 - e^{-\delta}$,
\begin{equation}
    \begin{split}
    \label{appxeqn:general result}
    \E \left[ \|\bthat(\bX) - \bttrue (\bX) \|_2^2 \right]        
    \leq C \left(  \frac{ Q L \log(Q) }{n}  \log n  + \frac{\log \log n + \delta}{n}  + \epsilon_n^2\right) \\
    \E_n \left[ \|\bthat(\bX) - \bttrue (\bX) \|_2^2 \right]        
    \leq C' \left(  \frac{ Q L \log(Q) }{n}  \log n  + \frac{\log \log n + \delta}{n}  + \epsilon_n^2\right),
    \end{split}
\end{equation}
for positive constants $C$ and $C'$ which do not depend on $n$ but depend on the constants given in Assumption \ref{asmpt:dgp dnn} and as well as the dimensionalities, including $\dtheta$.

The bound in Equation \eqref{appxeqn:general result} provide a result comparable to Theorem \ref{thm:normality} of \cite{Farrell-Liang-Misra2021_Ecma}. In the same way as in that paper, this bound can be specialized to different architectures. Here we focus on the case of a fully-connected, feed-forward network (multi-layer perceptron, MLP) and one where both the width and depth may diverge with $n$. Other possibilities may be useful in certain cases, for example in bounded-width, very deep networks \cite{Yarotsky2018_COLT,Hanin2019_Mathematics} or in sparse networks \cite{Kohler-Krzyzak-Langer2022_IEEE,Fan-Gu2024_JASA}. 

 Importantly, only $\dc$ will matter here. To see why, suppose $X_1$ is binary. Then for two $(\dx-1)$-dimensional functions $\ttrue_{k,1}$ and $\ttrue_{k,0}$, obeying the same smoothness assumption placed on the original $\ttrue_k$, it holds that $\ttrue_k(\bx) = x_{1} \ttrue_{k,1}(x_2, \ldots, x_{\dx}) + (1-x_{1}) \ttrue_{k,0}(x_2, \ldots, x_{\dx})$. Adding a single node to each hidden layer allows the network to pass forward the input $x_1$ and multiply it with two separate learned functions just prior to the parameter, giving exactly $\that_k(\bx) = x_1 \that_{k,1}(x_2, \ldots, x_{\dx}) + (1-x_1) \that_{k,0}(x_2, \ldots, x_{\dx})$. (Intuitively, this is similar to the combination of $\ahat(\bx)$ and $\bbhat(\bx)$ in Figure \ref{fig:arch-glm}, with $\bt$ there playing the role of the binary $x_1$ here.) The same argument can be applied to every category of the discrete data and to each function to be learned. Since $\dx$ is fixed, this results in only a constant increase in the width of the network. 

For the MLP case we leverage the approximation result of Theorem 3 of \cite{Tang2025_WP}. In the present notation, this result guarantees that there exists network $\btheta_n$ with width $J = J(\epsilon_n)  \leq C \epsilon_n^{-(\dc/2p)} $ and depth $L=L(\epsilon_n) \leq C \cdot ( \log (1/\epsilon_n) + 1)$ that yields an approximation error of at most $\epsilon_n$. Thus to obtain $\epsilon_n = n^{-\frac{p}{2 p+\dc}}$ we select $J \asymp  n^{\frac{\dc}{4p + 2\dc}} $ and $L \asymp \log n$, yielding the final result as $Q \leq C L J^2$. \qed

\begin{remark}[Approximation Improvement]
    \label{appxrem:approximation}
    The above proof uses the approximation of Theorem 3 of \cite{Tang2025_WP}, which is a derivation tailored directly to the MLP case and thus is able to improve upon the coarser method of \cite{Farrell-Liang-Misra2021_Ecma} that uses the approximation of \cite{Yarotsky2017_NN} plus an embedding into a suitably large MLP. That prior method, applied here, would yield a bound of $n^{-p/(p+\dc)} \log^8 n$ in the theorem statement, because for an approximation error of $\epsilon_n$ the network would need $J(\epsilon_n) \leq Q(\epsilon_n) L(\epsilon_n) \leq C^2  \epsilon_n^{-(\dc/p)} ( \log (1/\epsilon_n) + 1)^2$ and $ L(\epsilon_n) \leq C \cdot ( \log (1/\epsilon_n) + 1)$. See full discussion in \cite{Tang2025_WP}. Also, note that \cite{Fan-Gu2024_JASA} Theorem 4 can be specialized to standard nonparametric regression by removing all factor modeling and sparsity, and fixing the dimension. This would yield a similar approximation result. However, this result uses weights bounded by a certain polynomial in $n$, whereas we treat unbounded weights to match standard implementation and to avoid having to choose more tuning parameters. \cite{Fan-Gu2024_JASA} use fixed-depth, and although this could be padded to $\log(n)$ depth with only an increase in the $\log(n)$ factor of the final bound, it would not allow for the flexibility to consider network structures that are very deep, such as \cite{Yarotsky2018_COLT,Hanin2019_Mathematics,Farrell-Liang-Misra2021_Ecma} that can be used in conjunction with the general bound in Equation \eqref{appxeqn:general result}.
\end{remark}

\subsection{Second-Step Inference}
    \label{appx:dml}

This section contains derivations and proofs for Theorem \ref{thm:normality}. We first derive the influence function. Then asymptotic Normality is established by applying standard results in double machine learning (DML). Finally, we verify the the high-level rate conditions on first-stage estimation of the expected Hessian term $\bL$ using neural networks. A set of remarks connected to various literatures is given at the close of the section.

\subsubsection{Influence Function Derivation}
    \label{appx:IF proof}

Our derivation of the influence function applies \cite{Newey1994_Ecma}. For deeper treatments of influence functions, including efficiency bounds and conditions for existence, see \cite{Newey1990_JAE}, \cite{Newey1994_Ecma}, \citet[Chapter 25]{vanderVaart1998_book}, and \cite{Ichimura-Newey2022_QE}.

The starting point is a parametric submodel, indexed by a parameter $\eta$. Because our first stage (3.3) is explicitly based on enriching the parametric structural model (3.1), our submodels are as well-behaved as the original model and derivatives in the submodel are well-understood ordinary derivatives of the original structural model. For the calculation, distributions and other nonparametric objects are indexed by $\eta$, and thus we define $\btheta(\bx;\eta)$ and $\bmtrue(\eta)$ as
\begin{equation}
	\label{appxeqn:theta eta}
	\bttrue(\cdot; \eta) = \argmin_{\bb} \int  \ell\left(\bw, \bb(\bx) \right) f_{\bw} (\bw ;\eta ) d\bw 
\end{equation}
and
\begin{equation}
	\label{appxeqn:mu eta}
	\bmtrue(\eta) = \int \bH\left(\bx, \btheta(\bx;\eta); \btstar\right) f_{\bx}(\bx ; \eta) d\bx,
\end{equation}
where $f_{\bw}$ and $f_{\bx}$ are the distributions of $\bw = (\by', \bt', \bx')'$ and $\bx$ respectively. For notational simplicity, we will assume throughout the derivation that such densities exist. The true data generating process is obtained at $\eta = 0$. When evaluating at $\eta=0$ we will often omit the dependence on $\eta$, such as $f_{\bx}(\bx ; \eta) = f_{\bx}(\bx)$, $\btheta(\bx ; 0) = \bttrue(\bx)$, or $\E[\cdot]$ for expectations with respect to the true distribution.

The pathwise derivative approach proceeds, as in \cite{Newey1994_Ecma} and others, by finding a function $\bpsi(\bw)$ such that 
\begin{equation}
	\label{appxeqn:pathwise}
	\left. \frac{\partial \bmu(\eta)}{\partial \eta} \right\vert_{\eta=0} = \E[ \bpsi(\bW) S(\bW)],
\end{equation}
for the (true) score $S(\bw) = S(\bw ; \eta)\vert_{\eta=0}$. 

The first step is differentiating \eqref{appxeqn:mu eta} with respect to the parameter $\eta$, and evaluating this at $\eta = 0$. The product rule and the chain rule yield
\begin{align}
	& \left. \frac{\partial \bmu(\eta)}{\partial \eta} \right\vert_{\eta=0}  = \frac{\partial }{\partial \eta} \left.\left\{ \int \bH\left(\bx, \btheta(\bx;\eta); \btstar\right) f_{\bx}(\bx ; \eta) d\bx \right\} \right\vert_{\eta=0}   			\nonumber \\
	& \quad   =    \int \bH\left(\bx, \btheta(\bx ; 0); \btstar\right) \left. \frac{\partial f_{\bx}(\bx; \eta)}{\partial \eta} \right\vert_{\eta=0} d\bx   +    \int \left. \frac{\partial\bH\left(\bx, \btheta(\bx;\eta); \btstar\right)}{\partial \eta} \right\vert_{\eta=0} f_{\bx}(\bx ; 0) d\bx ,   			\nonumber \\
	& \quad   =    \int \bH\left(\bx, \bttrue(\bx); \btstar\right) \left. \frac{\partial f_{\bx}(\bx; \eta)}{\partial \eta} \right\vert_{\eta=0} d\bx   +    \int  \bH_{\btheta}(\bx, \bttrue(\bx ); \btstar) \btheta_{\eta}(\bx) f_{\bx}(\bx ) d\bx ,   			\label{appxeqn:chain rule}
\end{align}
where $\btheta_{\eta}(\bx) = \btheta_{\eta}(\bx ;0)$ is the $\dtheta$-vector gradient of $\btheta(\bx;\eta)$ with respect to $\eta$, evaluated at $\eta=0$, given by
\[
	\btheta_{\eta}(\bx ;0) = \left.\frac{\partial \btheta(\bx;\eta)}{\partial \eta} \right\vert_{\eta=0},
\]
and $\bH_{\btheta}(\bx, \bttrue(\bx ); \btstar)$ is the $\dmu \times \dtheta$ Jacobian of $\bH$ with respect to $\btheta$, evaluated at $\eta=0$, that is, the matrix with $\{h,k\}$ element, for $h = 1, \ldots, \dmu, k = 1, \ldots, \dtheta$, given by
\[
	\Big[ \bH_{\btheta}(\bx, \bttrue(\bx ); \btstar) \Big]_{h,k} = \left. \frac{\partial H_h (\bx, \bb; \btstar)}{\partial b_k}  \right\vert_{\bb = \btheta(\bx ; 0)},
\]
with $H_h$ the $h^{\rm th}$ element of $\bH$ and $b_k$ the $k^{\rm }$ element of $\bb$. For intuition, note that element $h = 1, \ldots, \dmu$ of the $\dmu$-vector $\bH_{\btheta}(\bx, \bttrue(\bx ); \btstar) \btheta_{\eta}(\bx)$ is 
\[
    \left.\frac{\partial \bH_h}{\partial \eta}\right\vert_{\eta=0} = \sum_{k = 1}^{\dtheta} \left. \frac{\partial \bH_h (\bx, \bb; \btstar)}{\partial \bb_k}  \right\vert_{\bb = \btheta(\bx ; 0)} \left.\frac{\partial \btheta_k(\bx ;\eta)}{\partial \eta} \right\vert_{\eta=0}.
\]

We will show that both terms of Equation \eqref{appxeqn:chain rule} above can be written as expectations of products with the full score $S(\bw)$, as required by \eqref{appxeqn:pathwise}. We will often use the standard facts that scores are mean zero and that 
\begin{equation}
	\label{appxeqn:score factors}
	S(\by,\bx,\bt) = S(\by,\bt \mid \bx) + S(\bx).
\end{equation}

The first term of Equation \eqref{appxeqn:chain rule} is 
\begin{align}
	\int \bH\left(\bx, \bttrue(\bx); \btstar\right) \left. \frac{\partial f_{\bx}(\bx; \eta)}{\partial \eta} \right\vert_{\eta=0} d\bx  & = \E\left[ \bH\left(\bX, \bttrue(\bx); \btstar\right) S(\bX)\right] 	  		\nonumber \\
	&  = \E\left[\bH\left(\bX, \bttrue(\bx); \btstar\right) S(\bY,\bX,\bT)\right],   		\label{appxeqn:term1}
\end{align}
where the first equality holds because the marginal score obeys $S(\bx) f_{\bx}(\bx) = \left. \partial f_{\bx}(\bx; \eta) / \partial \eta \right\vert_{\eta=0}$ and the second equality follows from the usual mean zero property of scores and \eqref{appxeqn:score factors}:
\[\E\left[\bH(\bX, \bttrue(\bx), \btstar) S(\bY,\bT \mid \bX) \right] = \E\Big[\bH(\bX, \bttrue(\bx), \btstar) \E\left[  S(\bY,\bT \mid \bX) \mid \bX \right] \Big] = 0.\]
This first term is then the standard ``plug-in'' portion of the influence function, that is, the term that would appear if $\bttrue(\bx)$ were known (or if $\bthat(\bx)$ were fixed). The second term of Equation \eqref{appxeqn:chain rule} will give rise to the correction factor that accounts for the nonparametric estimation.

To find this correction factor, we must find $\btheta_{\eta}(\bx) = \partial \btheta(\bx;\eta) / \partial \eta \vert_{\eta=0}$. This is a key step in the derivation, and crucially leverages the structure of the model $\ell$ and the fact that $\ell$ depends on $\btheta(\cdot)$ only through evaluation at a single point and only through $\bX$. We will use these facts to derive and expression for $\partial \btheta(\bx;\eta) / \partial \eta$, which involves the appropriate scores and then may be substituted into \eqref{appxeqn:chain rule} to yield the required form. 

We begin with the fact that the first order condition holds as an identity in $\eta$ and conditional on $\bX$. That is, as an identity in $\eta$, 
\begin{equation}
	\label{appxeqn:foc}
	\E_\eta \left[ \left. \bm{\ell}_{\btheta}(\bW, \btheta(\bx;\eta))   \right\vert \bX = \bx \right] \equiv 0,
\end{equation}
where $\bm{\ell}_{\btheta}$ is the $\dtheta$-vector gradient of $\ell$ with respect to $\btheta$, given by
\[
	 \bm{\ell}_{\btheta}(\bw, \btheta(\bx;\eta)) = \left. \frac{ \partial \ell\left(\bw, \bb \right)}{\partial \bb}\right \vert_{\bb = \btheta(\bx;\eta)}.
\]
The expectation is also indexed by $\eta$ in the submodel, as the density depends on $\eta$. To be explicit, as an identity in $\eta$ we have
\[
	\int  \left. \frac{ \partial \ell\left(\bw, \bb \right)}{\partial \bb}\right \vert_{\bb = \btheta(\bx;\eta)} f_{\by,\bt | \bx}(\by, \bt ; \eta \mid \bx) d\by d\bt \equiv 0.
\]
Define $\bm{\ell}_{\btheta\btheta}(\bw, \btheta(\bx;\eta))$ as the $\dtheta \times \dtheta$ matrix of second derivatives of $\ell\left(\bw, \bb \right)$ with respect to $\bb$, evaluated at $\bb = \btheta(\bx;\eta)$. That is, $\bm{\ell}_{\btheta\btheta}(\bw, \btheta(\bx;\eta))$ has $\{k_1, k_2\}$ element given by
\[
	\Big[\bm{\ell}_{\btheta\btheta}(\bw, \btheta(\bx;\eta))\Big]_{k_1,k_2} = \left. \frac{ \partial^2 \ell\left(\bw, \bb \right)}{\partial b_{k_1}\partial b_{k_2}}\right \vert_{\bb = \btheta(\bx;\eta)},
\]
where $b_{k_1}$ and $b_{k_2}$ are the respective elements of $\bb$. With this notation, differentiating the above identity with respect to $\eta$ and applying the chain rule we find
\begin{multline*}
	\int  \left. \frac{ \partial \ell\left(\bw, \bb(\bx) \right)}{\partial \bb}\right \vert_{\bb = \btheta(\bx;\eta)} \frac{ \partial f_{\by,\bt | \bx}(\by, \bt ; \eta \mid \bx)}{\partial \eta} d\by d\bt   
	\\ 
	+    \int  \bm{\ell}_{\btheta\btheta}(\bw, \btheta(\bx;\eta)) \btheta_{\eta}(\bx ;\eta) f_{\by,\bt | \bx}(\by, \bt ; \eta \mid \bx) d\by d\bt   = 0,
\end{multline*}
where the second term captures the derivatives of $\bm{\ell}_{\btheta}(\bw, \btheta(\bx;\eta))$ with respect to $\eta$, and recall, $\btheta_{\eta}(\bx ;\eta)$ is the $\dtheta$-vector gradient of $\btheta$ with respect to $\eta$, and is the key ingredient.

Evaluating this result at $\eta=0$, we obtain
\begin{equation}
	\label{appxeqn:partial foc wrt eta}
	 \E \left[ \left.  \bm{\ell}_{\btheta}(\bW, \bttrue(\bx)) S(\bY,\bT \mid \bX) \right\vert \bX \right]    	+   \E \left[ \left. \bm{\ell}_{\btheta\btheta}(\bW, \bttrue(\bx)) \btheta_{\eta}(\bx) \right| \bX \right]  = 0,
\end{equation}
where $S(\bY,\bT \mid \bX)$ is the conditional score and is obtained because $S(\by,\bt \mid \bx) f_{\by,\bt | \bx}(\by, \bt \mid \bx) = \left. \partial f_{\by,\bt | \bx}(\by, \bt ; \eta \mid \bx) / \partial \eta \right\vert_{\eta=0}$. Rearranging \eqref{appxeqn:partial foc wrt eta}, and using that $\btheta$ is only a function of $\bX$, gives
\[
	\E \left[ \left. \bm{\ell}_{\btheta\btheta}(\bW, \bttrue(\bx)) \right| \bX \right]  \btheta_{\eta}(\bx) =  - \E \left[ \left.  \bm{\ell}_{\btheta}(\bW, \bttrue(\bx)) S(\bY,\bT \mid \bX) \right\vert \bX \right].
\]
Then, because $\bL(\bx) := \E \left[ \left. \bm{\ell}_{\btheta\btheta}(\bW, \bttrue(\bx)) \right| \bX = \bx \right]$ is invertible, we have
\begin{align*}
	\btheta_{\eta}(\bx) & =  - \E \left[ \left. \bm{\ell}_{\btheta\btheta}(\bW, \bttrue(\bx)) \right| \bX \right]^{-1} \E \left[ \left.  \bm{\ell}_{\btheta}(\bW, \bttrue(\bx)) S(\bY,\bT \mid \bX) \right\vert \bX \right]   		\\
	& = - \E \left[ \left. \bL(\bx)^{-1} \bm{\ell}_{\btheta}(\bW, \bttrue(\bx)) S(\bY,\bT \mid \bX) \right\vert \bX \right].
\end{align*}
Substituting this into the second term of Equation \eqref{appxeqn:chain rule} and applying iterated expectations, we have
\begin{align*}
	\int  \bH_{\btheta}(\bx, \bttrue(\bx); \btstar) \btheta_{\eta}(\bx) f_{\bx}(\bx ) d\bx  
	& = - \E\Big[  \bH_{\btheta}(\bX, \bttrue(\bX); \btstar) \E \left[ \left. \bL(\bX)^{-1} \bm{\ell}_{\btheta}(\bW, \bttrue(\bx)) S(\bY,\bT \mid \bX) \right\vert \bX \right] \Big]  		\\
	& = - \E\Big[ \E \left[ \left. \bH_{\btheta}(\bX, \btheta(\bX); \btstar) \bL(\bX)^{-1} \bm{\ell}_{\btheta}(\bW, \bttrue(\bx)) S(\bY,\bT \mid \bX) \right\vert \bX \right] \Big]  		\\
	& = - \E\Big[  \bH_{\btheta}(\bX, \bttrue(\bX); \btstar) \bL(\bX)^{-1} \bm{\ell}_{\btheta}(\bW, \bttrue(\bx)) S(\bY,\bT \mid \bX)  \Big].
\end{align*}
Next, because the first order condition holds conditionally,
\begin{multline*}
	\E\Big[  \bH_{\btheta}(\bX, \bttrue(\bX); \btstar) \bL(\bX)^{-1} \bm{\ell}_{\btheta}(\bW, \bttrue(\bx)) S(\bX)  \Big]
	\\
	= \E\Big[  \bH_{\btheta}(\bX, \bttrue(\bX); \btstar) \bL(\bX)^{-1} \E\left[  \bm{\ell}_{\btheta}(\bW, \bttrue(\bx)) \mid \bX \right] S( \bX)  \Big].
\end{multline*}
Therefore, continuing from the previous display and applying \eqref{appxeqn:score factors}, the second term of Equation \eqref{appxeqn:chain rule} is of the required form:
\begin{equation}
	\label{appxeqn:term2}
	- \E\Big[  \bH_{\btheta}(\bX, \bttrue(\bX); \btstar) \bL(\bX)^{-1} \bm{\ell}_{\btheta}(\bW, \bttrue(\bx)) S(\bY,\bT, \bX)  \Big]
\end{equation}

Combining Equations \eqref{appxeqn:term1} and \eqref{appxeqn:term2} with \eqref{appxeqn:chain rule}, we find that
\begin{multline}
	\label{appxeqn:pathwise2}
	\left. \frac{\partial \bmu(\eta)}{\partial \eta} \right\vert_{\eta=0}
	= 
	\E\left[\bH\left(\bX, \bttrue(\bx); \btstar\right) S(\bY,\bX,\bT)\right]
	\\ 	
	- \E\Big[  \bH_{\btheta}(\bX, \bttrue(\bX); \btstar) \bL(\bX)^{-1} \bm{\ell}_{\btheta}(\bW, \bttrue(\bx)) S(\bY,\bT, \bX)  \Big].
\end{multline}
Thus we have verified Equation \eqref{appxeqn:pathwise} with
\begin{equation}
	\label{appxeqn:IF}
	\bpsi(\bw) =  \bH\left(\bx, \bttrue(\bx); \btstar\right)   - \bH_{\btheta}(\bx, \bttrue(\bX); \btstar) \bL(\bx)^{-1} \bm{\ell}_{\btheta}(\bw, \bttrue(\bx)).
\end{equation}
This is not an influence function as it lacks the appropriate centering, but of course $\E[\bmtrue S(\bW)] = \bmtrue \E[ S(\bW) ] = 0$, and thus we can freely center this $\bpsi(\bt)$ and still obey \eqref{appxeqn:pathwise}. \qed

\subsubsection{Asymptotic Normality via DML}
	\label{appx:clt}

The asumptotic Normality of Theorem \ref{thm:normality} follows from Theorems 3.1 and 3.2 of \cite{Chernozhukov-etal2018_EJ} upon verifying Assumptions 3.1 and 3.2 therein. Assumption 3.1(a) holds for $\bpsi(\by,\bt,\bx, \btheta, \bL) - \bmtrue$ given in Equation \eqref{eqn:if flm2}: the first term of $\bpsi$ has mean $\bmtrue$ by definition in \eqref{eqn:mu flm2} while the second is (conditionally) mean zero as assumed in Assumption \ref{asmpt:dgp if}, with $\bL(\bx)^{-1}$ uniformly bounded. Assumption 3.1(b), linearity, holds by definition of \eqref{eqn:mu flm2} and the form of the score in Equation \eqref{eqn:if flm2}. Assumption 3.1(c) holds by Assumption \ref{asmpt:dgp if}, in particular the assumed smoothness and the nonsingularity of $\bL(\bx)$. Assumption 3.1(d), Neyman orthogonality, is verified directly by the calculation of the influence function. Assumption 3.1(e) holds trivially as the matrix $J_0$ therein is the identity. Assumption 3.2, parts (b) and (d) follow directly from the moment conditions imposed. Conditions (a) and (c) follow from Equations (3.7) and (3.8) of \cite{Chernozhukov-etal2018_EJ} and the assumed convergence of the first stage estimates in Assumption \ref{asmpt:first step}.\qed

\subsubsection{DNN Estimation of Expected Hessian}
	\label{appx:estimation of Lambda}

Theorem \ref{thm:normality} is stated under generic conditions on the first stage estimators, as discussed in the text. For $\bttrue(\bx)$, these are verified directly by Theorem \ref{thm:dnn rates}. Here we show, under standard sufficient conditions, that $\bLtrue(\bx)$ can also be estimated using deep neural networks using least squares loss. This is important practical assurance of reliability in empirical work, and may be useful in other contexts. It may be possible to weaken the estimation requirements $\bLtrue(\bx)$ by extending \cite{Chernozhukov-etal2022_Ecma} to the present setting; we defer this to future research.

Recall that $\bL(\bx; \btheta) = \E [  \bm{\ell}_{\btheta\btheta}(\by,\bt, \btheta(\bx)) \mid \bX = \bx ]$, with $\bLtrue(\bx) = \bL(\bx; \bttrue)$ in particular, and $\lambda_{j,m}(\bx; \btheta) = [\bL(\bx; \btheta)]_{j,m}$ and $\ltrue_{j,m}(\bx) = \lambda_{j,m}(\bx; \bttrue) = [\bLtrue(\bx; \btheta)]_{j,m}$ as the respective $(j,m)$ elements. For each $j, m \in \{1,\ldots,\dtheta\}$ and a generic $\btheta(\bx)$, define the estimator
\begin{align}
	\label{appxeqn:lambda hat}
	\lhat_{j,m}(\cdot; \btheta) = \argmin_{g \in \cG_{\rm DNN} } \frac{1}{n}\sum_{i=1}^n \left( \left[\bm{\ell}_{\btheta\btheta}(\by_i,\bt_i, \btheta(\bx_i))\right]_{j,m}  -  g(\bx_i)\right)^2,
\end{align}
where $\cG_{\rm DNN}$ is an MLP class of neural networks, assumed to have width and depth proportional to $n^{\frac{\dc}{4p + 2\dc}} \log^2 n$ and $\log n$, respectively, with unbounded parameters and bounded output functions $g(\cdot)$. This class is not structured, as this is a standard regression problem. For simplicity we use a specific architecture and width and depth choices to state a concrete result as proof of concept. For other choices and their consequences for approximation biases and convergence, see \cite{Farrell-Liang-Misra2021_Ecma} for complete discussion.

We then impose the following conditions for estimation of $\ltrue_{j,m}(\bx)$. This assumption collects standard regularity conditions for nonparametric least squares regression with ``outcome'' $\left[\bm{\ell}_{\btheta\btheta}(\by,\bt, \btheta(\bx))\right]_{j,m}$. In particular, these conditions are identical to \cite{Farrell-Liang-Misra2021_Ecma} and \cite{Tang2025_WP}, upon which we rely here.
\begin{assumption}
	\label{asmpt:lambda}
	For all $j, m \in \{1,\ldots,\dtheta\}$ and any $\btheta(\bx)$, Assumption \ref{asmpt:SMOOTH} of the text holds for $\bW = \left( \left[\bm{\ell}_{\btheta\btheta}(\bY,\bT, \btheta(\bX))\right]_{j,m} , \bX'\right)'$ and $\lambda_{j,m}(\bx; \btheta)$ in place of $\ttrue_k(\bx)$, where $\lambda_{j,m}(\bx; \btheta) \in \cG$ is uniquely identified as
		\[	\lambda_{j,m}(\bx; \btheta) = \argmin_{g \in \cG } \E \left[\left( \left[\bm{\ell}_{\btheta\btheta}(\bY, \bT, \btheta(\bX))\right]_{j,m}  -  g(\bX)\right)^2 \right]. \]
\end{assumption}

Under this assumption, from \cite{Farrell-Liang-Misra2021_Ecma} and \cite{Tang2025_WP}, we immediately obtain the following result for a fixed $\btheta(\bx)$.
\begin{lemma}
	\label{appxlem: lambda hat rate 1}
Let $(\by_i', \bt_i', \bx_i')$, $i=1,\ldots, n$ be a random sample obeying Assumption \ref{asmpt:lambda}. Then the estimator of Equation \eqref{appxeqn:lambda hat} obeys
\[
	\left\| \lhat_{j,m}(\cdot; \btheta)  - \lambda_{j,m}(\bx; \btheta) \right\|^2_{L_2(\bX)} = O_P \left( n^{-2p/(2p+\dc)} \log^4 n \right).
\]
\end{lemma}
\begin{proof}
	Immediate from Theorem \ref{thm:normality} of \cite{Farrell-Liang-Misra2021_Ecma} and the approximation result of \cite{Tang2025_WP}. 
\end{proof}
For appropriate $p$ and $\dc$, this rate is sufficiently fast for inference, and we will assume such is the case throughout this section. In particular, the above result justifies deep neural network estimation of $\bLtrue(\bx)$ as a first step for inference whenever $\bLtrue$ does not depend on $\bttrue$. The leading case here is an enriched linear model, that is, Equation \eqref{eqn:glm flm2} below with $G(u) = u$. For the further special case of a scalar binary treatment, this result is the propensity score as explicitly discussed in \cite{Farrell-Liang-Misra2021_Ecma}. Using Theorem \ref{thm:dnn rates} of \cite{Farrell-Liang-Misra2021_Ecma}, albeit one that may still be sufficient for inference.

However, in the general case when $\bLtrue(\bx)$ depends on $\bttrue$, the above result no longer applies. We instead seek to estimate $\ltrue_{j,m}(\bx) = \lambda_{j,m}(\bx; \bttrue)$ using $\lhat_{j,m}(\bx; \bthat)$. Here a simple argument holds if sample splitting is used so that $\bthat$ is obtained on data independent from $\lhat_{j,m}(\bx; \bthat)$. We then can demonstrate validity under the same assumptions as above with an additional smoothness restriction on the loss function. See Remark \ref{rem:sample splitting}. This condition is standard in structural modeling and obeyed by leading examples. The existence of a $\bthat$ obeying the convergence below is immediate from Theorem \ref{thm:dnn rates}, but we leave the requirement generic in the statement below to match the generality of the first-step requirements in Theorem \ref{thm:normality}. Recall that we assume that $p$ and $\dc$ are such that the nonparametric rate is $o(n^{-1/4})$.
\begin{lemma}
	\label{appxlem: lambda hat rate 2}
Let $(\by_i', \bt_i', \bx_i')$, $i=1,\ldots, n$ be a random sample obeying Assumption \ref{asmpt:lambda}. Assume $\left[\bm{\ell}_{\btheta\btheta}(\bY,\bT, \btheta)\right]_{j,m}$ is bounded and continuously differentiable in $\btheta$ with bounded derivative. Further, assume that from an independent sample an estimator $\bthat$ is available that obeys $\left\| \that_j - \ttrue_j\right\|^2_{L_2(\bX)} = o_P \left(n^{-1/4} \right)$. Then the estimator $\lhat_{j,m}(\cdot; \bthat)$ of Equation \eqref{appxeqn:lambda hat} obeys
\[
	\left\| \lhat_{j,m}(\cdot; \bthat)  - \lambda_{j,m}(\bx; \bttrue) \right\|^2_{L_2(\bX)} = O_P \left(n^{-2p/(2p+\dc)} \log^4 n\right) + o_P \left(n^{-1/4} \right).
\]
\end{lemma}
\begin{proof}
	From Lemma \ref{appxlem: lambda hat rate 1},
	\[
		\left\| \lhat_{j,m}(\cdot; \bthat)  - \lambda_{j,m}(\bx; \bthat) \right\|^2_{L_2(\bX)} = O_P \left(n^{-2p/(2p+\dc)} \log^4 n\right),
	\]
	and so by the triangle inequality it remains to show that for the population objects
	\begin{equation}
		\label{appxeqn: lambda proof 1}
		\left\| \lambda_{j,m}(\bx; \bthat) - \lambda_{j,m}(\bx; \bttrue) \right\|^2_{L_2(\bX)} = O_P \left(n^{-2p/(2p+\dc)} \log^4 n\right) + o_P \left(n^{-1/4} \right).
	\end{equation}
	For a generic $\btheta(\bx)$, the population object $\lambda_{j,m}(\bx; \btheta)$ can be defined as the solution to the population analogue of Equation \eqref{appxeqn:lambda hat}, as follows:
	\[
		\lambda_{j,m}(\cdot; \btheta) = \argmin_{g \in \cG} R(\btheta, g)   , \qquad  R(\btheta, g) = \E \left[ \left( \left[\bm{\ell}_{\btheta\btheta}(\bY,\bT, \btheta(\bX))\right]_{j,m}  -  g(\bX)\right)^2 \right],
	\]
	where $\cG$ is the class of functions obeying Assumption \ref{asmpt:lambda}. 
	
	First use the convergence of $\bthat$. Using the difference of squares, Jensen's inequality, and the conditions on the outcome $\left[\bm{\ell}_{\btheta\btheta}(\bY, \bT, \btheta(\bX))\right]_{j,m}$, and Cauchy-Schwarz, we have for any $g(\cdot)$ the following bound.
	\begin{align*}
		\left| R\big(\bthat, g \big) - R\big(\bttrue,  g \big) \right|  & =  \left| \E \left[ \left( \left[\bm{\ell}_{\btheta\btheta}(\bY,\bT, \bthat(\bX))\right]_{j,m}  -  g(\bX)\right)^2 - \left( \left[\bm{\ell}_{\btheta\btheta}(\bY,\bT, \bttrue(\bX))\right]_{j,m}  -  g(\bX)\right)^2  \right] \right| 		 \\
		& = \Bigg| \E \Bigg[ \left( \left[\bm{\ell}_{\btheta\btheta}(\bY,\bT, \bthat(\bX))\right]_{j,m}  - \left[\bm{\ell}_{\btheta\btheta}(\bY,\bT, \bttrue(\bX))\right]_{j,m} \right)   		 \\
		& \qquad \qquad \times \left( \left[\bm{\ell}_{\btheta\btheta}(\bY,\bT, \bthat(\bX))\right]_{j,m}  +  \left[\bm{\ell}_{\btheta\btheta}(\bY,\bT, \bttrue(\bX))\right]_{j,m} - 2 g(\bX) \right)\Bigg] \Bigg|    		 \\
		& \leq  \E \Bigg[ \left| \left[\bm{\ell}_{\btheta\btheta}(\bY,\bT, \bthat(\bX))\right]_{j,m}  - \left[\bm{\ell}_{\btheta\btheta}(\bY,\bT, \bttrue(\bX))\right]_{j,m} \right|   		 \\
		& \qquad \qquad \times \left| \left[\bm{\ell}_{\btheta\btheta}(\bY,\bT, \bthat(\bX))\right]_{j,m}  +  \left[\bm{\ell}_{\btheta\btheta}(\bY,\bT, \bttrue(\bX))\right]_{j,m} - 2 g(\bX) \right| \Bigg]  		 \\
		& \leq C \E \left[ \left| \left[\bm{\ell}_{\btheta\btheta}(\bY,\bT, \bthat(\bX))\right]_{j,m}  - \left[\bm{\ell}_{\btheta\btheta}(\bY,\bT, \bttrue(\bX))\right]_{j,m} \right| \right]  		 \\
		& = O_P\left( \left\| \bthat - \bttrue\right\| \right) = o_p(n^{-1/4}).  
	\end{align*}
	Using this result twice and the fact that each $\lambda_{j,m}(\cdot; \btheta)$ minimizes $R(\btheta, g)$, we find 
	\begin{align*}
		R(\bthat, \lambda_{j,m}(\bx; \bthat)) \leq R(\bthat, \lambda_{j,m}(\bx; \bttrue)) & = R(\bttrue, \lambda_{j,m}(\bx; \bttrue))  + o_p(n^{-1/4})  		\\
		& \leq R(\bttrue, \lambda_{j,m}(\bx; \bthat))  + o_p(n^{-1/4})  		\\
		& \leq R(\bthat, \lambda_{j,m}(\bx; \bthat))  + o_p(n^{-1/4}),
	\end{align*}
	showing that the risks converge. Then by definition $\lambda_{j,m}(\bx; \bthat)$ is the true conditional expectation, and so using the property of least squares loss, we have
	\[
		R\left(\bthat, \lambda_{j,m}(\bx; \bttrue) \right) - R\left(\bthat, \lambda_{j,m}(\bx; \bthat) \right) = \E\left[ \left(\lambda_{j,m}(\bx; \bttrue) -0 \lambda_{j,m}(\bx; \bthat) \right)^2 \right],
	\]
	which completes the proof.
\end{proof}

\subsubsection{Remarks}

\begin{remark}[Other Uses of Influence Functions]
    \label{rem:other uses}
    Influence functions have appeared in many different areas in statistics and our results can potentially be used to extend these contexts. Here we list a few examples. (i) \cite{Athey-Wager2021_Ecma} study policy optimization and show that using an orthogonal score yields faster reminder rates in terms of welfare just as for inference. Our score could be used to bring their insight into new areas. (ii) \cite{Koh-Liang2017_ICML} use influence functions to try to understand ``black-box'' ML methods. Extending this to economic contexts would be valuable in applied research and policy evaluation. (iii) \cite{Robins-etal2008_IMS} use higher-order influence functions to obtain refined semiparametric inference. We conjecture that using automatic differentiation could be used to obtain higher order inference just as with our first order results. (iv) \cite{Firpo-Fortin-Lemieux2009_Ecma} rely on influence functions for distributional statistics, and could potentially be generalized to other models. (v) \cite{Semenova-Chernozhukov2021_EJ} and \cite{Colangelo-Lee2023_WP} use orthogonal scores for two-step \emph{non}parametric inference (following ML) and we conjecture that the same could be done in the enriched structural models considered here. This would be valuable for future research. See also Remark B.3. 
\end{remark}

\begin{remark}[High Dimensional Parametric Approach]
    \label{rem:series approach}
    An alternative approach in two-step inference would be to consider $\bthat(\bx_i)$ as a parametric model (where the weights and biases of the deep net are the parameters) and apply parametric two-step estimation. This is shown to be a valid approximation to the semiparametric case in some contexts by \cite{Ackerberg-etal2012_REStat}. Applying this idea to deep learning may be valid, but is practically infeasible as the number of parameters is too large and the estimator too complex. For example, the equivalent of $\bL(\bx)$ would be a square matrix of dimension equal to the number of parameters in the deep net, which can be extremely large. Computing and inverting such a matrix may be impossible.
\end{remark}

\begin{remark}[Use of Sample Splitting]
	\label{rem:sample splitting}
	Sample splitting (or cross fitting for full efficiency) is not a necessary condition for asymptotic normality. Rather, it is used in semiparametric/DML contexts to provide valid and feasible asymptotic normality under simple and intuitive conditions on first-step estimators. Without sample splitting, asymptotic normality for semiparametric estimation has been obtained under stronger conditions on first step estimators (for just some examples see \cite{Newey1994_Ecma,Hirano-Imbens-Ridder2003_Ecma,Imbens-Newey-Ridder2007_MSE,Cattaneo-Farrell2011_chapter}. Importantly, these conditions difficult to verify for machine learning estimators such as neural networks. A notable feature of these works, compared to the present case and those below, is that these consider the plug-in estimator (not an orthogonal score or influence function), which makes the conditions even stronger. Using influence functions to reduce first-step sensitivity, valid inference has been obtained without sample splitting in some special cases. See \cite{Cattaneo2010_JoE} for series estimators, \cite{Farrell2015_arXiv} for group LASSO estimation, and \cite{Farrell-Liang-Misra2021_Ecma} for deep neural networks. The assumptions in some cases are stronger than what is required by generic DML strategies, as in \cite{Chernozhukov-etal2018_EJ,Chernozhukov-etal2022_Ecma}. See \cite{Chernozhukov-etal2018_EJ} for further discussion, and simulation evidence, for the role of sample splitting in DML contexts. Also, \cite{Newey-Robins2018_WP} establish that cross-fitting has certain optimality properties in that the remainder terms in the stochastic expansion of $\sqrt{n}\left( \bmhat - \bmtrue\right)$ vanish at optimal or near-optimal rates.
	
	In the present context, it is not clear if the approach of \cite{Farrell-Liang-Misra2021_Ecma} can be generalized to the present setting to remove the need for either two- or three-way sample splitting. Inspection of the proof of Lemma \ref{appxlem: lambda hat rate 2} shows that one must establish certain uniformity in $\btheta$ for estimation of $\bLtrue$. This would necessitate more delicate theoretical arguments, taking us beyond the scope of this paper and its current goal of providing DML-based inference for two-step estimation in heterogeneity-enriched structural models, including hewing to the principle of establishing inference under simple conditions.
\end{remark}

\begin{remark}[First-Step Estimation Conditions and Invariance]
Double machine learning provides a method for obtaining a valid first-order distributional approximation under weak, and importantly, simple, conditions on first-step estimators. As discussed in the text, this is crucial when the first step uses machine learning methods about which relatively little is known mathematically. The first-order distributional approximation is invariant to the specific first step estimators. This is a classical feature of semiparametric inference \citep{Newey1994_Ecma,Newey-McFadden1994_handbook} and an important theoretical result. However, this means that (a) by virtue of being general and simple, the assumptions are not minimal and (b) the invariance to the first stage can be a poor finite sample approximation, because different nonparametric estimators naturally perform differently. Addressing (a) and/or (b) requires more delicate calculations that are specific to either the first stage estimator or the inference target. \cite{Cattaneo-Jansson-Ma2019_RESTUD} show that in some problems computationally intensive procedures can be used to weaken first step assumptions. More refined approximations, which account for the first step explicitly, have been obtained in some cases \citep{Cattaneo-Crump-Jansson2014_ET_small,Cattaneo-Jansson-Newey2018_JASA,Cattaneo-Farrell-Jansson-Masini2025_JoE}. Extending any of these to machine learning contexts, and neural networks specifically, would be useful for future research.
\end{remark}

\section{Generalized Linear Models}
    \label{appx:glm}

To help illustrate our results, and because it is a leading use case, this section specializes to modeling the conditional mean with a linear index. This model will also help us link to prior work, both in nonparametrics and semiparametrics. We first show the enriched model and how structural deep learning is done in this case. This is helpful for intuition because it builds so directly on the standard prediction case for neural networks. We then verify the conditions for Theorem \ref{thm:dnn rates} using standard assumptions for these models, as well as showing that the conditional expectation function itself is recovered at the same rate. The influence function of Theorem \ref{thm:normality} is specialized to the enriched generalized linear model, which helps connect with various other literatures as this nests many settings from past work (Appendix C below). Finally a set of remarks connected to various literatures is given at the close of the section.

The parametric structural model in this case is determined by the conditional mean restriction on a scalar outcome $Y$ given a vector of explanatory variables $\bT$. Assume that
\begin{equation}
    \label{eqn:glm parametric}
    \E[Y \mid \bT = \bt] = G(\atrue + {\bbtrue}'\bt ),
\end{equation}
for an inverse link function $G(u): \R \to \R$. The full model may come with assumptions about variances and other parts of the data generating process. Standard examples include linear and logistic regression. Slope coefficients $\bbtrue$ in such models, along with marginal effects in nonlinear models, are among the most commonly studied objects in empirical research. 

To enrich this model with ML, we change the intercept and slope to the parameter functions $\bttrue(\bx) = (\atrue(\bx), \bbtrue(\bx)')'$ and assume that
\begin{equation}
    \label{eqn:glm flm2}
    \E[Y \mid \bT = \bt, \bX=\bx] = G(\atrue(\bx) + \bbtrue(\bx)'\bt ).
\end{equation}
This formulation maintains the structural relationship between $\bT$ and $Y$ but allows for rich heterogeneity in $\bX$. In contrast to our structured approach, the naive ML approach would treat $\bX$ and $\bT$ as equally informative about $Y$, and assume that for an unknown $\eta(\bt,\bx)$ to be estimated
\begin{equation}
    \label{eqn:glm naive}
    \E[Y \mid \bT = \bt, \bX=\bx] = G(\eta(\bx,\bt)),
\end{equation}
or often simply $E[Y \mid \bT = \bt, \bX=\bx] = \eta(\bx,\bt)$, without even the structure of the inverse link.

Deep learning architectures to estimate these two models are shown in Figure \ref{fig:arch-glm}. The loss function is the same in both cases. Panel (a) is the standard ML approach as available in standard software. All information in $(\bX', \bT')$ is fed into the hidden layers. Panel (b) specializes Figure \ref{fig:arch-general} in the text to force the expressivity to learn the slope and intercept parameter functions in order to reduce the loss. The figure illustrates how easy it is to implement the structural approach, with only an extra line or two of code. As an aside, Figure \ref{fig:arch-glm} also illustrates how our method (and theory) applies to generalized additive models, where the different components of $\bttrue$ are known to rely on different subsets of $\bX$: we simply sever the appropriate links, so that separate networks feed into the parameter layer nodes according to the model. 

\begin{figure}
	\centering
	\begin{subfigure}[b]{0.4\textwidth}
		\centering
		\hspace{-0.5in}\scalebox{0.8}{			\def\layersep{1.75cm}
\begin{tikzpicture}[shorten >=1pt,->,draw=black!50, node distance=\layersep]
	\tikzstyle{every pin edge}=[<-,shorten <=1pt]
	\tikzstyle{neuron}=[circle,fill=black!25,minimum size=20pt,inner sep=0pt]
	\tikzstyle{input neuron}=[neuron, fill=blue!50];
	\tikzstyle{hidden neuron}=[neuron, fill=gray!50];
	\tikzstyle{parameter neuron}=[neuron,minimum size=40pt,fill=green!50];
	\tikzstyle{output neuron}=[neuron, fill=red!50];
	
	\tikzstyle{annot} = [text width=4em, text centered]
	
	\node[input neuron] (I-1) at (0,-1) {$x_1$};
	\node[input neuron] (I-2) at (0,-2) {};
	\node[input neuron] (I-3) at (0,-3) {$x_d$};
	\node[output neuron] (I-4) at (0,-4) {$\bt$};
	
	\foreach \name / \y in {1,...,5}
	\path[yshift=0.5cm]
	node[hidden neuron] (Ha-\name) at (\layersep,-\y cm) {};
	
	\foreach \name / \y in {1,...,5}
	\path[yshift=0.5cm]
	node[hidden neuron,right of=Ha-\name] (Hb-\name) at (\layersep,-\y cm) {};

    \node[output neuron, pin={[pin edge={->}]right:$\widehat{y} = G(\cdot) $}, right of=Hb-3] (O) { \ $\widehat\eta(\bx,\bt)$ \ };
    
	\foreach \source in {1,...,4}
	\foreach \dest in {1,...,5}
	\path (I-\source) edge (Ha-\dest);
	
	\foreach \source in {1,...,5}
	\foreach \dest in {1,...,5}
	\path (Ha-\source) edge (Hb-\dest);

	\foreach \source in {1,...,5}
	\path (Hb-\source) edge (O);

	\node[annot,above of=Ha-1, xshift=\layersep/2, node distance=1cm] {Hidden layers};
	\node[annot,above of=Ha-1, node distance=1cm] (hal) {};
	\node[annot,above of=Hb-1, node distance=1cm] (hbl) {};
	\node[annot,left of=hal] {Inputs};
	\node[annot,right of=hbl] {Prediction};
	
\end{tikzpicture}}
		\caption{Standard prediction architecture}
	\end{subfigure}
	\begin{subfigure}[b]{0.5\textwidth}
	\centering
		\scalebox{0.8}{			\def\layersep{1.75cm}
			\begin{tikzpicture}[shorten >=1pt,->,draw=black!50, node distance=\layersep]
			    \tikzstyle{every pin edge}=[<-,shorten <=1pt]
			    \tikzstyle{neuron}=[circle,fill=black!25,minimum size=20pt,inner sep=0pt]
			    \tikzstyle{input neuron}=[neuron, fill=blue!50];
			    \tikzstyle{hidden neuron}=[neuron, fill=gray!50];
			    \tikzstyle{parameter neuron}=[neuron,minimum size=40pt,fill=green!50];
			    \tikzstyle{output neuron}=[neuron, fill=red!50];
			
			    \tikzstyle{annot} = [text width=4em, text centered]
			
				\node[input neuron] (I-1) at (0,-1.5) {$x_1$};
				\node[input neuron] (I-2) at (0,-2.5) {};
				\node[input neuron] (I-3) at (0,-3.5) {$x_d$};
				
			    \foreach \name / \y in {1,...,5}
			        \path[yshift=0.5cm]
			            node[hidden neuron] (Ha-\name) at (\layersep,-\y cm) {};
			
			    \foreach \name / \y in {1,...,5}
			        \path[yshift=0.5cm]
			            node[hidden neuron,right of=Ha-\name] (Hb-\name) at (\layersep,-\y cm) {};
			
				\node[yshift=10pt, parameter neuron, right of=Hb-2] (P-1) { \ $\ahat(\bx)$ \ };
				\node[yshift=-10pt, parameter neuron, right of=Hb-4] (P-2) { \ $\bbhat(\bx)$ \ };

			    \node[output neuron, right of=Hb-2, xshift=\layersep] (T) {$\bt$};
                \node[output neuron, pin={[pin edge={->}]right:$\widehat{y} = G(\cdot) $}, right of=Hb-4, xshift=\layersep] (O) { \  \ };
			
			    \foreach \source in {1,...,3}
			        \foreach \dest in {1,...,5}
			            \path (I-\source) edge (Ha-\dest);
			            
			    \foreach \source in {1,...,5}
			        \foreach \dest in {1,...,5}
			            \path (Ha-\source) edge (Hb-\dest);

			    \foreach \source in {1,...,5}
			        \foreach \dest in {1,2}
			            \path (Hb-\source) edge (P-\dest);
			
			        
				\path (P-1) edge (O);
				\path (P-2) edge (O);
				\path (T) edge (O);
				
			    \node[annot,above of=Ha-1, xshift=\layersep/2, node distance=1cm] {Hidden layers};
			    \node[annot,above of=Ha-1, node distance=1cm] (hal) {};
			    \node[annot,above of=Hb-1, node distance=1cm] (hbl) {};
			    \node[annot,left of=hal] {Inputs};
			    \node[annot,right of=hbl] (pl) {Parameter layer};
			    \node[annot,right of=pl] {Model layer};
			
			\end{tikzpicture}}
		\caption{Structured deep learning}
	\end{subfigure}
	\caption{Comparing the standard prediction-focused architecture to learning parameter functions using the structured approach, matching the models of \eqref{eqn:glm naive} and \eqref{eqn:glm flm2} respectively.}
	\label{fig:arch-glm}
\end{figure}

Interpreting $\bbtrue(\bx)$ in \eqref{eqn:glm flm2} is the same as interpreting the original (homogeneous) slope $\bbtrue$ in \eqref{eqn:glm parametric}. In contrast, \eqref{eqn:glm naive} is unstructured and uninterpretable. Here $\eta(\bx,\bt)$ is truly a nuisance function, where ``nuisance'' is taken literally to mean annoying and uninteresting. Economically, as demonstrated in Section \ref{sec:structure}, the lack of structure makes this output not useful. It is difficult, if not impossible, to recover many second stage objects without structure, beyond (weighted) average derivatives. Further, from a statistical point of view, $\eta(\bx,\bt)$ can only be learned at a much slower rate, governed by $\dx + \dt$, compared to learning $\bbtrue(\bx)$ which depends only upon $\dx$. Though generally assumed away in theory since all dimensions are finite, this difference matters in applications. All in all, it is better to view \eqref{eqn:glm flm2} as an enriched version of \eqref{eqn:glm parametric} rather than a restricted version of \eqref{eqn:glm naive}.

\subsection{Deep Neural Networks for Generalized Linear Models}

The theoretical results for deep neural networks (this subsection) and inference (next subsection) for the special case of the model \eqref{eqn:glm flm2} are useful to illustrate the required assumptions and compare to prior work. Further, these special cases are directly useful in many applications, given the popularity of the model.

For first step estimation, we can verify the high level conditions using the following familiar, primitive assumptions. 
\begin{assumption}
    \label{asmpt:glm}
    (i) The conditional expectation $G(\atrue(\bx) + \bbtrue(\bx)'\bt )$ enters the loss through a known, real-valued transformation $g(\cdot)$, where (i) $g$ and $G$ are continuously invertible and $g/\|g\|_\infty$ and $G/\|G\|_\infty$ belong to $\cW^{p, \infty}([-1, 1])$, for $p\geq 3$. (ii) Assumption \ref{asmpt:dgp dnn} holds with $\ell(\by,\bt,\btheta(\bx) )$ replaced by $\ell(\by,g)$, and the conditions therein apply to the scalar argument $g$. (iii) The eigenvalues of $\E[\bT\bT' \mid \bX=\bx]$ are bounded and bounded away from zero uniformly in $\bx$.
\end{assumption}
Condition (i) ensures that the loss function is sufficiently smooth while (ii) and (iii) ensures the curvature through the standard positive variance condition. These conditions are familiar from the parametric case, where, for example, assuming that $\E[\bT\bT']$ is positive definite would be standard. 

Specializing Theorem \ref{thm:dnn rates} to this case, we have the following result.
\begin{corollary}
    \label{cor:glm rates}
    Let the conditions of Theorem \ref{thm:dnn rates} and Assumption \ref{asmpt:glm} hold. Then for a DNN structured according to Figure \ref{fig:arch-glm}, $\| \ahat - \atrue \|_{L_2(\bX)}^2 = O(n^{-2p/(2p+\dc)} \log^4 n)$ and $\| \bhat_j - \btrue_j \|_{L_2(\bX)}^2 = O(nn^{-2p/(2p+\dc)} \log^4 n)$ for $j=1,\ldots,\dt$. Further $\| G(\ahat(\bx) + \bbhat(\bx)'\bt) - G(\atrue(\bx) + \bbtrue(\bx)'\bt ) \|_{L_2(\bX)}^2 = O(n^{-2p/(2p+\dc)} \log^4 n)$.
\end{corollary} 
Here we give two simple results (cf the more full statement in Theorem \ref{thm:dnn rates}). First, we show that we can estimate the heterogeneous intercept and slope parameters at the appropriate rate, depending on the (continuous) dimension of the heterogeneity. This is direct from Theorem \ref{thm:dnn rates}. This is required as economic constructs depend on these parameters, rather than on the conditional expectation function as a whole. But we also give a result for the mean function in the enriched model \eqref{eqn:glm flm2}. This result may be of independent interest, as it establishes that structural deep learning has good statistical properties in varying coefficient models, additive models, and other such cases. It is also useful for comparing to the more typical use of inference after ML, where the (prediction) function $\E[Y \vert \bx, \bt]$ would be unstructured and would be learnable at a rate dependent on $\dx + \dt$. For example, it is much easier (statistically) to estimate $\E[\bbtrue(\bX)]$ than the average derivatives $\E[\partial G(\eta(\bX,\bT)) / \partial \bt]$, even though both are linear summaries of the dependence of $Y$ on $\bT$.

\subsection{Influence Function for Generalized Linear Models}

Specializing our influence function to case of \eqref{eqn:glm flm2} helps connect with past work. To state the result compactly and clearly, define $\bT_1 = (1,\bT')'$ and $\bt_1 = (1,\bt')'$, so that $G(\btheta(\bx)'\bt_1) = G(\alpha(\bx) + \bbeta(\bx)'\bt)$. The derivative with respect to the scalar argument is denoted $\dot{G}(\btheta(\bx)'\bt_1) = d G(u) / d u$ at $u = \btheta(\bx)'\bt_1 = \alpha(\bx) + \bbeta(\bx)'\bt$.

Assume the standard estimation approach is taken so that $\bm{\ell}_{\btheta}(y,\bt, \btheta(\bx)) =  \bT_1(G(\btheta(\bx)'\bt_1)-y)$. Then $\bL(\bx; \btheta) = \E[\dot{G}(\btheta(\bx)'\bt_1)\bT_1\bT_1'\mid\bX=\bx]$, making the influence function in this case $\bpsi - \bmtrue$, with
\begin{align}
    \begin{split}
    \label{eqn:glm IF}
        & \bpsi(y,\bt,\bx, \btheta, \bL)  =  \bH\left(\bx, \btheta(\bx); \btstar\right)  
        \\
        & \qquad  + \bH_{\btheta}\left(\bx, \btheta(\bx ); \btstar\right) \E\left[\dot{G}(\btheta(\bx)'\bt_1)\bT_1\bT_1'\mid\bX=\bx\right]^{-1} \bT_1 (y - G(\btheta(\bx)'\bt_1)),
    \end{split}
\end{align}
Further, if $\dt = \dmu = 1$, so that $\bttrue = (\atrue,\btrue)'$, this can be written
\begin{align}
    \begin{split}
    \label{eqn:IF univariate}			
		& \psi(y, t, \bx, \btheta, \bL) 
         = H\left(\bx, \btheta(\bx); \tstar \right)     		\\
			& \quad    +    \frac{H_\alpha(\bx; \btheta, \tstar) \Big(\lambda_2(\bx; \btheta) - \lambda_1(\bx; \btheta) t \Big) +    H_\beta(\bx; \btheta, \tstar) \Big(\lambda_0(\bx; \btheta) t - \lambda_1(\bx; \btheta) \Big)}{\lambda_2(\bx; \btheta)\lambda_0(\bx; \btheta) - \lambda_1(\bx; \btheta)^2} \Big(y - G(\btheta(\bx)'\bt_1) \Big),
		\end{split}
\end{align}
where $H_\alpha(\bx; \btheta, \tstar) = \partial H(\bx, \btheta(\bx); \tstar)/\partial \alpha$ and $H_\beta(\bx; \btheta, \tstar) = \partial H(\bx, \btheta(\bx); \tstar)/\partial \beta$, and we define the short-hand $\lambda_k(\bx; \btheta) = \E [ \dot{G}(\btheta(\bx)'\bt_1)  T^k | \bX = \bx ]$, for $k=\{0,1,2\}$, i.e., $\lambda_{j+m-2}(\bx; \btheta) = \lambda_{j,m}(\bx; \btheta) = \E [ \dot{G}(\btheta(\bx)'\bt_1)  T^{j+m-2} | \bX = \bx ]$.

In nonlinear models the use of three-way sample splitting is clear: the regressand of $\bL(\bx; \btheta) = \E[\dot{G}(\btheta(\bx)'\bt_1)\bT_1\bT_1'\mid\bX \!=\! \bx]$ depends on the unknown $\bttrue(\bx)$ through $\dot{G}(\btheta(\bx)'\bt_1)$, which must be estimated in a first step. For linear models, three splits are not necessary. See Section \ref{appx:estimation of Lambda}, Remark \ref{rem:sample splitting}, and discussion in the text for more regarding estimation of $\bLtrue(\bx)$. Note however that this is not true of 
$H_\alpha(\bx; \btheta, \tstar)$ and $H_\beta(\bx; \btheta, \tstar)$ (or $\bH_{\btheta}\left(\bx, \btheta(\bx ); \btstar\right)$ more generally) which depend on $\btheta(\bx)$ also, but these do not require further estimation: $\bH$ is known and there is not additional projection/conditioning as there is with $\bL(\bx; \btheta)$, so $\bH_{\btheta}\left(\bx, \bthat(\bx ); \btstar\right)$ can be used directly, whether the functional form is known or the values are obtained via automatic differentiation.

This result recovers several known influence functions. For linear models, it matches \cite{Hahn1998_Ecma} for binary treatments and \cite{Graham-Pinto2022_JoE} for continuous treatments. For mean square projections in general, see \cite{Newey1994_Ecma}. For semiparametric regression, where $\bbtrue$ is assumed constant, our result gives valid inference but does not match the efficient influence function \citep{Mammen-vandeGeer1997_AoS,Vansteelandt-Dukes2022_JRSSB} because we do not impose the constancy of the slope.

We can also effortlessly obtain new results. For example, in Section \ref{sec:application} we conduct inference on the fully flexible average marginal effects in a logistic regression, where $\bmtrue = \E[\dot{G}(\bttrue(\bx)'\bt_1)\bbtrue(\bX)]$. In this case $\bH$ and its derivatives are available in closed form, and the derivatives of the loss are well known, but again, this is not necessary. We also use \eqref{eqn:glm IF} for a function $\bH$ that is not available in closed form \citep{Spall1986_CSTM,Jorgensen1993_Bmka}.

\subsection{Proof of Corollary \ref{cor:glm rates}}

For this derivation, recall that $\bT_1 = (1, \bT')'$ and $\bttrue = (\atrue, {\bbtrue}')'$ (and similarly for realizations, estimators, etc). First, consider identification. Since $G$ is invertible and conditional expectations are always identified, the quantity $G^{-1} ( \E[Y \mid \bX = \bx, \bT = \bt] )$ is identified. Suppose that $\bttrue(\bx)$ is not identified. Then there exists $\btheta_1(\bx)$ and $\btheta_2(\bx)$ such that $G^{-1} \left( \E[Y \mid \bX = \bx, \bT = \bt] \right) = \btheta_1(\bx)'\bT_1 = \btheta_2(\bx)'\bT_1$ a.e.\ or equivalently that for $\btheta_*(\bx)= \btheta_1(\bx) - \btheta_2(\bx)$, $\btheta_*(\bx)'\bT_1 = 0$.
But $\btheta_*(\bx)'\bT_1 = 0$ a.e.\ implies that
\[
    0 = \E\left[ \left(\btheta_*(\bx)'\bT_1\right)^2\mid \bX = \bx \right] = \btheta_*(\bx)' \E[\bT_1\bT_1' \vert X] \btheta_*(\bx),
\]
but because the middle matrix is positive definite, this means that $\btheta_*(\bx)$ is zero. For linear $G$, this argument is given in \citet{Huang-Shen2004_SJS} and elsewhere.

The estimation bounds follow immediately from Theorem \ref{thm:dnn rates}, given the conditions of Assumption \ref{asmpt:glm}. The fact that $\E[\bT_1\bT_1' \mid \bX]$ is (uniformly) positive yields
\begin{align*}
	\E_{\bX,\bT}  \left[ \left( \bthat(\bX)'\bT_1  - \bttrue(\bX)'\bT_1 \right)^2 \right] \textbf{}
    & = \E_{\bX}\left[ \left(\bthat(\bX) - \bttrue(\bX)\right)'\E[\bT_1\bT_1' \mid \bX]\left(\bthat(\bX) - \bttrue(\bX)\right) \right] \\
    & \geq C \E_{\bX}\left[ \left(\bthat(\bX) - \bttrue(\bX)\right)'\left(\bthat(\bX) - \bttrue(\bX)\right) \right].
\end{align*}
This verifies the curvature condition on the loss function. The continuity condition holds because the loss is smooth in $g$ and the linear index can be recovered from $g(G(\btheta(\bx)'\bT_1))$. The structure of the network ensures that the network and the smoothness of the loss imply that the approximation and bounds immediately apply to the function $g(G(\btheta(\bx)'\bT_1))$, and the smoothness of these functions mean that the linear index $\btheta(\bx)'\bT_1$ can be recovered.   \qed

\subsection{Remarks}

\begin{remark}[Origins of Model \eqref{eqn:glm flm2}]
    \label{rem:model origins}
    Models of the form \eqref{eqn:glm flm2} are common in the literature, variously referred to as ``varying coefficient'' models \citep{Cleveland-Grosse-Shyu1991_BookChap,Hastie-Tibshirani1993_JRSSB}, ``functional coefficient'' models \citep{Chen-Tsay1993_JASA}, or ``smooth coefficient'' models \citep{Li-etal2002_JBES}, and fall into the class of ``extended linear models'' as in \cite{Stone-Hansen-Kooperberg-Truong1997_AoS}. \cite{OHagan1978_JRSSB} may be the earliest treatment. See \cite{Vansteelandt-Dukes2022_JRSSB} for recent related results for inference and heterogeneity. Our results apply to all these cases as well as other similar models including generalized additive models or further restrictions such as partially linear models. See also Remarks \ref{rem:adaptivity} and \ref{rem:CATE} and other discussion in Appendix \ref{appx:examples}.
\end{remark}

\begin{remark}[Adaptivity]
    \label{rem:adaptivity}
    Specific forms of deep neural networks are known to be adaptive to structures like \eqref{eqn:glm flm2} in the sense that if the structure holds, then even if estimation is done under the generic \eqref{eqn:glm naive}, the estimator will still obtain the same rate as though \eqref{eqn:glm flm2} was imposed \citep{Bach2017_JMLR,Bauer-Kohler2019_AoS,Schmidt-Hieber2020_AoS,Kohler-Krzyzak-Langer2022_IEEE,Fan-Gu2024_JASA,Fan-Jana-Kulkarni-Yin2025_WP}. This is perhaps not surprising, since neural networks are based on compositions and \eqref{eqn:glm flm2} is based on a composition of simpler functions. This is still not useful for our purpose because $\bttrue(\bx) = (\atrue(\bx), \bbtrue(\bx))'$ cannot be recovered from an adaptive procedure and used in the second stage. The estimator \emph{statistically} adapts to the structure, but not \emph{economically}. Economic structure is enforced on the data, not discovered from the data, as demonstrated in Section \ref{sec:structure}. Further, experience shows that, despite the (asymptotic) theory of adaptivity, finite sample performance is improved by structure.
\end{remark}

\begin{remark}[CATE Estimation]
    \label{rem:CATE}
    A large recent strand of machine learning literature studies the conditional average treatment effect (CATE) function, which corresponds to $\btrue(\bx)$ in \eqref{eqn:glm flm2} with identity $G(u)$ and a scalar, binary $\bT$ (so that \eqref{eqn:glm flm2} is without loss of generality, i.e. equivalent to \eqref{eqn:glm naive}). \cite{Farrell-Liang-Misra2021_Ecma} study this case. Under certain conditions $\btrue(\bx)$, being the difference between two prediction functions, can be recovered faster than the typical nonparametric bound of \cite{stone1982optimal}, which is itself faster than Corollary \ref{cor:glm rates}. Generally, this is possible when $\beta(\bx)$ is somehow ``simpler'' than its components, such as being smoother or lower dimensional. See \cite{Kennedy-Balakrishnan-Robins-Wasserman2024_AoS} for optimal rates, further discussion, and many more references. Some of these methods rely on orthogonal scores, and so it would be interesting to see if our score and neural networks could be adapted to provide similar rate improvements in other economic contexts.
\end{remark}

\section{Examples and Discussion}
    \label{appx:examples}

In this section we discuss a few special cases of our methodology to build intuition and connect to prior work, particularly other work on semiparametric inference. These examples help place our work in the literature, illustrate how familiar models can be enriched with deep learning, and show how the identification assumptions required on parametric models carry over.

\subsection{Average Effect of a Binary Treatment}
    \label{appx:ATE}

Perhaps the most well known and commonly used influence function is for the average effect of a binary treatment in observational data. The history of this influence function is illustrative: it was characterized precisely first for the purposes of efficiency considerations \citep{Hahn1998_Ecma}, later used to show that certain plug-in estimators could be efficient \citep{Hirano-Imbens-Ridder2003_Ecma,Imbens-Newey-Ridder2007_MSE,Cattaneo-Farrell2011_chapter} under strong assumptions, then recently used to obtain valid (and in this case efficient) inference under weaker first-stage conditions \citep{Cattaneo2010_JoE,Farrell2015_arXiv}. This influence function matches the doubly robust estimator \citep{Robins-Rotnitzky-Zhao1994_JASA,Robins-Rotnitzky-Zhao1995_JASA} and has been used in other contexts to obtain sharper results, see Remarks \ref{rem:other uses} and \ref{rem:CATE}.

Here we have a scalar outcome and $\bT = T = \{0,1\}$ is the scalar binary treatment indicator. Let $Y(t)$ be the potential outcome under treatment $T = t$ and assume that the observed outcome $Y = T Y(1) + (1-T) Y(0)$ along with unconfoundedness. The parameter of interest is $\mtrue = \E[Y(1) - Y(0)]$. In a randomized experiment without additional covariates, one estimates $\mtrue$ with a difference in means, which is equivalent to running a regression of the observed outcome $Y$ on the dummy variable $T$. The enriched structural version is then \eqref{eqn:glm flm2}, with $G(u) = u$, so that $\E[Y \mid \bx, \bt] = \atrue(\bx) + \btrue(\bx) \bt$. In this notation, $\btrue(\bx) = \E[Y(1) - Y(0) \mid \bx]$ is the conditional average treatment effect (CATE) function (Remark \ref{rem:CATE}). 

The average treatment effect $\mtrue = \E[\btrue(\bX)]$ is defined via $\bH(\bx,\bttrue; \tstar) = \btrue$. Equation \eqref{eqn:IF univariate} matches \cite{Hahn1998_Ecma}, because in this case $H_\alpha = 0$, $H_\beta = 1$, $\lambda_0(\bx; \btheta) = 1$, and $\lambda_1(\bx; \btheta) = \lambda_2(\bx; \btheta) = \P[T = 1 \vert \bX = \bx] := p(\bx)$, the propensity score, and so by adding and subtracting $p(\bx)$ and using the fact that $(1-t)t = 0$, we have
\begin{align*}
    & \psi(\bw, \btheta, \bL) \\ 
    & = \beta(\bx) +  \frac{H_\alpha(\bx; \btheta, \tstar) (\lambda_2(\bx; \btheta) - \lambda_1(\bx; \btheta) t ) +    H_\beta(\bx; \btheta, \tstar) (\lambda_0(\bx; \btheta) t - \lambda_1(\bx; \btheta) )}{\lambda_2(\bx; \btheta)\lambda_0(\bx; \btheta) - \lambda_1(\bx; \btheta)^2} (y - G(\btheta(\bx)'\bt ) )       \\
    & = \beta(\bx) +  \frac{ ( t - p(\bx) ) (y - \alpha(\bx) - \beta(\bx) t ) }{p(\bx) - p(\bx)^2}          \\
    & = \beta(\bx) +  \frac{ [ (1-p(\bx))t - p(\bx)(1-t) ] ( y - \alpha(\bx) - \beta(\bx) t)  ) }{p(\bx)(1 - p(\bx))}          \\
    & = \beta(\bx) +  \frac{ (1-p(\bx))t ( y - \alpha(\bx) - \beta(\bx) t)  ) }{p(\bx)(1 - p(\bx))}       -  \frac{  p(\bx)(1-t)  ( y - \alpha(\bx) - \beta(\bx) t)  ) }{p(\bx)(1 - p(\bx))}          \\
    & = \beta(\bx) +  \frac{ t ( y - \alpha(\bx) - \beta(\bx) t)  ) }{p(\bx)}       -  \frac{  (1-t)  ( y - \alpha(\bx) )  ) }{(1 - p(\bx))} .
\end{align*}
In this example, the standard overlap assumption, that the propensity score is bounded away from zero and one, ensures that $\bLtrue(\bx)^{-1}$ is well behaved: the determinant of $\bLtrue(\bx) = p(\bx)(1-p(\bx))$, the initial denominator above.

It is straightforward to extend this example in a number of directions. Additional mean parameters could be added to cover average treatment effects for specific treatment groups or multi-valued treatments (see \cite{Cattaneo2010_JoE} and \cite{Cattaneo-Farrell2011_chapter} for inference using classical nonparametrics (series) and \cite{Farrell2015_arXiv} for machine learning (group lasso) results). 

Beyond means, the framework makes it easy to consider other objections. For example one might be interested in the variance of $Y(1)$ versus that of $Y(0)$, to assess riskiness of treatments over the population, by taking a quasi-likelihood approach, i.e., letting $\ell(\by,\bt,\btheta(\bx))$ include the variance instead of simply fitting least squares. The conditions for convexity of this loss are well-known from likelihood theory and can be directly used here.

\subsection{Partially Linear Models}

Partially linear models are a common case for semiparametric inference, dating to the seminal work of \cite{Robinson1988_Ecma}. Here \eqref{eqn:glm flm2} holds, but with $\bbtrue(\bx) = \bbtrue$ constant. Restricting the slope to be constant rules out all treatment effect heterogeneity and is a strong assumption that should be used with caution. 

For simplicity, consider the case with a scalar treatment variable. If $\btrue$ is the parameter of interest, \eqref{eqn:IF univariate} shows that $\psi(\bw, \bttrue, \bL) - \btrue$ is
\[
	  \left[\lambda_2(\bx; \btheta) - \frac{\lambda_1(\bx; \btheta)^2}{\lambda_0(\bx; \btheta)}\right]^{-1}\left(t - \frac{\lambda_1(\bx; \btheta)}{\lambda_0(\bx; \btheta)}\right) \Big(y - G(\atrue(\bx) + \btrue t ) \Big).
\]

We must assume that $\lambda_2(\bx; \btheta)\lambda_0(\bx; \btheta) \neq \lambda_1(\bx; \btheta)^2$, which for identity $G$ requires positive conditional variance of $T$. In nonlinear models the conditional moments will be weighted by $\dot{G}$ if we have used the appropriate loss. In some cases the nonsingularity will follow from other regularity conditions, such as for the logistic link, where $\dot{G} = G(1-G)$ and the Hessian is invertible under bounded covariates and we use the log-likelihood.

Partially linear models have featured prominently in the recent literature on high dimensional and ML settings, particularly with idenity link. The pioneering work of \cite{Belloni-Chernozhukov-Hansen2014_REStud} established valid inference after lasso selection in this context. \cite{Chernozhukov-etal2018_EJ} use it as the leading example of their generic results, and present several different Neyman orthogonal scores that could be used, none of which agree with ours due to the fact that we do not impose constant slope.
\cite{Cattaneo-Jansson-Newey2018_JASA} study high dimensional linear models, including many-terms series settings, establish a more refined distributional approximation, and study standard error (in)validity. For the case of nonlinear link function, \cite{Carroll-Fan-Gijbels-Wand1997_JASA}, \cite{Mammen-vandeGeer1997_AoS}, and \cite{Vansteelandt-Dukes2022_JRSSB} are closest to our work, while \cite{Belloni-Chernozhukov-Wei2016_JBES} study high-dimensional sparse models. Not imposing the constant slope means we do not attain the efficiency bound \citep{Mammen-vandeGeer1997_AoS,Vansteelandt-Dukes2022_JRSSB} in general, but only under restrictions on the variance, such as $Y$ being homoskedastic in $\bT$ in the linear case. 

However, it is trivial to impose a constant slope by changing the architecture in Figure \ref{fig:arch-glm} so that only $\alpha(\bx)$ is flexible. Establishing that such a slope estimate is root-$n$ consistent is beyond the scope here, but appears natural as the functional approximation holds without essential change. Deriving the statistical properties of this estimate would be interesting future work.

Finally, our results could be used to study other components of the partially linear setting. For example, in both empirical finance \cite{Cattaneo-Crump-Farrell-Schaumburg2020_REStat} and applied microeconomics \cite{Cattaneo-Crump-Farrell-Feng2024_AER} the function $\alpha(\bx)$ is of direct interest and we could conduct inference on the average.

\subsection{Continuous Treatment Variables}

\cite{Wooldridge2004_WP} and \cite{Graham-Pinto2022_JoE} consider the linear model case and discuss conditions for a causal interpretation of $\E[\bbtrue(\bX)]$. Our result in \eqref{eqn:glm IF} matches the locally efficient influence function of \cite{Graham-Pinto2022_JoE}. \cite{Chernozhukov-Demirer-Lewis-Syrgkanis2019_NIPS} use the model with the goal of policy targeting.

Nonlinear models with continuous treatments are common in many contexts. Th binary choice problem considered in the main paper is but one case. Our method could be used to enrich other structural models in this context. As a first example, consider the so-called Berry logit \citep{Berry1994_RAND} model for demand. Here the outcome is the market share distribution across firms. The researcher observes $\{Y_{jm}\}$ which represent a collection of $j=0, \ldots, J$ market shares across $m=1 \dots M$ markets. The objective is then to model these as a function of firm (marketing) decisions $\bT_{jm}$ (see e.g. \cite{Nevo2001}). We can introduce heterogeneity across markets by allowing for the marketing effects to be moderated by consumer characteristics $\bx_{m}$, so that we can write a collection of $(J-1)$ equations as follows
	\[
		\E\left[ \left. \log \left(\frac{Y_{jm}}{Y_{0m}} \right) \right| \bX = \bx_m, \bT_{jm} = \bt_{jm} \right] = \atrue_j(\bx_m) + \bbtrue(\bx_m)'(\bt_{jm} - \bt_{0m}).
	\]
Stacking these equations and the corresponding data allows us to construct an estimator for $\alpha_{0j}(\bx_m)$ and $\bbtrue(\bx_m)$. We could extend this to include instruments \citep{Okui_etal2012}. 

Relatedly, the first version of this paper on arXiv \citep{FLM2_2020_arXiv} does explicit calculations for multinomial logistic regression, both conditional/McFadden and unrestricted. This is a special case of Section \ref{sec:framework}, but is notationally involved, so those derivations may be of interest. Similar to the above, it involves a scalar, discrete outcome $Y$ taking on values $j=0, \ldots, J$, with $\P[Y = j \mid \bX = \bx, \bT = \bt] = G_j\left(u_1, u_2, \ldots, u_J \right)$, where $G_j = \exp\{u_j\} / (\exp\{u_0\} + \sum_{j'=1}^J \exp\{u_{j'}\})$, with utilities $u_0 = 0$ and $u_j(\bx, \bt) = \ttrue_{0,j}(\bx) + \bttrue_{1,j}(\bx)'\bt$ or $u_j(\bx, \bt) = \ttrue_{0,j}(\bx) + \bttrue_{1}(\bx)'\bt$ for the McFadden logit.

Second, consider the Cobb-Douglas production function with heterogeneous parameters, which is given by $Y=CK^{\ttrue_1(\bx)}L^{\ttrue_2(\bx)}$. The standard approach is to take logs and assume 
\[
	\E\left[ \log {Y}\mid \bX = \bx, K = k,L=l \right] = \log C + \ttrue_1(\bx) \log k + \ttrue_2(\bx)\log l.
\]
We can then estimate the structural parameters using our deep learning approach and conduct inference to decide if on average the technology exhibits increasing, constant, or decreasing returns to scale by computing $\bmtrue = \E[\ttrue_1(\bx)+ \ttrue_2(\bx)]$. The Cobb-Douglas specification has also been used in demand settings and marketing mix models and the framework described above would be readily applicable there as well.

\subsection{Fractional Outcomes}

The case of nonlinear \eqref{eqn:glm flm2} (nonidentity $G(u)$) is less well studied. The empirical application in Sections 2 and 4 uses a logistic link. To give another example, consider a fractional outcome model where $Y$ is continuous but restricted to lie in $[0,1]$. Following the seminal treatment of \cite{Papke-Wooldridge1996_JAE}, we take a quasi-likelihood approach, using logistic distribution with mean given by Equation \eqref{eqn:glm flm2}: $\E[Y \mid \bT \!=\! \bt, \bX \!=\! \bx] = G(\atrue(\bx) + \btrue(\bx)'t)$. \cite{Papke-Wooldridge1996_JAE} explicitly advocate the use of structure to ensure that the outcomes remain on the unit interval and argue that this specification is valid even at the endpoints and is more practically relevant then transformations of the dependent variable.

In the application of \cite{Papke-Wooldridge1996_JAE}, the data is at the firm level, the outcome is 401(k) participation, and the policy variable is the firm's rate of contribution matching. The quantities of interest are the marginal effect of the match rate on participation and the degree to which this marginal effect exhibits diminishing patterns. Our framework makes it trivial to enrich this model with heterogeneity across firms and then conduct inference on average marginal effects or the average change in the marginal effect, given by
\[
    \text{AME}\left(\tstar\right)=\mathbb{E}\left[\left.\frac{\partial \mathbb{E}[Y\mid\mathbf{X},t]}{\partial t}\right|_{t=\tstar}\right]
    \quad \text{ and } \quad
    \text{ACME}\left(\tstar\right)=\mathbb{E}\left[\left.\frac{\partial^{2}\mathbb{E}[Y\mid\mathbf{X},t]}{\partial t^{2}}\right|_{t=\tstar}\right].
\]
Because of the structure of the model, these are easily recovered in the form of $\bmtrue$, 
by taking $H_{\text{AME}}(\bx, \btheta;\tstar) = \beta G(\btheta(\bx)'\bt_1) (1-G(\btheta(\bx)'\bt_1))$ and $H_{\text{ACME}}(\bx, \btheta;\tstar) = {\beta}^2 G(\btheta(\bx)'\bt_1)(1-G(\btheta(\bx)'\bt_1))(1-2G(\btheta(\bx)'\bt_1))$, respectively.

It is useful to contrast our model with the unstructured, naive ML approach, as in \eqref{eqn:glm naive}. We impose structure on the nature of the effect of $\bT$ on $\bY$. For unrestricted effects of continuous treatments using influence functions, see \cite{Kennedy-Ma-McHugh-Small2017_JRSSB} and \cite{Colangelo-Lee2023_WP}. The unrestricted model may increase the generality of the results but can make inference and interpretation more difficult. A common parameter of interest in such cases is the average derivative \citep{Powell-Stock-Stoker1989_Ecma,Newey-Stoker1993_Ecma}. This represents the average of a linear approximation of an unstructured relationship of $\bT$ to $Y$. Our approach is perhaps more direct and transparent: if a linear summary is of interest in the end, we directly enrich the linear approximation, rather than recover it from a more complex object. This contrast can also be seen in the fractional outcome models, where recovering the second derivative of a complex $G(\widehat\eta(\bx,\bt))$ could be challenging but our approach is transparent and simple.

\subsubsection{Application -- American Community Survey}

To illustrate this idea, we revisit the American Community Survey data studied by \cite{Cattaneo-Crump-Farrell-Feng2024_WP}. Replication files are available at \url{https://github.com/maxhfarrell/FLM2}. This is a simple, and narrow, empirical exercise, intended only as an example. We will apply our method to the setting of Figure 4(a) in \cite{Cattaneo-Crump-Farrell-Feng2024_WP}. The data is at the zip code tabulation area level and has $n=27,985$ samples. The outcome $Y$ is the percentage of individuals without health insurance and the treatment variable $T$, which is of course not randomized, is a binary indicator for low and high population density states, defined as those with population densities below or above 100 people per square mile, respectively. Density is defined as the average population per square mile, and the data is available from the Census Bureau. The heterogeneity variable $X$ is per capita income. 

\cite{Cattaneo-Crump-Farrell-Feng2024_WP} are concerned with nonparametric properties and inference of the functions $G(\atrue(x))$ and $G(\atrue(x) + \btrue(x))$, in the notation of \eqref{eqn:glm flm2}. Here we will conduct semiparametric inference on the average effect of density:
\[
    \mtrue = \E[G(\atrue(X) + \btrue(X)) - G(\atrue(X))].
\]
With $X$ being scalar and continuous and nearly 30,000 observations, this is a fairly simple nonparametric problem. We estimate $\atrue(x)$ and $\btrue(x)$ using structured neural networks, as in Figure \ref{fig:arch-glm} (or Figure \ref{fig:arch-general} in the text more generally) using two hidden layers with 20 nodes each. For $\bL(x; \btheta)$ we use same network, but unstructured as this is regression. We additionally implement short stacking of \cite{Ahrens-Hansen-Schaffer-Wiemann2025_JAE} using neural nets, random forests, and linear regression, as an example. We use 20-fold cross fitting with three-way splitting and two-way splitting (where $\ahat(x)$, $\bhat(x)$, and $\bLhat(x; \bthat)$ are fit on the same data), but also no splitting at all. 

\begin{figure}
    \centering 
    \includegraphics[scale=0.5]{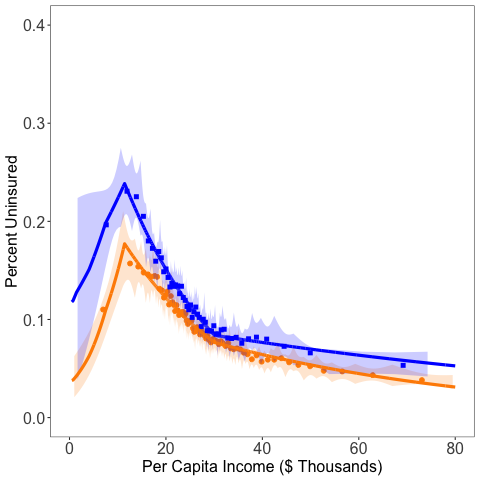}
    \caption{
        \label{fig:acs}
        {\bf Fractional Outcome Model with Binary Treatment -- First Stage.} This figure uses the ACS data of \cite{Cattaneo-Crump-Farrell-Feng2024_WP} to compare areas in low density states (blue) and high density states (orange). Low density states are defined as those with average population per square mile below 100. The dots and shaded region are the point estimates and robust bias corrected 95\% confidence bands from the original paper. The solid lines show structured neural network estimates.
    }
\end{figure}

The results are shown in Figure \ref{fig:acs} and Table \ref{table:acs}. Figure \ref{fig:acs} shows the first stage: the estimated $G(\ahat(x))$ and $G(\ahat(x) + \bhat(x))$ using neural networks (solid lines) and binscatter regressions (dots). This figure is made using the full sample. The shaded regions are robust bias corrected confidence bands for the binscatter regressions. The binscatter results are obtained using {\tt binsreg} {\sf R} package \citep{Cattaneo-Crump-Farrell-Feng2025_Stata} and are identical to Figure 4(a) of \cite{Cattaneo-Crump-Farrell-Feng2024_WP}. We see that the neural networks and binscatter regressions are similar estimates of the unknown functions. 

Table \ref{table:acs} shows the semiparametric inference results. The point estimates are consistent across the different approaches. The standard errors vary, and we see much smaller standard errors without cross fitting, most likely reflecting invalidity of this approach. It is possible that the cross-fitting based standard errors are larger due to the extra variation (across the samples) which might be reduced with repeated splitting and median aggregation. Note that the short-stacking standard errors are also valid for the row above, i.e., without short stacking. The bottom two rows, using three-way splitting, are the only results that are fully theoretically justified, however, as above, the others may be as well.

\begin{table}[h]
    \centering
    \caption{
        {\bf Fractional Outcome Model with Binary Treatment -- Semiparametric Inference.} This table shows point estimates and standard errors for semiparametric inference on $\mtrue = \E[G(\atrue(X) + \btrue(X)) - G(\atrue(X))]$, comparing uninsuredness rates between low and high density states, with heterogeneity in per capita income. Low density states are defined as those with average population per square mile below 100.
    }
    \begin{tabular}{lrr}
        \hline
        \textbf{Method} & \textbf{Point Estimate} & \textbf{Standard Error} \\
        \hline
        Plug-in  & 0.020      &      ---         \\
        No cross fitting    & 0.020     &      0.0015    \\
        2-way cross fitting     & 0.027    &      0.0041    \\
        \quad \ldots w/ short stacking  & 0.015     &      0.0041    \\
        3-way cross fitting     & 0.018     &      0.0045    \\
        \quad \ldots w/ short stacking  & 0.022     &      0.0045    \\
        \hline
    \end{tabular}
    \label{table:acs}
\end{table}

\subsection{Tobit}

The type I Tobit model is well-studied in the parametric case, and so it serves as a useful example to illustrate how existing knowledge can be used to interpret the assumptions required in the enriched case. 

The observed outcome is $Y = \max(0,Y^*)$, where in the parametric, homogeneous case $Y^*$ is Gaussian given $\bT=\bt$ with mean $\atrue + {\bbtrue}'\bt$ and variance $\sigma^2$. The enriched version has mean $\atrue(\bx) + \bbtrue(\bx)'\bt$ and variance $\sigma^2(\bx)$ (in practice it can help to fit $\sigma^2(\bx) = \exp\{\tilde{\sigma}(\bx)\}$). As usual, we work with the transformed parameters $\bttrue(\bx) = (\bttrue_1(\bx)',\ttrue_2(\bx))'$, with $\bttrue_1(\bx) = (\atrue(\bx), \bbtrue(\bx)')'/\sigma(\bx)$ and $\ttrue_2(\bx) = \sigma^{-1}(\bx)$. See \cite{Amemiya1985_book} and \cite{Wooldridge2010_book} for textbook treatments.

The gradient and Hessian are cumbersome but known. These can be used both for understanding the assumptions required but also, if desired, in the computation. Let $\One_0 = \One\{Y^* \leq 0\}$ and $\One_1 = \One\{Y^* > 0\}$. Let $\phi$ and $\Phi$ denote the Gaussian density and distribution functions. With $\bT_1 = (1,\bT')'$, the gradient (score) terms are
\[
    \bm{\ell}_{\btheta_1}(\bw, \btheta(\bx)) = \One_0 \frac{\phi(\btheta_1(\bx)'\bT_1) \bT_1 }{1 - \Phi(\btheta_1(\bx)'\bT_1)}  -  \One_1 \Big(\theta_2(\bx) y - \btheta_1(\bx)'\bT_1\Big)\bT_1'
\]
and 
\[
    \bm{\ell}_{\theta_2}(\bw, \btheta(\bx)) =  -  \One_1 \theta_2(\bx)^{-1}  +  \One_1 (\theta_2(\bx) y - \btheta_1(\bx)'\bT_1)y.
\]
The second derivatives are
\begin{gather*}
    \bm{\ell}_{\btheta_1 \btheta_1}(\bw, \btheta(\bx)) = - \One_0 \frac{\phi(\btheta_1(\bx)'\bT_1) (\btheta_1(\bx)'\bT_1)\bT_1\bT_1'}{1 - \Phi(\btheta_1(\bx)'\bT_1)}  +  \One_0 \frac{\phi(\btheta_1(\bx)'\bT_1)^2\bT_1\bT_1\bT_1'}{[1 - \Phi(\btheta_1(\bx)'\bT_1)]^2} + \One_1  \bT_1\bT_1' ,             \\
    \ell_{\theta_2 \theta_2}(\bw, \btheta(\bx)) = - \One_1 \theta_2(\bx)^{-2} + \One_1 y^2 ,
    \qquad \text{ and } \qquad
    \bm{\ell}_{\theta_1 \theta_2}(\bw, \btheta(\bx)) =  \One_1 y \bT_1.
\end{gather*}
That the gradients are conditionally mean zero can be directly verified. The matrix $\bL(\bx)^{-1}$ exists because $\btheta_1(\bx)'\bT_1  - \phi(\btheta_1(\bx)'\bT_1) / [1 - \Phi(\btheta_1(\bx)'\bT_1)] >0$, using exactly the logic from parametric models \citep{Donald_1990,Olsen1978_Ecma,Amemiya1985_book}. The remaining assumptions required would be standard for nonparametrics/ML, such as smoothness and boundedness.

\subsection{Instrumental Variables}

For multiple first stage objects, the influence function correction terms are generally additive \citep{Newey1994_Ecma}. A good example is when $\bT$ is endogenous and instrumental variables are available. Suppose there is single endogenous variable $T$ and a single instrument $Z$ and the researcher is applying two stage least squares. We enrich this to allow for fully flexible observed heterogeneity in the effects of the instrument and the endogenous variable, arriving at the two-equation model 
\begin{align}
    Y & = \ttrue_1(\bX) + \ttrue_2(\bX)T  + V,           \label{appxeqn:iv structural}\\
    T & = \ztrue_1(\bX) + \ztrue_2(\bX)Z  + U,           \label{appxeqn:iv first stage}
\end{align}
where $\E[V \mid \bX, Z] = \E[U \mid \bX, Z] = 0$. For estimation, and moreover, derivation of an orthogonal score, we simply plug \eqref{appxeqn:iv first stage} into \eqref{appxeqn:iv structural} to obtain the reduced form equation
\begin{align}
    \begin{split}
        \label{appxeqn:iv reduced form}
        & Y  = \atrue(\bX) + \btrue(\bX)Z  + \tilde{V},        \\
        & \atrue(\bx) = \ttrue_1(\bx) + \ttrue_2(\bx) \ztrue_1(\bx),    \quad    \btrue(\bx) = \ttrue_2(\bx) \ztrue_2(\bx),    \quad     \tilde{V} = \ttrue_2(\bX)U + V.
    \end{split}
\end{align}
This approach directly generalizes two-stage least squares to handle high-dimensional, complex heterogeneity, but notice that the linearity structure is maintained. To estimate the first stage parameter functions we simply apply (3.3) where the loss is the sum of two squared losses and the architecture in Figure \ref{fig:arch-glm} is used twice. The leading case for inference would be the average partial effect $\mtrue = \E[\ttrue_2(\bX)] = \E[\btrue(\bX)/\ztrue_2(\bX)]$. We again see the familiarity of the assumptions required: we need strong instruments, which here means we need $\ztrue_2(\bX)$ to be nowhere zero. This is a strong assumption, and may not be tenable in applications. A constant slope can be assumed to make this assumption more plausible, at the cost of restricting heterogeneity. Any other function $H(\bX, \atrue, \btrue, \ztrue_1, \ztrue_2; \btstar)$ could be used. 

The influence function will be of the standard form in this case, with essentially two copies of the linear case above. Define $\bttrue = (\atrue, \btrue, \ztrue_1, \ztrue_2)'$, $\bw = (y,t,z)$, $\bt_1 = (1,t)'$, $\bz_1 = (1,z)'$ and $\bm{I}_2$ the $2\times2$ identity matrix. Then
\begin{equation*}
    \setstretch{1}
    \bm{\ell}_{\btheta}(\bw, \bttrue(\bx)) = 
        - \begin{pmatrix}
            y - \atrue(\bx) - \btrue(\bx)z           \\
            t - \ztrue_1(\bx) - \ztrue_2(\bx)z  
	    \end{pmatrix} \otimes \bz
    \quad \text {and } \quad 
    \bm{\ell}_{\btheta\btheta}(\bw, \bttrue(\bx)) = \bm{I}_2 \otimes \bz_1\bz_1' .
\end{equation*}
Therefore $\bL(\bx) = \bm{I}_2 \otimes \bL_Z(\bx)$, where $\bL_Z(\bx) = \E[\bz_1\bz_1' \mid \bX = \bx]$. 

As before, our score is not the only possibility for estimation and inference IV models. Our approach aims for ease of use and transparency, both by sticking to the two stage least squares and by the structural compatibility of deep learning. Restricting to homogeneous effects, \cite{Chernozhukov-etal2018_EJ} study partially linear IV models and study three different possibly orthogonal scores, each requiring different functions in the first step and in the correction term. The same could be done in our unrestricted model. As in partially linear models, if any parameters are constant this can be enforced in the architecture. It is not obvious which approach is best.

\section{Simulation Study}
    \label{appx:simuls}

We conducted a simulation exercise to validate the automatic differentiation approach, study the finite-sample coverage of the confidence intervals, and compare performance with both generalized random forests \citep{Athey-Tibshirani-Wager2019_AoS} and auto-DML (see \cite{Chernozhukov-etal2024_WP} and references therein, and Remark \ref{rem:auto DML} in the main text for more discussion). 
To make these comparisons and validations feasible, so that all approaches are applicable and the influence function is known, we study the case of average effects of a binary treatment under the standard conditions of overlap, unconfoundedness, SUTVA, and consistency. See Section \ref{appx:ATE} for more details.

\begin{remark}
    \cite{Hetzenecker-Osterhaus2024_WP} conduct a thorough simulation study of our methodology in context of discrete choice models. The use the special case of multinomial logistic regression and study estimation, inference, and implementation issues. A major lesson of that work is the important role played by regularization, which as discussed in the main text, can be a key ingredient in semiparametric estimation and inference in general and is one reason why characterizing $\bLtrue(\bx)$ can be useful. 
\end{remark}

\subsection{Model and Set Up}

We consider a scalar outcome $Y$ and binary treatment variable $T = \{0,1\}$. Let $Y(t)$ be the potential outcome under treatment $T = t$. Under these assumptions, it is actually without loss of generality that $\E[Y \mid \bx, t] = \atrue(\bx) + \btrue(\bx) t$, so we use this model (this is \eqref{eqn:glm flm2}, with $G(u) = u$) where the parameter functions are $\bttrue(\bx) = (\atrue(\bx), \btrue(\bx))'$ and in particular $\btrue(\bx) = \E[Y(1) - Y(0) \mid \bx]$ is the conditional average treatment effect (CATE) function (Remark \ref{rem:CATE}). The second-step parameter of interest is the average treatment effect $\mtrue = \E[Y(1) - Y(0)] = \E[\bH(\bx,\bttrue; \tstar)]$, with $\bH(\bx,\btheta; \tstar) = \beta$. In this case, the influence function in our framework matches the efficient one of \cite{Hahn1998_Ecma} that is most commonly used. The overlap assumption ensures that $\bLtrue(\bx)^{-1}$ is well behaved: the determinant of $\bLtrue(\bx) = p(\bx)(1-p(\bx))$, with the propensity score $p(\bx) := \P[T = 1 \vert \bX = \bx] = \lambda_1(\bx; \btheta) = \lambda_2(\bx; \btheta)$. 

For discussion comparing the approaches, recall that the influence function is $\psi(\by,\bt,\bx, \btheta, \bL) - \mtrue$ where, respectively in the general form and specific to this problem (see Section \ref{appx:ATE}),
\begin{align*}
    & \psi(\by,\bt,\bx, \btheta, \bL)  = \beta(\bx)   + \bH_{\btheta}\left(\bx, \btheta(\bx ); \btstar\right) \bL(\bx; \btheta)^{-1} \bm{\ell}_{\btheta}(y,\bt, \btheta(\bx))           \\
    & = \beta(\bx) +  \frac{ ( t - p(\bx) )  }{p(\bx) - p(\bx)^2}  (y - \alpha(\bx) - \beta(\bx) t ) ,
\end{align*}
Between these two sit the forms given in Equations \eqref{eqn:glm IF} and \eqref{eqn:IF univariate}, which we do not explicitly use in the simulations. 

With these two forms in hand, we can elucidate the comparisons we make, which differ in both the first and second stage. We mainly study four different approaches, with some additional variants as discussed below. 
\begin{enumerate}

    \item Use the second form of the influence function, i.e., the known form, with structural deep neural networks for the parameter functions $\bttrue(\bx) = (\atrue(\bx), \btrue(\bx))'$ and standard networks for estimation of $\lambda_1(\bx; \btheta) = \lambda_2(\bx; \btheta) = p(\bx)$. This follows our work, and \cite{Farrell-Liang-Misra2021_Ecma}, without automatic differentiation. That is, the correction term is estimated directly using
    \[\frac{ ( t - \widehat{p}(\bx) )  }{\widehat{p}(\bx) - \widehat{p}(\bx)^2}  (y - \ahat(\bx) - \bhat(\bx) t ),\]
    for neural network estimators $\widehat{p}$, $\ahat$, and $\bhat$.

    \item Use the first form of the influence function, i.e., the fully generic format, and use automatic differentiation to obtain both $\bH_{\btheta}\left(\bx, \btheta(\bx ); \btstar\right)$ and the regressands required for estimation of $\bL(\bx; \btheta)$. All nonparametric estimation remains neural networks. This approach differs from the above only in the use of automatic differentiation instead of using the known form of the influence function, and is thus, from an inference point of view, fully automatic. That is, the correction term is estimated as
    \[
        \bH_{\btheta}\left(\bx, \bthat(\bx ); \btstar\right) \bLhat(\bx; \bthat)^{-1} \bm{\ell}_{\btheta}(y,\bt, \bthat(\bx)),
    \]
    where the first-step estimators $\bthat = (\ahat, \bhat)'$ are the same, $bH_{\btheta}\left(\bx, \bthat(\bx ); \btstar\right)$ is computed using these and automatic differentiation, and the four entries $\lambda_{j,m}$, $(j,m) \in \{1,2\}^2$, are obtained using automatic differentiation and nonparametrically regressed on $\bx$.

    \item Use the auto-DML method to obtain the influence function correction term. The first-step estimators $\bthat = (\ahat, \bhat)'$ are again the same, but now, following \cite{Chernozhukov-etal2024_WP}, it is known that $\psi(\by,t,\bx, \btheta, \bL)$ can be writted as $\beta(\bx) + R(\bx,t) (y - \alpha(\bx) - \beta(\bx) t)$ where the Riesz representer $R(\cdot)$ is known to obey
    \[
        R = \argmin_{r \in \mathcal{R}} \E\left[ -2 H(\bX, r(\bx, \bt)) + r(\bx,\bt)^2 \right],
    \]
    for an appropriate function class $\mathcal{R}$. We then solve the empirical analogue of this problem using neural networks, so that for a class of networks $\cG_{\rm DNN}$, which is not structured,
    \[
        \widehat{R} = \argmin_{r \in \cG_{\rm DNN}} \frac{1}{n}\sumi \left[ -2 H(\bX, r(\bx, t)) + r(\bx,t)^2 \right].
    \]
    Then the correction term is estimated as
    \[
        \widehat{R}(\bx,t) (y - \ahat(\bx) - \bhat(\bx) t ).
    \]
    To connect to the above, recall that $\bm{\ell}_{\btheta}(y,\bt, \btheta(\bx)) =  \bT_1(G(\btheta(\bx)'\bt_1)-y)$, and so
    \[
        R(\bx,t) = \bH_{\btheta}\left(\bx, \btheta(\bx ); \btstar\right) \bL(\bx; \btheta)^{-1}\bt_1 = \frac{ ( t - p(\bx) )  }{p(\bx) - p(\bx)^2}.
    \]
    These first three methods use the different forms above to obtain the influence function correction term.

    \item Use the second form of the influence function, i.e., the known form, with random forests for nonparametric estimation, instead of neural networks as in the above three. The generalized random forest package \citep{Athey-Tibshirani-Wager2019_AoS} has the built-in function \verb|average_treatment_effect| for estimation of the ATE, which uses forests in the first step, fit via the \verb|causal_forest| function. 
    
\end{enumerate}

\subsection{Data Generating Process and Estimation}

The data generating process for the simulations is summarized as follows. We draw $n=5,000$ i.i.d. samples according to the following.
\begin{itemize}
    \item $X$ is scalar and distributed uniformly from -1/2 to 1/2.
    \item $T \in \{0,1\}$ with $p(x) = \Phi(4x^3)$, where $\Phi(\cdot)$ is the standard normal distribution function.
    \item $Y = \atrue(X) + \btrue(X)T + \varepsilon$, where $\varepsilon$ is normal with mean zero and heteroskedastic variance $(1 + [x^2 + t^2]/10)$.
    \item The parameter functions are $\atrue(x) = \sin(5x) + 0.5$, and $\btrue(x) =  -\cos(5x) - 0.5$.
\end{itemize}

Simulation and estimation is done in {\sf R}. All first-step neural networks (for $\bthat$, $\bLhat$, and $\widehat{R}$) use two hidden layers with twenty nodes each, trained for 500 epochs, implemented in \verb|torch| \citep{Keydana2023_book}. Generalized random forests use \verb|grf| with default settings \citep{Athey-Tibshirani-Wager2019_AoS}. We use $K=5$ folds for cross-fitting, when applicable. All simulations are done with 2,000 replications. Similar results are found for other architectures and longer training, using more folds for cross-fitting, other sample sizes, and with short stacking \citep{Ahrens-Hansen-Schaffer-Wiemann2025_JAE}.

\subsection{Results}

The main results are shown in Table \ref{table:simuls table}. Here we show the empirical (Monte Carlo) coverage and length of nominal 95\% confidence intervals from each method as described above. We see that all neural network based methods attain close to nominal coverage and have similar length. This validates the automatic differentiation approach and shows it is competitive with auto-DML. Cross fitting is not required for validity in this case, which is borne out by the results, although the cross-fitted versions are very slighly closer to nominal 95\%. The generalized random forests have slightly above nominal coverage and slightly longer intervals, but are still quite accurate. 

\begin{figure}
    \begin{center} 
    \begin{subfigure}[t]{\columnwidth}
         \includegraphics[scale=0.55]{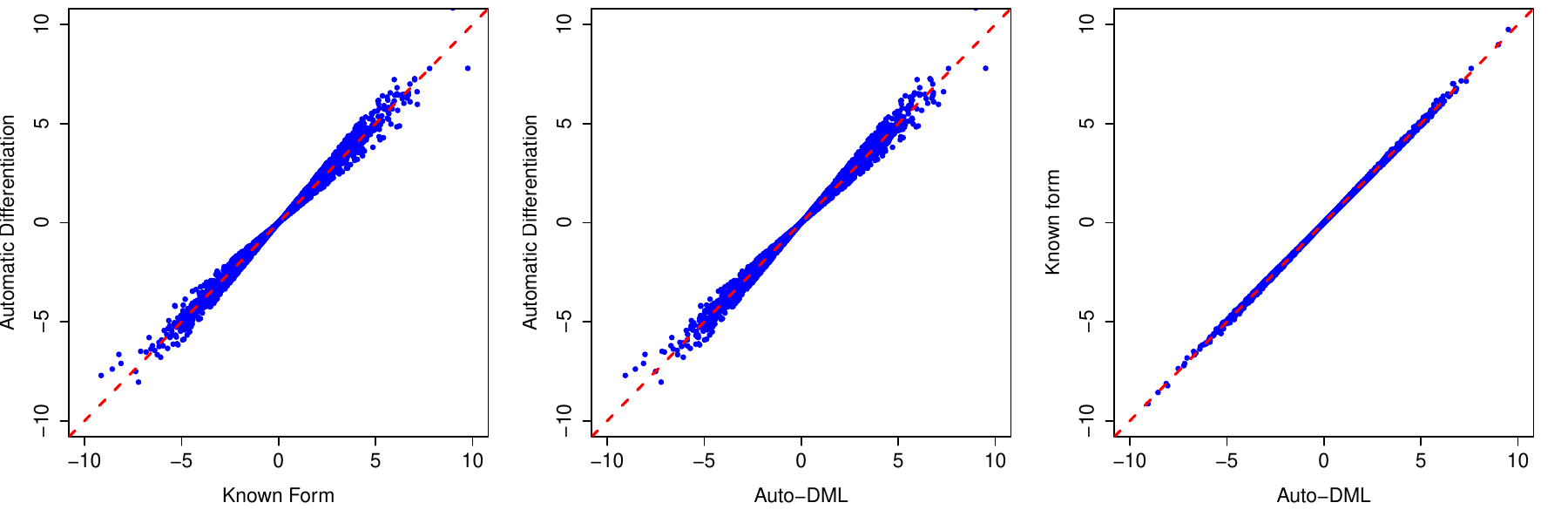}
    \end{subfigure}
    \begin{subfigure}[t]{\columnwidth}
         \includegraphics[scale=0.55]{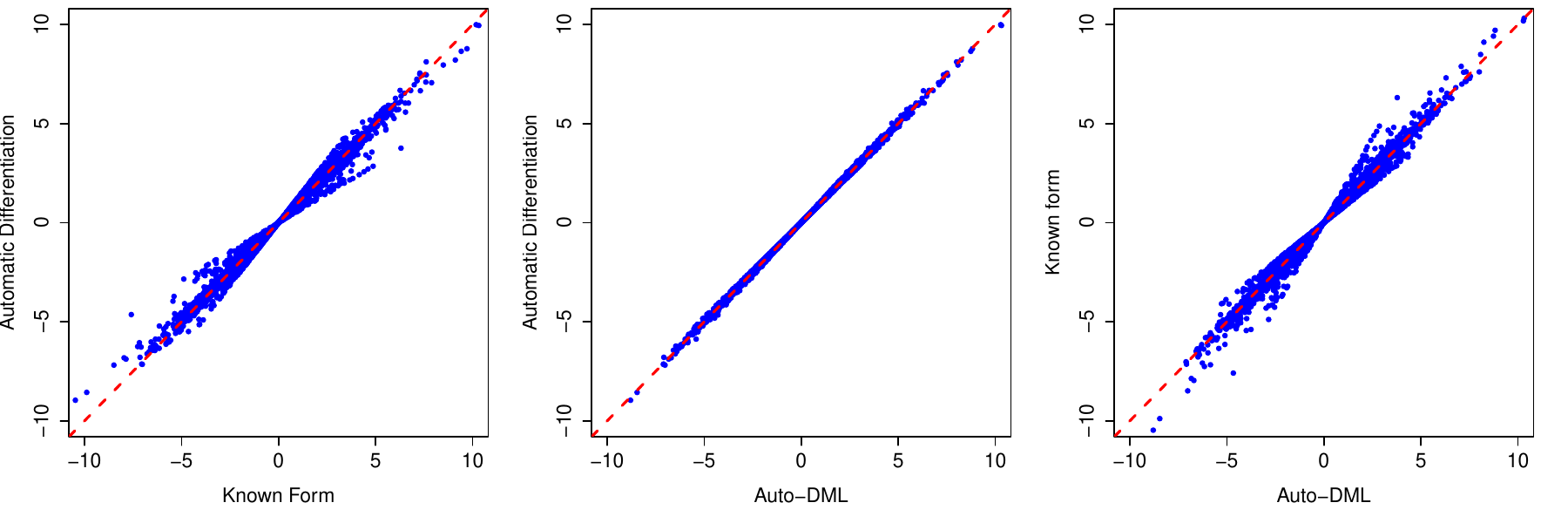}
    \end{subfigure}
    \begin{subfigure}[t]{\columnwidth}
         \includegraphics[scale=0.55]{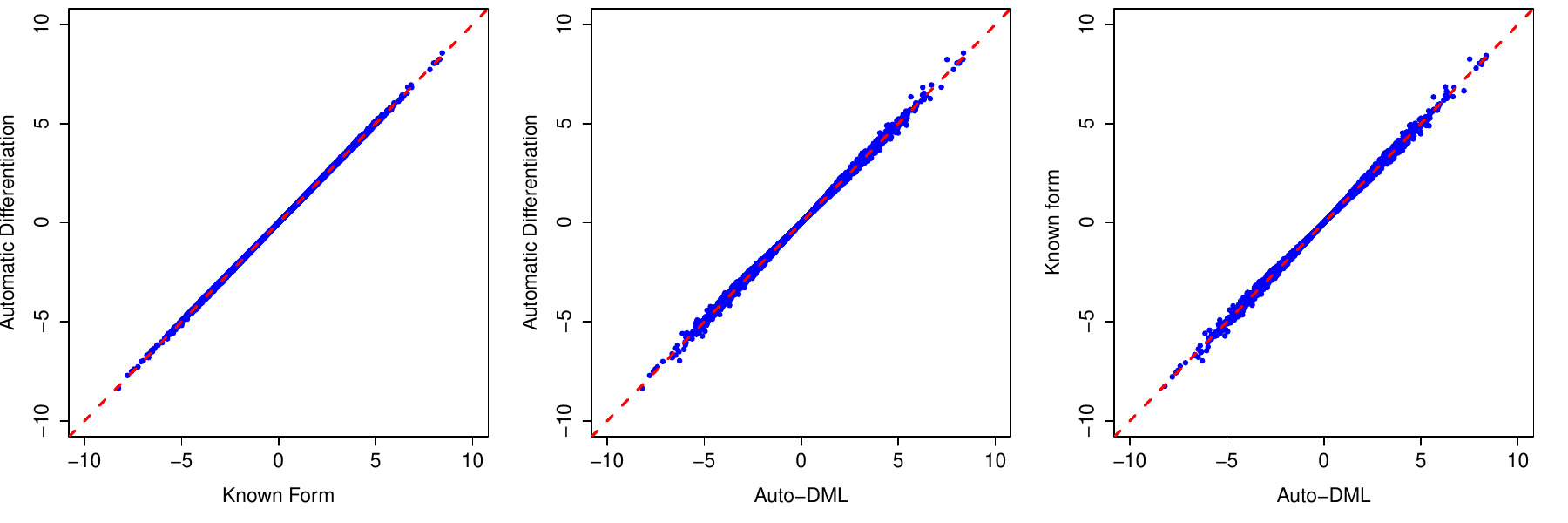}
    \end{subfigure}
    \caption{
        \label{fig:simuls fig compare}
        {\bf Comparing Correction Term Methods.} This figure compares the three different neural network based estimates of the correction term of the influence function for three representative Monte Carlo replications (i.e. row is from a single data set of size $n=5,000$). The full sample is used. The (red) dashed line is the ``y = x'' line indicating perfect agreement.
    }
    \end{center}
\end{figure}

Figure \ref{fig:simuls fig compare} compares the three neural network based methods in how well they capture the correction term of the influence function, as described above. Each row shows one data set (i.e. one Monte Carlo draw), and the three in total are representative of the patterns across draws. In this case, the first entry of $\bLtrue(\bx)$ is exactly one, i.e. $\lambda_{1,1}(\bx, \btheta) = \lambda_0(\bx; \btheta) \equiv 1$, but the automatic differentiation method does not use this: the automatic differentiation in this case returns a column of ones which is then nonparametrically regressed on the covariates, slightly increasing the noise. However, this has no impact on coverage as above. This can be seen best in the first row/replication. In the second row, automatic differentiation nearly perfectly agrees with auto-DML, and both are noisy relative to the known form.  In the third replication, the agreement is near exact across the methods.

\begin{figure}
    \centering 
    \includegraphics[scale=0.5]{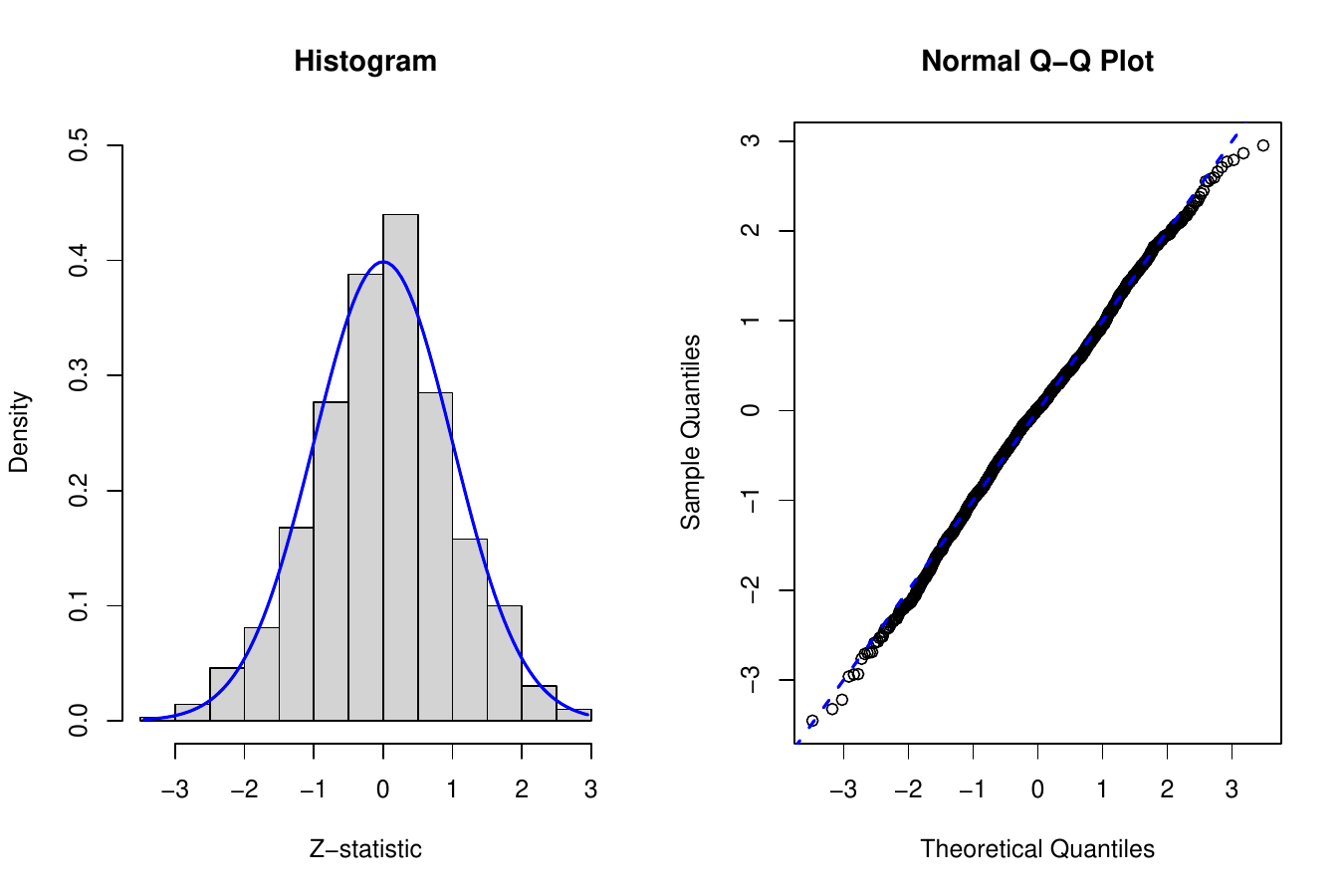}
    \caption{
        \label{fig:simuls fig dist}
        {\bf Sampling Distribution.} This figure shows the Monte Carlo sampling distribution for $\sqrt{n}\Psihat^{-1/2}(\mhat - \mtrue)$ over 1000 replications. Estimation is based on automatic differentiation and 5-fold cross-fitting. See full details in text. The left panel shows the histogram with the standard normal reference in blue while the right compares the quantiles of each in a Q-Q plot. 
    }
\end{figure}

Figure \ref{fig:simuls fig dist} shows an example of the sampling distribution of the centered and scaled estimator $\sqrt{n}\Psihat^{-1/2}(\mhat - \mtrue)$, where estimation and inference are based on automatic differentiation and five-fold cross fitting. This figure confirms that Theorem \ref{thm:normality} yields a good approximation. The sampling distribution is nearly identical to the standard normal benchmark, as seen in both the histogram and quantile-quantile comparison. 

In sum, the proposed estimator performs well. The simulations confirm that automatic differentiation provides a valid and competitive tool when implementing double machine learning.

\begin{table}[ht]
    \renewcommand{\arraystretch}{1.12}
    \begin{center}
    \caption{
        {\bf Empirical Coverage and Length}
        \label{table:simuls table}
    }
    \begin{tabular}{lrr}
    Method & Coverage & Length \\
    \hline
    \multicolumn{3}{l}{\bfseries Neural Network First Step} \tabularnewline
    ~~\emph{Full Sample}&&\tabularnewline
    ~~~Known Form & 0.940 & 0.124\tabularnewline
    ~~~Automatic Differentiation & 0.9405 & 0.124\tabularnewline
    ~~~Auto DML & 0.938 & 0.123\tabularnewline
    ~~\emph{Cross Fitting}&&\tabularnewline
    ~~~Known Form & 0.940 & 0.126\tabularnewline
    ~~~Automatic Differentiation & 0.944 & 0.126\tabularnewline
    ~~~Auto DML & 0.9395 & 0.124\tabularnewline
    \hline
    \multicolumn{3}{l}{\bfseries Random Forest First Step} \tabularnewline
    ~~~Generalized Random Forests & 0.9575 & 0.135\tabularnewline
    \hline
    \end{tabular} 
    \end{center}
\end{table}

\end{appendices}

\end{document}